%% file: sn-article.tex
\documentclass[pdflatex,sn-mathphys]{sn-jnl}

\jyear{2023}%

\theoremstyle{thmstyleone}%

\theoremstyle{thmstyletwo}%

\theoremstyle{thmstylethree}%

\usepackage{siunitx}
\usepackage{amsmath,amstext,color,tabu,mathtools}
\usepackage{gensymb}
\usepackage{natbib}
\usepackage{hyperref}
\usepackage{longtable}
\usepackage{graphicx}
\usepackage{tablefootnote}
\usepackage{mathrsfs}
\usepackage{comment}
\usepackage[mathlines,displaymath]{lineno}

\begin{document}

\input{journal-abbrev}

\title[A bright MeV line in GRB~221009A]{A bright megaelectronvolt emission line in $\gamma$-ray burst GRB 221009A}

\author*[1,2]{\fnm{Maria Edvige} \sur{Ravasio}}\email{mariedvige.ravasio@ru.nl}

\author*[2,5]{\fnm{Om Sharan} \sur{Salafia}}\email{om.salafia@inaf.it}
\author*[3,4]{\fnm{Gor} \sur{Oganesyan}}\email{gor.oganesyan@gssi.it}
\author[3,4]{\fnm{Alessio} \sur{Mei}}
\author[2,5]{\fnm{Giancarlo} \sur{Ghirlanda}}
\author[6,7]{\fnm{Stefano} \sur{Ascenzi}}
\author[3,4]{\fnm{Biswajit} \sur{Banerjee}}
\author[3,4]{\fnm{Samanta} \sur{Macera}}
\author[3,4]{\fnm{Marica} \sur{Branchesi}}
\author[1,9]{\fnm{Peter G.} \sur{Jonker}}
\author[1,10]{\fnm{Andrew} \sur{Levan}}
\author[1,11,12]{\fnm{Daniele Bj{\o}rn} \sur{Malesani}}
\author[1,13]{\fnm{Katharine B.} \sur{Mulrey}}
\author[8]{\fnm{Andrea} \sur{Giuliani}}
\author[14,2,15]{\fnm{Annalisa} \sur{Celotti}}
\author[2]{\fnm{Gabriele} \sur{Ghisellini}}

\affil[1]{\orgdiv{Department of Astrophysics/IMAPP}, \orgname{Radboud University}, \orgaddress{\street{6525 AJ}, \city{Nijmegen}, \country{The Netherlands}}}

\affil[2]{\orgname{INAF} -- \orgdiv{Osservatorio Astronomico di Brera}, \orgaddress{\street{via E. Bianchi 46}, \postcode{23807}, \city{Merate (LC)}, \country{Italy}}}

\affil[3]{Gran Sasso Science Institute, Viale F. Crispi 7,I-67100,L’Aquila (AQ), Italy}

\affil[4]{\orgname{INFN} -- \orgdiv{Laboratori Nazionali del Gran Sasso}, I-67100, L’Aquila (AQ), Italy}

\affil[5]{\orgname{INFN} -- \orgdiv{Sezione di Milano-Bicocca}, \orgaddress{\street{Piazza della Scienza 3}, \postcode{20146}, \city{Milano (MI)}, \country{Italy}}}

\affil[6]{Institute of Space Sciences (ICE, CSIC), Campus UAB, Carrer de Can Magrans s/n, E-08193, Barcelona, Spain}

\affil[7]{Institut d’Estudis Espacials de Catalunya (IEEC), Carrer Gran Capità 2-4, E-08034 Barcelona, Spain}

\affil[8]{\orgname{INAF} -- \orgdiv{Istituto di Astrofisica Spaziale e Fisica Cosmica}, \orgaddress{\street{Via Alfonso Corti, 12}, \postcode{I-20133}, \city{Milano}, \country{Italy}}}

\affil[9]{SRON, Netherlands Institute for Space Research, Niels Bohrweg 4, 2333~CA, Leiden, The Netherlands}

\affil[10]{Department of Physics, University of Warwick, Coventry, CV4 7AL, UK}

\affil[11]{Cosmic Dawn Center (DAWN), Denmark}

\affil[12]{Niels Bohr Institute, University of Copenhagen, Jagtvej 128, 2200
Copenhagen N, Denmark}

\affil[13]{\orgname{Nikhef}, \orgaddress{Science Park Amsterdam}, \postcode{1098 XG }, \city{Amsterdam}, \country{The Netherlands}}

\affil[14]{\orgname{SISSA}, \orgaddress{\street{via Bonomea, 265}, \postcode{I-34136}, \city{Trieste}, \country{Italy}}}

\affil[15]{\orgname{INFN} -- \orgdiv{Sezione di Trieste}, \orgaddress{\street{Via Valerio, 2}, \postcode{I-34127}, \city{Trieste}, \country{Italy}}}

\abstract{
The highly variable and energetic pulsed emission of a long gamma--ray burst (GRB) is thought to originate from local, rapid dissipation of kinetic or magnetic energy within an ultra-relativistic jet launched by a newborn compact object, formed during the collapse of a massive star.
The spectra of GRB pulses are best modelled by 
power-law segments, indicating the dominance of non-thermal radiation processes.
Spectral lines in the X-ray and soft $\gamma$-ray regime for the afterglow have been searched for intensively, but never confirmed. No line features ever been identified in the high energy prompt emission.
Here we report the discovery of a highly significant ($> 6 \sigma$)  
narrow emission feature at around $10$ MeV in the brightest ever GRB 221009A. By modelling its profile with a Gaussian, we find a roughly constant width $\sigma \sim 1$ MeV and temporal evolution
both in energy ($\sim 12$ MeV to $\sim 6$ MeV) and luminosity ($\sim 10^{50}$ erg/s to $\sim  2 \times 10^{49}$ erg/s) over 80 seconds. We interpret this feature as a blue-shifted annihilation line of relatively cold ($k_\mathrm{B}T\ll m_\mathrm{e}c^2$) electron-positron pairs, which could have formed within the jet region where the brightest pulses of the GRB were produced. 
A detailed understanding of the conditions that can give rise to such a feature could shed light on the so far poorly understood GRB jet properties and energy dissipation mechanism.

}


\maketitle
Gamma-ray bursts (GRBs) are powerful cosmological transient phenomena appearing as brief (from a fraction of a second up to several hundreds of seconds) energetic flashes of kiloelectronvolt to megaelectronvolt (keV-MeV) radiation at random positions in the sky.  
During the intense and highly variable $\gamma$-ray radiation phase, termed `prompt emission', an enormous amount of energy is released, typically of the order of a few $\times 10^{52} - 10^{53}$ erg, if emitted isotropically.
Tens of years of data collection and theoretical studies paved the way for the elaboration of a widely accepted scenario, that links these powerful sources to a newborn stellar-mass black hole.
The extraction of rotational energy from the black hole powers a  
relativistic jet: the prompt emission is thought to be produced by the conversion of a small fraction of the jet kinetic or magnetic energy into radiation \cite{Rees1994,Sari1996}.

Despite being intensively studied for tens of years, the physics of the prompt emission remains poorly understood. Among many open and highly debated questions, the dominant form of energy in the relativistic jet is still unknown, 
as is the nature of the radiative process responsible for the observed radiation. 
The prompt emission spectrum is typically fitted by two power-laws with slopes $\alpha$ and $\beta$ smoothly connected at a `peak' photon energy $E_{\rm peak}$ (where most of the power is emitted). A detailed broadband modeling of the whole spectral shape, when possible, can reveal fundamental deviations from the typical continuum double power-law spectrum, 
which can help in unveiling key features of the underlying physical processes. The recently discovered presence of an additional low-energy spectral break provided strong observational support for the synchrotron origin of the emission (e.g.\ \cite{Oganesyan2018} and \cite{Ravasio2019}).
On the high-energy side, the study of the spectrum up to MeV-GeV energies allowed the discovery of a high-energy spectral softening (in the form of an exponential cutoff) from a few tens to hundreds of MeV (e.g.\ \cite{Ackermann2012, Vianello2018}), which has been interpreted as a sign of photon--photon absorption (producing electron-positron pairs) in the prompt emitting region.\\

On the 9th of October 2022, the Gamma-Ray Burst Monitor onboard the \textit{Fermi} spacecraft (\textit{Fermi}/GBM) was triggered by GRB 221009A, the brightest GRB observed so far (with a reported fluence of $F \sim0.2 \, \rm erg/cm^{2}$ \cite{Frederiks2023,An2023,Burns2023}). The redshift of the host galaxy of GRB 221009A ($z=0.151$) has been measured thanks to the X-shooter spectrograph at the Very Large Telescope \cite{Malesani2023}. The distance of this GRB and the measured fluence and flux at the brightest pulse of GRB 221009A result in extreme values for the intrinsic isotropic equivalent energy and peak luminosity of $E_{\rm iso} \sim 10^{55}$ erg and $L_{\rm peak, iso} \sim 10^{54}$ erg/s\, \cite{Frederiks2023,An2023,Burns2023}. Given the unprecedented flux of this GRB, the data of most $\gamma$-ray satellites, including \textit{Fermi}/GBM, have been affected by saturation (dead-time and pile-up) effects. While the analysis of \textit{Fermi}/GBM data at the brightest pulses has been discouraged, the relatively less bright portions of the prompt emission offer a unique probe of the spectral properties of GRBs beyond the standard double power-law function, owing to the broad spectral range covered by GBM (8 keV - 40 MeV) and to the high signal-to-noise ratio.  

\begin{figure}[ht]
    \centering
    \includegraphics[width=\textwidth]{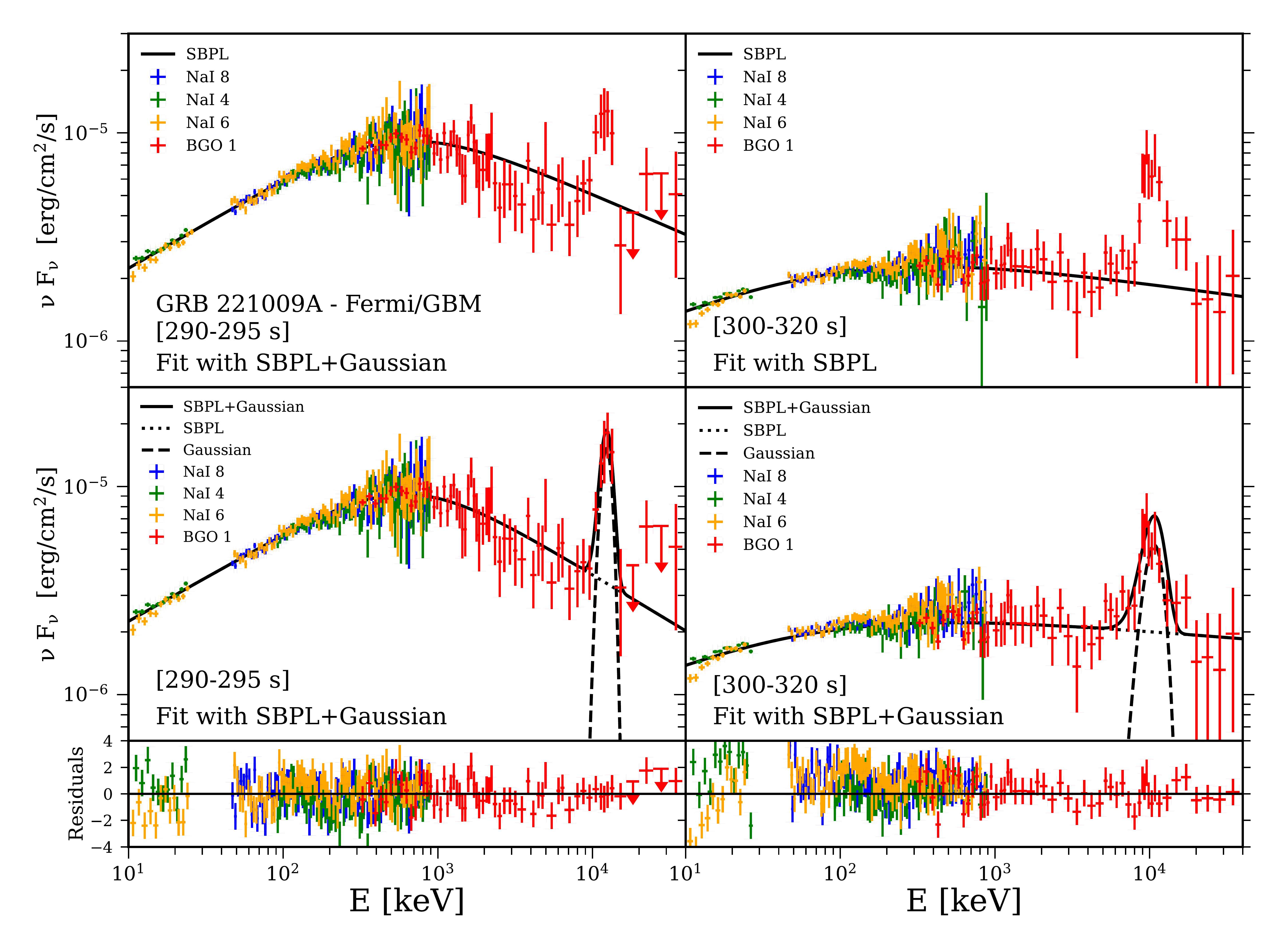}
    \caption{\textbf{Top: } Spectra of GRB 221009A in the time intervals 290-295 s and 300-320 s, fitted with a standard GRB empirical continuum model (SBPL), in the $\nu F_{\nu}$ representation. The narrow feature appears as an excess around $\sim 12-10$ MeV in the Bismuth Germanate (BGO) detector data (red crosses).
    \textbf{Bottom: } Same spectra fitted with the SBPL plus the addition of a Gaussian component, to model the observed excess. In both panels, data points have been re-binned for graphical purposes. Data points refers to multiple NaI detectors (as shown in the legend) and one BGO. 
    }
    \label{fig:spectrum5}
\end{figure}

We performed a time-resolved analysis of \textit{Fermi}/GBM data from 0 up to 460 s after the GBM trigger time. Following the Fermi Collaboration caveats regarding the analysis of this burst\footnote{\url{https://fermi.gsfc.nasa.gov/ssc/data/analysis/grb221009a.html}}, we excluded the time interval affected by saturation (called Bad Time Interval, BTI), namely 219-277 s.  
Our analysis of the spectral data ranging from 280 s to 320 s reveals the presence of a distinct, bright, narrow emission feature at around 10 MeV, which is well modelled by a Gaussian, on top of the typical prompt emission spectrum, modelled with a smoothly broken power-law (SBPL, see Methods). Figure \ref{fig:spectrum5} shows, as an example, the stark improvement in the description of the data obtained by  adding the Gaussian component on top of the SBPL in the two time bins 290-295 s and 300-320 s. 

\begin{figure}[ht]
    \centering
    \includegraphics[width=\textwidth]{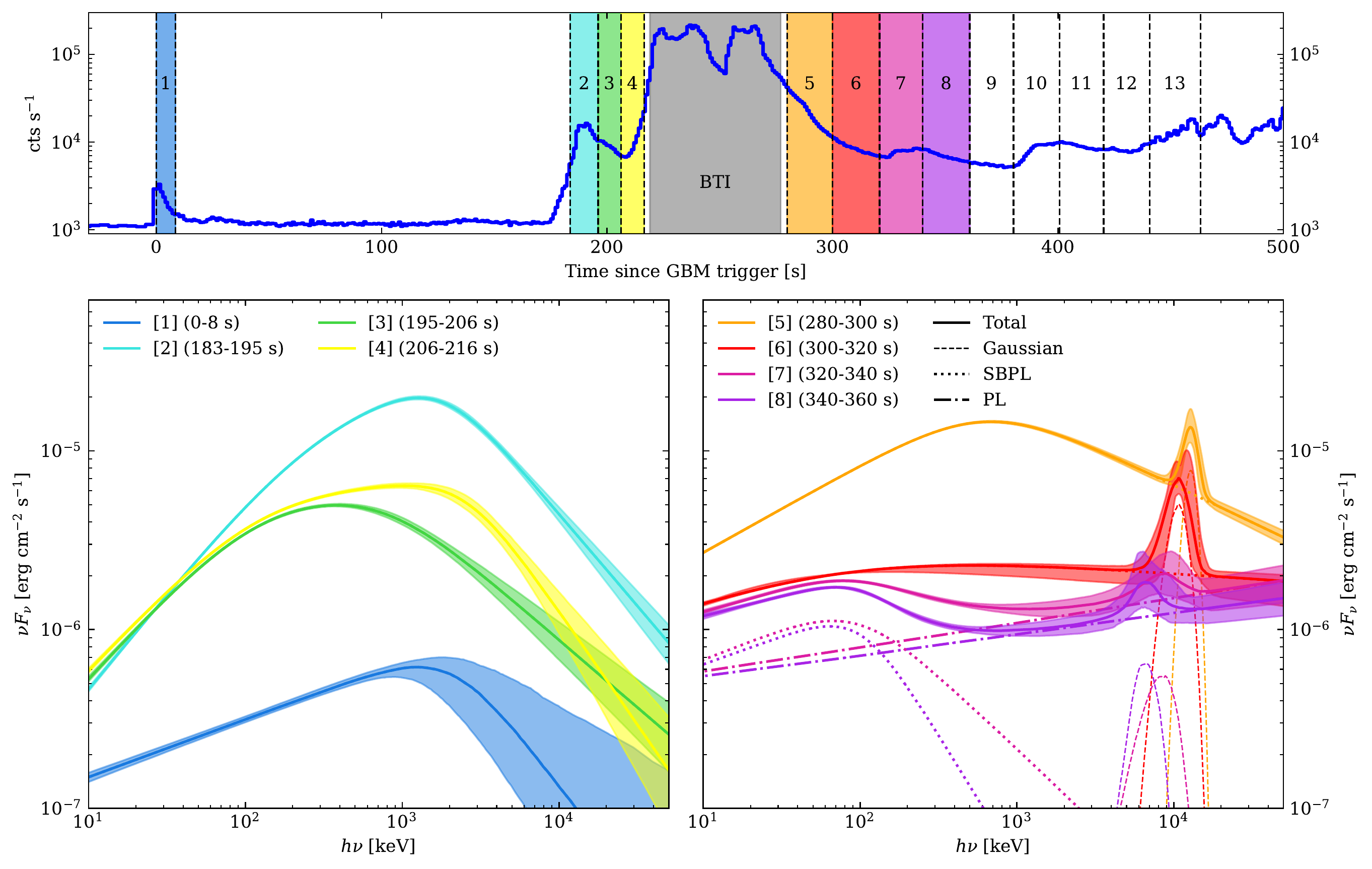}
        \caption{\textbf{Top:} Count rate light curve of GRB~221009A (blue solid line), as detected by \textit{Fermi}/GBM in the energy band 8-900 keV, along with the 13 time intervals analysed in this work (dashed black vertical lines) and the time interval excluded from the analysis (i.e.\ BTI, represented by the grey area, see text for details). The first eight time intervals analysed have been color-coded to match the corresponding colored spectra shown in the bottom panels.
    \textbf{Bottom Left:} medians (solid lines) and 90\% credible intervals (shaded areas) of the posterior distributions of $\nu F_{\nu}$ at each fixed frequency $\nu$, for the first 4 time intervals (spectra 1, 2, 3 and 4, color-coded as shown in the top panel), characterized by the typical GRB continuum prompt spectral shape (either 2SBPL or SBPL models, see text and Extended Data Table~\ref{tab2:prompt}).
    \textbf{Bottom Right: } 
    same as the left-hand panel, but for the 4 time intervals after the BTI (i.e.\ spectra 5, 6, 7 and 8), showing the evolution of the Gaussian model of the narrow feature.}
    \label{fig:lc_and_spectra}
\end{figure}

Fig.~\ref{fig:lc_and_spectra} (top panel) shows the raw light curve of GRB 221009A, as recorded by one of the sodium iodide (NaI) detectors, along with 13 selected time intervals (see also Extended Data Table~\ref{tab2:prompt}), separated by black dashed vertical lines and highlighted in different colours. The locations and widths of the intervals have been chosen based on the behaviour of the variable emission (e.g.\ ignoring quiescent times). We fit the corresponding spectra 
using a range of models (see the Methods). 
The evolution of the 8 selected spectra covering the first 360 s of emission (except for the BTI) in the 10 keV - 40 MeV range is shown in two panels in the bottom part of Fig.~\ref{fig:lc_and_spectra}.   
The left-hand panel displays the evolution of the model spectra in four time intervals before the brightest part of the light curve, where we find no evidence for the presence of the narrow emission feature. The right-hand panel shows the evolution of the spectrum in the 4 time intervals after the BTI, with dashed lines showing the contribution of the Gaussian component that models the narrow feature. The shape of the continuum component is similar to that observed in the time intervals before the BTI (dotted lines). The central photon energy of the narrow feature evolves over time, shifting down in energy, from an initial value of $12.56\pm0.30$ MeV to $6.12_{-0.59}^{+0.74}$ MeV, while its luminosity decreases from $(1.12 \pm 0.20)\times 10^{50}$ erg/s to $(2.1\pm 0.10)\times 10^{49}$ erg/s. No trend is observed in the evolution of the width over time.

The presence of the additional narrow emission feature
is strongly favoured by the Akaike Information Criterion \cite{Akaike1974} in both the 280-300 s and the 300-320 s time intervals analysed: the difference in the value of the criterion ($\Delta \rm AIC = 49$ and $141$, respectively, between the model with and without the Gaussian component) corresponds, in the context of model selection \cite{Burnham2004}, to a 6.6 and 11.6 sigma-equivalent significance. Moreover, the fact that the feature is found in multiple time bins further increases its significance.
In the following two time intervals up to 360 s, number 7 and 8, the prompt emission is fading, revealing also the presence of a sub-dominant power-law component, which we attribute to the rising GRB afterglow, as already seen in the BGO energy range of other bursts \cite{Meszaros1993,Sari1998,Ravasio2019b}.
While the spectral feature appears to be still present in the data of these two time intervals, the presence of the extra power law and the possible fading of the narrow component act as confounding factors, and the AIC test does not yield a clear support for the presence of the narrow feature anymore (see Table \ref{tab1:line} and Methods). On the other hand, assuming the feature to be present, the fit procedure yields parameters that are still relatively well constrained (see Methods) and show a consistent evolution towards lower photon energies and luminosities. In particular, the luminosity must have decreased by around (at least) a factor of two over 20 s from time interval 6 to 7. 
The properties of the Gaussian model of the feature derived from our fits (the luminosity $L_{\rm gauss}$, the central photon energy $E_{\rm gauss}$ and the width $\sigma_{\rm gauss}$) are reported in Table~\ref{tab1:line}.

In order to investigate further the presence of the discovered feature and further characterize its evolution, we sub-divided the two time intervals where the line is most significant, namely 280-300~s and 300-320~s. Given the higher signal-to-noise of the spectrum in the 5$^{\rm th}$ time interval analyzed (280-300 s), we subdivided this 20~s time bin into four bins of 5~s, while the 6$^{\rm th}$ time interval (300-320 s) was subdivided into two 10 s bins. 
In each of the six finer time intervals the feature remains clearly visible, and it shifts from around 14 MeV down to 9.8 MeV. With the exception of the first two time bins, 
the model with the Gaussian component over the SBPL is preferred with high significance also in these finer bins ($\Delta \rm AIC = 42, 5, 45, 36$, respectively).

The narrow feature is found in data outside the BTI flagged by the Fermi Collaboration. The spectral data in a similar time interval, from 277 s to 324 s,  have been analyzed also by \cite{Lesage2023}, and an excess above the fitted continuum is indeed visible in the BGO data (see the top right panel of their Fig. 5).
Nonetheless,
we investigated a possible instrumental origin (see Supplementary Information for details), 
 finding no evidence supporting an instrumental effect able to explain the bright narrow feature observed to evolve in the spectrum of GRB 221009A. 
The feature is found also in the data of the second BGO detector of GBM, with spectral parameters in agreement with those found in the more on-axis BGO (see Supplementary Information).
As far as we know, among the plethora of currently operational $\gamma$-ray satellites that detected GRB 221009A, none has collected usable data during the relevant period and within the relevant spectral band (see Supplementary Information). This unfortunately precludes an independent check of this exciting discovery.

Lines in emission and absorption over the typical continuum GRB spectral shape have been reported over the years, 
but none was ever confirmed as a statistically significant detection ($>5\sigma$).
 During the prompt GRB phase, absorption lines at 30-70 keV and emission lines at 400-460 keV were revealed by the Konus experiment onboard the \textit{Venera} 11 and 12 missions \cite{Mazets1981} and independently by the Ginga satellite \cite{Murakami1988}. 
The search for lines, mostly in absorption and $<100$ keV \cite{Band1996} in the first three year sample of GRBs detected by the Burst Alert \& Transient Source Experiment (BATSE) onboard the Compton Gamma-Ray Observatory (\textit{CGRO}) satellite, led to no convincing evidence of any spectral line \cite{Palmer1994}. {\it Beppo}SAX revealed a possible transient Fe absorption feature during the prompt emission of GRB~990705 \cite{Amati2000} and GRB~021211 (reaching the 2.8--3.1$\sigma$ significance level - \cite{Frontera2004}). 
Line searches extending to the afterglow emission phase revealed possible features in the soft X--ray data of {\it Beppo}SAX \cite{Piro1999,Antonelli2000}, ASCA \cite{Yoshida2001}, XMM-Newton \cite{Reeves2002,Watson2003} and \textit{Chandra} \cite{Piro2000,Butler2003}. A detailed re-analysis of these cases \cite{Sako2005} argues against the relatively high significance initially claimed. An extensive search for emission or absorption lines in the XMM-Newton spectra of GRB afterglows \cite{Campana2016} also did not reveal any significant feature.

GRB~221009A represents the first burst where a bright and narrow, statistically significant, emission line is identified at several MeV energies and in the \textit{Fermi}/GBM spectral data of a GRB.
The excellent quality of the BGO data, due to the exceptional brightness of this GRB, certainly played a major role in revealing this emission feature at such a high signal-to-noise ratio in the data. Indeed, by simulating spectra similar to our spectrum 6 (where the feature is found with the highest significance) but with progressively lower flux, we found that a reduction in flux by a factor of 20-40 would have been sufficient for the feature to disappear completely in the noise (see Methods and Extended Data Figure \ref{fig:simul}).  

We further investigated whether a similar narrow spectral feature could have been detected in other bright GRBs. To this end, we analyzed the spectra of the three
most fluent bursts ever detected by \textit{Fermi}/GBM (in terms of fluence in the conventional energy band 10-1000 keV, obviously excluding GRB~221009A), namely GRB~130427A, GRB~160625B and the recent GRB~230307A (see Methods for details). For each of them, we extracted the spectrum corresponding to the peak of the lightcurve and a few spectra (minimum of 2) during the decaying phase of the pulse with the highest count rate. 
Despite the large fluences and peak counts rates of their light curves (see Extended Data Figure~\ref{fig:lc_comparison}), none of these GRBs shows evidence for a similar narrow and bright excess in the spectra analyzed. \\

A transient, narrow MeV spectral component is usually not predicted by the standard prompt emission models \cite{Rees1994,Drenkhahn2002,Lazzati2009,Zhang2011}. In order to interpret our findings, we explored several scenarios, with the assumption that the narrow spectral component is produced within the GRB jet. 
The large bulk Lorentz factors inferred for GRB jets imply a very low baryon content and, even so, such baryons are not expected to participate in significant nucleosynthesis. Hence, they remain in the form of free protons, deuterium and, at most, $\alpha$-particles \cite{Beloborodov2003}. This prevents the production of observable narrow lines by, for instance, fluorescent recombination within the jet. Conceivably, cold electrons within the jet might interact with nearly monochromatic photons from some sort of `narrow line region' that surrounds the progenitor, up-scattering them by means of bulk Comptonization \cite{Sikora1994,Vietri2001}, which would result in a blue-shifted and Doppler boosted line: we discuss this possibility and the related difficulties in the Supplementary Information. 

More naturally, a narrow spectral feature could arise in the form of a blue-shifted electron-positron pair annihilation line. The conditions for efficient creation of pairs within the jet are likely met in regions where  energy dissipation processes (internal shocks and/or magnetic reconnection events) take place \cite{Peer2004}. As we show in Methods, during the brightest pulse in GRB\,221009A a sufficient amount of electron-positron pairs could have formed through two-photon annihilation within a region of the jet moving at a moderate bulk Lorentz factor $\Gamma\sim 20$ and located at $R\sim 10^{15}$ cm from the central engine. Their subsequent annihilation gives rise to a spectral feature with duration, luminosity and spectrum consistent with that observed. The moderate Lorentz factor $\Gamma\sim 20$, required to place the line at  $\sim$10 MeV as observed, is lower than the typical $\Gamma\gtrsim 100$ expected in powerful GRB jets. Such relatively low $\Gamma$ could arise temporarily (see Methods) during the collision of a very fast portion of the jet with a slower one, which is a requirement for efficient energy dissipation in the leading GRB prompt emission mechanisms \cite{Rees1994,Zhang2011}. 

A slight modification of such scenario, which would allow for a larger Lorentz factor (possibly more in line with the expectations, given the large luminosity) and would naturally accommodate the relatively fast evolution of the narrow feature, is one where the pair annihilation line sweeps the GBM band during the steep decline of one of the brightest pulses of the GRB, due to the high-latitude emission effect \cite{Kumar2000,Oganesyan2020,Ascenzi2020} (see Methods).

\begin{table}
\begin{center}
\begin{minipage}{\textwidth}
\caption{Spectral parameters of the Gaussian function modelling the bright narrow line observed in GRB 221009A, for each time interval analyzed in this work.}\label{tab1:line}%
\begin{tabular*}{\textwidth}{@{\extracolsep{\fill}}lcccc@{\extracolsep{\fill}}}
\toprule
Time interval [s] & $\rm L_{gauss}$  [$10^{50}$ erg/s] & $\rm E_{gauss}$ [MeV] & $\rm \sigma_{gauss}$ [MeV] & $\Delta \rm AIC$\\
\midrule
280 - 300 [5]   & ${1.12}_{-0.19}^{+0.19}$   & $12.56_{-0.31}^{+0.30}$ & $1.31_{-0.30}^{+0.31}$ & 49 \\

\,\,\, 280 - 285 [5.1]    & \,\,\,\,\,\, ${0.77}_{-0.42}^{+0.42}$   &  \,\,\,\,\,\, $14.40_{-0.87}^{+0.86}$ &  \,\,\,\,\,\, $0.99_{-0.57}^{+0.66}$ & 2.4\\
\,\,\, 285 - 290 [5.2]   &  \,\,\,\,\,\, ${0.43}_{-0.28}^{+0.33}$   &  \,\,\,\,\,\, $13.21_{-1.51}^{+6.36}$ &  \,\,\,\,\,\, $1.14_{-0.62}^{+0.59}$  & -1.2\\
\,\,\, 290 - 295 [5.3]   &  \,\,\,\,\,\, ${1.84}_{-0.33}^{+0.36}$   &  \,\,\,\,\,\, $12.16_{-0.30}^{+0.30}$ &  \,\,\,\,\,\, $1.08_{-0.30}^{+0.34}$  & 42 \\
\,\,\, 295 - 300 [5.4]   &  \,\,\,\,\,\, ${0.63}_{-0.27}^{+0.28}$   &  \,\,\,\,\,\, $12.55_{-1.45}^{+0.47}$ &  \,\,\,\,\,\, $0.79_{-0.45}^{+0.81}$  & 5\\

300 - 320 [6]    & $1.14_{-0.18}^{+0.20}$  & $10.19_{-0.28}^{+0.29}$  & $1.70_{-0.42}^{+0.52}$ & 141 \\
\,\,\, 300 - 310 [6.1]    & \,\,\,\,\,\, ${1.08}_{-0.17}^{+0.19}$   &  \,\,\,\,\,\, $10.42_{-0.30}^{+0.31}$ &  \,\,\,\,\,\, $1.14_{-0.29}^{+0.36}$ & 45\\
\,\,\, 310 - 320 [6.2]\footnotemark[1]   &  \,\,\,\,\,\, ${0.75}_{-0.19}^{+0.21}$   &  \,\,\,\,\,\, $9.77_{-0.49}^{+0.42}$ &  \,\,\,\,\,\, $1.24_{-0.21}^{+0.25}$ & 30\\
320 - 340 [7]\footnotemark[1]    & $0.23_{-0.13}^{+0.15}$  & $7.22_{-1.72}^{+1.63}$  & $2.38_{-0.83}^{+0.45}$  & -2 \\
340 - 360 [8]\footnotemark[1]   & $0.21_{-0.10}^{+0.12}$  & $6.12_{-0.59}^{+0.74}$  & $1.35_{-0.74}^{+1.08}$  & 0\\
\botrule
\end{tabular*}
\footnotetext[1]{These spectra require the presence of an extra power-law component.} 
\end{minipage}
\end{center}
\end{table}

\section*{Methods}\label{methods}

\bmhead{Data analysis} The GBM instrument is composed of 12 NaI (sensitive for photons over the energy range of 8 keV to 900 keV) and two bismuth germanate (BGO, 300 keV - 40 MeV)
scintillation detectors \cite{Meegan2009}. Following the suggestions of the Fermi Collaboration regarding the exploitation of GBM data for this burst\footnote{\url{https://fermi.gsfc.nasa.gov/ssc/data/analysis/grb221009a.html}}, we initially analysed the data from two NaI, namely NaI 8 and NaI 4, and one BGO detector, BGO 1. These NaI detectors registered the highest count rates and they observed the source under an angle of less than 60$^\circ$ (35$^\circ$ and 51$^\circ$, respectively), for which the effective area is maximised, while the BGO 1 detector has a smaller viewing angle (80$^\circ$) with respect to BGO 0. At a later stage, we also analysed data from one additional NaI detector, NaI 6 (source viewing angle of 46$^\circ$), and also the second BGO 0 (100$^\circ$), in order to further investigate the observed spectral shape (see the Supplementary Information).  

We retrieved spectral data files and the corresponding most updated response matrix files (rsp2) from the HEASARC online archive\footnote{\url{https://heasarc.gsfc.nasa.gov/W3Browse/fermi/fermigbrst.html}}. As part of the standard data analysis procedure, we selected energy channels in the range 10--900\,keV for NaI detectors, and 0.3--40\,MeV for BGO detectors, and excluded channels in the range 25--45\,keV from our analysis due to the presence of the Iodine K-edge at 33.17\,keV\footnote{\url{https://fermi.gsfc.nasa.gov/ssc/data/analysis/GBM\_caveats.html}}. In addition to the standard procedure, we also excluded energy channels corresponding to the 45--90\,keV energy range from the NaI8 detector data, due to their systematically different behaviour with respect to the other two NaI detectors (NaI8 and NaI6). 
We used inter-calibration constant factors among NaI and BGO detectors, scaled to the brightest NaI and free to vary within 30\%. We made use of CSPEC data, which have 1024\,ms time resolution. To model the background, we selected time intervals (see Supplementary Information) before and after the burst and fitted them with a polynomial function up to the fourth order. Spectra were extracted with the public software {\sc gtburst}\footnote{\url{https://fermi.gsfc.nasa.gov/ssc/data/analysis/scitools/gtburst.html}} and analysed with {\sc{xspec}}\footnote{\url{https://heasarc.gsfc.nasa.gov/xanadu/xspec/}}. We used the PG-Statistic, valid for Poisson data with a Gaussian background, in the fitting procedure.

We extracted time-resolved spectra excluding the BTI time intervals. Before the start of the  BTI at 219~s, we selected four time intervals with $\sim$ 10 s width, according to the pulse trend in the light curve and excluding the quiescent time. To assess the spectral evolution after the BTI end at $t=277$~s, we extracted a sequence of 20-second-wide time bins, up to 460 s after the trigger time. The time intervals selected for the spectral analysis are reported in the first column of Extended Data Table~\ref{tab2:prompt}. The evolution of the spectral shape of this burst after 460~s is beyond the scope of this work.

\bmhead{Model comparison}
We fitted the extracted spectra with different combinations of five models. 
The simplest is the smoothly broken power-law (SBPL) model, which is composed of two power laws with low-energy photon index $\alpha$ and high-energy photon index $\beta$, smoothly connected at one break energy, representing the peak energy $E_{\rm peak}$ in the $\nu F_{\nu}$ representation. The SBPL, together with the Band function, is a standard function used to model the non-thermal spectral shape of the GRB prompt emission (e.g. \cite{Gruber2014}). 

The second model is a double smoothly broken power-law (2SBPL), a modified version of the SBPL model which allows for the presence of an additional spectral break at low-energy. The 2SBPL is composed of three power-law segments (with photon indices $\alpha_1$, $\alpha_2$ and $\beta$) smoothly connected at two break energies, $E_{\rm break}$ and $E_{\rm peak}$. The 2SBPL model has been found to fit the spectra of the brightest \textit{Fermi} GRBs significantly better than the standard single-break function \cite{Ravasio2019, Gompertz2023}. 

Whenever the fit with these models was inadequate to capture the complexity of the spectrum, as revealed by a poor fit statistic and/or by the presence of systematic trends in the residuals, we added a \texttt{Gaussian} component, an extra power law component (PL), or both. In order to model the excess flux observed at MeV energies, we also tried to add a blackbody to the fit function. However, a blackbody 
 is not enough narrow to  properly fit the emission feature.
We compared the different models through the AIC and, in particular, we accepted a more complex model to be preferred over a simpler one whenever the difference in AIC was greater than 4 \cite{Burnham2004}. The parameter space has been explored through a Markov Chain Monte Carlo (MCMC) approach, by means of the built-in \texttt{chain} command in \textsc{xspec}.
The median values of the marginalized posterior probability densities and the symmetric one-sigma errors for each model parameter are reported in Table~\ref{tab1:line} and Extended Data Table~\ref{tab2:prompt}. Their evolution over time is displayed in Extended Data Figure~\ref{fig:evolution_params}.

In the time intervals analyzed before the BTI, up to 216 s (spectra number 1, 2, 3 and 4), there is no evidence for the presence of narrow features. 
In the first time interval, corresponding to the first pulse of the light curve, the spectrum is best fit by a simple SBPL. In this time bin, the low-energy photon index $\alpha = -1.68_{-0.01}^{+0.01}$ is close to the theoretical predictions from synchrotron emission in fast cooling regime ($\alpha_{\rm syn} = -1.5$) and there is no evidence for an additional spectral break in the continuum at low energies.
In the following three time intervals, characterized by more than an order of magnitude brighter emission, the 2SBPL model provides a statistical improvement over the simpler SBPL fit function. The three spectra  statistically require the presence of an additional low-energy spectral break, besides the usual peak energy. The break energy slowly evolves from $\sim$ 230 keV to $\sim$ 115 keV over 20 s. Overall, the prompt emission spectra at T$<216$~s are consistent with the marginally fast cooling synchrotron model.
Independent modelling of the time-resolved spectra of GRB 221009A found them to be consistent with the synchrotron radiation model\, \cite{Yang2023}.

After the BTI ending at 277 s, GBM data are again usable for spectral analysis. The first spectrum extracted from 280 to 300 s (spectrum number 5) shows a strong excess in the BGO data with respect to the SBPL high-energy power-law. We modelled this excess by adding a Gaussian to the SBPL model, finding it well-constrained at $E_{\rm gauss} = 12.56_{-0.31}^{+0.30}$ MeV with a width $\sigma = 1.31_{-0.30}^{+0.31}$ MeV. The addition of the Gaussian is strongly favoured by a $\Delta \rm AIC = 49$ over the simpler SBPL fit. To check if the peculiar feature is also present on smaller timescales and study its evolution with time, we further split this 20 s time interval into four 5-second-long time bins (called 5.1, 5.2, 5.3 and 5.4). The addition of the Gaussian line component improves the fit statistic in all of the four shorter-integration spectra analysed, but its presence is statistically preferred only in spectra 5.3 and 5.4, with $\Delta \rm AIC = 42$ and $5$, respectively. Figure~\ref{fig:spectrum5} shows the spectrum during time interval 5.3.

In the following time bin (300-320 s, spectrum number 6), the best-fitting model is the SBPL+Gaussian, with the presence of the Gaussian line significantly required with respect to the simple SBPL model ($\Delta \rm AIC = 141$). The prompt emission spectrum has now decreased by more than $75\%$ in luminosity, allowing for the line to be clearly visible as the highest peak in the GBM energy range (the line luminosity is comparable to what found in the previous time bin). The central energy of the Gaussian has decreased to $10.19_{-0.28}^{+0.29}$ MeV, while its width has not significantly changed with respect to the previous time bin.
As done for the 5$^\mathrm{th}$ time bin, we performed a finer time resolved analysis by splitting this bin into two 10-second-long finer bins (referred to as 6.1 and 6.2 in Table~\ref{tab1:line}). In both time intervals, the prompt spectrum shows an unusual higher flux between the peak energy and the Gaussian line (where previously the spectrum was showing the typical decaying trend with a photon index $\beta \in {(-3,-2.5)} $). The fit tends to model the higher flux with a quite flat high-energy photon index $\beta \sim -2$. For this reason, we tested the addition of a PL over the SBPL+Gaussian model in both time intervals. While the PL is not statistically required in the 6.1 time bin, which is instead well modelled by SBPL+Gaussian, the spectrum in the second time bin 6.2 is best-fitted by the SBPL+Gaussian+PL model (with a $\Delta \rm AIC = 12$ with respect to the SBPL+Gaussian one). The addition of the PL allows for a better modelling of the spectrum between the two peaks and, as suspected, yields a $\beta$ parameter similar to the previous values ($\beta = -2.17 \pm 0.06$). The PL has a photon index of $\Gamma_{\rm PL} = 1.84_{-0.01}^{+0.04}$ and a luminosity in the 10 keV--40 MeV energy band of $L_{\rm PL} = 0.39_{-0.13}^{+0.08} \times 10^{51}$ erg/s. 

In the next two time intervals, 320-340 s and 340-360 s (spectrum 7 and 8), the emission line is still potentially visible as a small excess with respect to the fainter prompt spectrum. The addition of a Gaussian still represents a much better model with respect to the SBPL alone ($\Delta \rm AIC = 53$ and $91$, for spectrum 7 and 8, respectively). However, given the similar high flux between peak energy and the Gaussian line and the presence of the PL in the previous time bin, we tested its addition over the SBPL also in these two time intervals. We found that the SBPL+PL fit allows to better model the flux between the two peaks ($\Delta \rm AIC = 82$ and $172$ with respect to the SBPL model, for spectrum 7 and 8, respectively). Given the AIC values, the SBPL+PL is the best-fitting model for both spectra, also over the SBPL+Gaussian one. 
The PL was found to have a photon index consistent with the one found in the time bin 6.2 ($\Gamma_{\rm PL} = -1.86 ; - 1.88$) and a similar luminosity in the 10 keV--40 MeV energy band ($L_{\rm PL} = 0.40 - 0.47 \times 10^{51}$ erg/s).
Although the Gaussian line is not statistically required ($\Delta \rm AIC = 0-2$) by the data in these two time intervals, having clearly assessed its strong significance in previous bins, we worked under the assumption that the line is still present in the spectrum, in order to fully capture its evolution. 
The fit yields well-constrained parameters, suggesting the line central energy to have decreased down to $E_{\rm gauss} = 7.22_{-1.72}^{+1.63}$ MeV and $E_{\rm gauss} = 6.12_{-0.59}^{+0.74}$ MeV, in the 7th and 8th spectrum, respectively. The line luminosity during these two time intervals is constant around $L_{\rm gauss} = 0.21-0.23 \times 10^{50}$ erg/s, revealing a fading by at least 50\% with respect to the previous time intervals.

From 360 s onwards (spectra number 9, 10, 11, 12, and 13), there is no evidence for the presence of the Gaussian line. The four spectra, covering the time interval 360-460 s, are best-fitted by the SBPL model. The luminosity of the prompt spectrum is steadily increasing during this period, although in the beginning it reaches values similar to those found in spectrum 6, thus potentially allowing for the line to be visible, if it were present and bright enough.
The low-energy photon index $\alpha$ of the SBPL model is distributed in the range -1.42 -- -1.64, as observed in the previous time intervals, thus not requiring the presence of an additional spectral break at low energy.
Although the spectrum has a typical SBPL shape during the whole 360-460 s period, we report the presence of a clear excess, in the form of a harder rising component, above an energy of $\sim 5$ MeV in the time interval 400-420 (spectrum 11).

\bmhead{Possible scenarios} \label{sec:interpretation}

Electron-positron pairs are naturally expected to form within a powerful GRB jet: in the co-moving frame of the plasma, a fraction of the photons produced as a consequence of energy dissipation (internal shocks and/or magnetic reconnection, \cite{Rees1994,Zhang2011}) is above the threshold $h\nu>m_\mathrm{e}c^2$ for Breit-Wheeler two-photon annihilation \cite{Breit1934,Jauch1976}, and can thus form electron-positron pairs. Pe'er and Waxman \cite{Peer2004} discussed in detail this process within the internal shock scenario. Following their treatment, we define the compactness parameter

\begin{equation}
    \ell'=\frac{\sigma_\mathrm{T}\epsilon_\pm L \Delta'}{4\pi R^2 \Gamma^2 m_\mathrm{e}c^3},
\end{equation}

\noindent
where $\sigma_\mathrm{T}$ is the Thomson cross section, $m_\mathrm{e}$ is the electron rest mass, $c$ is the speed of light, $L$ is the observed GRB luminosity, $R$ is the radius of the region of interest within the jet, $\Gamma$ is its bulk Lorentz factor, $\Delta'=\xi R/\Gamma$ is its co-moving width ($\xi$ is a dimensionless parameter, which is typically $\xi\sim 0.1-1$ if the region of interest is downstream of an internal shock) and $\epsilon_\pm$ is  the fraction of the GRB luminosity in photons that are above the pair-production threshold in the co-moving frame. Adopting the usual notation $Q_x = Q_\mathrm{cgs}/10^x$, where $Q$ is any quantity and $Q_\mathrm{cgs}$ is its value in cgs units, we have 

\begin{equation}
 \ell'=8.1\times 10^{3} \xi_{-0.5}\epsilon_{\pm,-1} L_{54} R_{15}^{-1} \Gamma_{1.3}^{-3},   
\end{equation}

\noindent
where $\Gamma_{1.3}\approx\Gamma/20$.  This shows that a region moving with $\Gamma\sim 20$ within the jet, at a radius $R\sim 10^{15}\,\mathrm{cm}$, that was illuminated by photons produced during the very bright peak of the emission of GRB\,221009A, had a very large compactness parameter. In such a condition, pairs are copiously and continuously created and their number density $n_\pm'$ is set by the balance between the creation and annihilation rate \cite{Peer2004,Rees2005}, 

\begin{equation}
    n_\pm'\sim {\ell'}^{1/2}\Gamma/\sigma_\mathrm{T}R\approx 2.7\times 10^{12} \xi_{-0.5}^{1/2}\epsilon_{\pm,-1}^{1/2} L_{54}^{1/2} R_{15}^{-3/2} \Gamma_{1.3}^{-1/2}\,\mathrm{cm^{-3}}.
\end{equation}

\noindent
Such a pair number density causes the shell to become optically thick to Thomson scattering,

\begin{equation}
    \tau_\mathrm{T,\pm}\sim \sigma_\mathrm{T}n_\pm'\xi R/\Gamma \approx 27 \xi_{-0.5}^{3/2}\epsilon_{\pm,-1}^{1/2} L_{54}^{1/2} R_{15}^{-1/2} \Gamma_{1.3}^{-3/2}.
\end{equation}

\noindent
Pair-annihilation photons, on the other hand, can escape the shell if they are born within a small external layer of thickness equal to the photon mean free path. Assuming that most photons annihilate at the threshold, which is consistent with the assumption of cold pairs, and is the case when the prompt emission spectral slope $\beta$ above threshold (defined through $L_\nu\propto \nu^{-\beta}$) is $\beta>1$, the mean free path is $\lambda'\sim (\sigma_{\gamma\gamma}n_\gamma' + \sigma_\mathrm{T}n_\pm')^{-1}\sim\Delta'/\eta_{\gamma\gamma}\ell'$. Here $n_\gamma'$ is the number density of photons with $h\nu \sim m_\mathrm{e}c^2$, $\eta_{\gamma\gamma}=\sigma_{\gamma\gamma}/\sigma_\mathrm{T}\sim 0.1$ \cite{Svensson1982} is the ratio of the (angle-averaged) Breit-Wheeler cross section to the Thomson cross section, and the last approximate equality is valid when the mean free path for Breit-Wheeler annihilation is shorter than that for Thomson scattering, which is the case for our reference parameters.  The escaping luminosity, in such high-compactness, saturated case, is thus a fraction $\lambda'/\Delta'$ of the luminosity that is converted into pairs,

\begin{equation}
L_\pm \sim (\lambda'/\Delta')\epsilon_\pm L\sim 1.2\times 10^{50}\eta_\mathrm{\gamma\gamma,-1}^{-1}\xi_{-0.5}^{-1}R_{15}\Gamma_{1.3}\,\mathrm{erg/s},    
\end{equation}

\noindent
which, as expected due to saturation, does not depend on $\epsilon_\pm L$. After the end of the bright pulse, when the compactness parameter drops, the formed pairs just keep annihilating over an observer-frame annihilation time scale $t_\mathrm{ann}\sim n_{\pm}'/\Gamma\dot{n}_\mathrm{\pm,ann}'\sim 1/\Gamma \sigma_\mathrm{T}c n_\pm' \approx 209\,\xi_{-0.5}^{-1/2}\epsilon_{\pm,-1}^{-1/2} L_{54}^{-1/2} R_{15}^{3/2} \Gamma_{1.3}^{-1/2}\,\mathrm{s}$, which represents the time scale over which the annihilation line fades away after pair creation stops.

These order-of-magnitude analytical estimates suggest that the feature we observed can be reasonably interpreted as due to pairs formed during the brightest emission phase of GRB\,221009A, within a relatively slow shell with bulk Lorentz factor $\Gamma\sim 20$ that was located at a radius $R\sim 10^{15}\,\mathrm{cm}$ and participated in the production of the prompt emission photons. The existence of such a slow shell is not unexpected in the internal shock scenario: to show this, we consider a jet shell with kinetic luminosity $L_4$ and Lorentz factor $\Gamma_4$ that collides with another shell with luminosity $L_1\ll L_4$ and Lorentz factor $\Gamma_1\ll\Gamma_4$. The collision gives rise to a forward shock (FS) that propagates from the contact discontinuity (CD) into shell 1, and a reverse shock (RS) that propagates backwards (as seen by an observer co-moving with the CD) into shell 4. We can therefore identify the region between the CD and the FS as region 2, and the region between the RS and the CD as region 3. Regions 2 and 3 move approximately at the same speed, with a Lorentz factor $\Gamma$ such that $\Gamma_4\gg \Gamma \gg \Gamma_1$. The value of $\Gamma$ can be estimated by imposing pressure balance between regions 2 and 3 at the CD \cite{Sari1995}: this leads to $\Gamma \sim (L_4/L_1)^{1/4}\Gamma_1$. Since a high efficiency of conversion of kinetic energy requires a large Lorentz factor contrast $\Gamma_4/\Gamma_1$, a low $\Gamma_1<20$ would be a favourable condition to produce the high luminosity we observed. Hence, the Lorentz factor $\Gamma\sim 20$ provides a possible explanation for the large observed luminosity as well. Given the observer-frame dynamical time $t_\mathrm{dyn}\sim R/\Gamma^2 c\approx 83\,R_{15} \Gamma_{1.3}^{-2}\,\mathrm{s}$, such a region does not need to be long-lived, but it can disappear as soon as the forward shock crosses shell 1 (after which the reverse shock would weaken, and regions 2 and 3 would accelerate), and still produce the line over a time comparable to that during which we see the excess.

Within this scenario, the observed time evolution of the line energy could be ascribed to a variation in $(L_4/L_1)$ during the propagation of the FS and RS, since the observed line photon energy is $E_\pm \sim \Gamma m_\mathrm{e}c^2$. The luminosity evolution could also involve the decay in the number density of pairs as they annihilate while the shell expands. A detailed assessment of the predictions of such a model, including the time evolution of the spectral feature, requires in-depth computations that are beyond the scope of this work.

We describe in the Supplementary Information the other two alternative scenarios explored.

\backmatter

\section*{Declarations}

\bmhead{Acknowledgments}

This research has made use of data obtained through the High Energy Astrophysics Science Archive Research Center Online Service provided by the NASA/Goddard Space Flight Center, and specifically this work has made use of public \emph{Fermi}-GBM data. 

MRE acknowledges support from the research programme Athena with project number 184.034.002, which is financed by the Dutch Research Council (NWO). S.A. is supported by the H2020 ERC Consolidator
Grant “MAGNESIA” under grant agreement No. 817661 (PI:
Rea) and National Spanish grant PGC2018-095512-BI00. AJL and DBM are supported by the European Research Council (ERC) under the European Union’s Horizon 2020 research and innovation programme (grant agreement No.~725246). The Cosmic Dawn Center is supported by the Danish National Research Foundation.
The research leading to these results has
received funding from the European Union’s Horizon 2020 Programme under
the AHEAD2020 project (grant agreement n. 871158).
BB and MB acknowledge financial support from MIUR (PRIN 2017 grant 20179ZF5KS). 

\bmhead{Data Availability}

The \emph{Fermi} data analysed in this study are all publicly available and can be downloaded from the the online HEARSAC archive at \url{https://heasarc.gsfc.nasa.gov/W3Browse/fermi/fermigbrst.html}. 

\bmhead{Code Availability}

The codes used in this publication are all publicly available. Fitting was performed in {\sc xspec}, available from \url{https://heasarc.gsfc.nasa.gov/xanadu/xspec/}. \emph{Fermi} tools are available from \url{https://fermi.gsfc.nasa.gov/ssc/data/analysis/scitools/gtburst.html}. The 2SBPL model is published in \cite{Ravasio2018}.

\bmhead{Authors' contributions}
M.E.R. extracted Fermi data, performed the spectral analysis and wrote the text.
O.S.S. and G.O. provided the theoretical interpretations of the spectral feature and co-wrote the text. 
A.M. extracted and analyzed Fermi data to search for the spectral feature in other bright bursts and co-wrote the text.
G. G. extracted and analyzed Fermi data for both GRB 221009A and some of the bright bursts, and co-wrote the text.
S.A. provided contributions to the theoretical interpretations of the spectral feature.
B.B, S.M., M.B. P.G.J, A.J.L., D.B.M. K.B.M. and A.G. contributed to the discussions and to the writing of the paper.
A.C. and G.G. contributed vital insights into the direction of the theoretical interpretations and to the writing of the paper.

\bmhead{Conflict of interest}

We declare no conflicts of interests.

\bmhead{Supplementary information}
Supplementary Information is available for this paper.

\bmhead{Search for potential instrumental effects} \label{sec:instrumentaleffects}

GRB 221009A has been detected by several $\gamma$-ray instruments and in almost all of them the extreme brightness of the burst induced severe instrumental effects (e.g. saturation and deadtime), including \textit{Fermi}/GBM \cite{Veres2022,Lesage2023,Bissaldi2022,Pillera2022}, Konus-Wind \cite{Frederiks2023}, AGILE \cite{Ursi2022,Piano2022}, INTEGRAL (SPI-ACS; \cite{Gotz2022}), Insight-HXMT \citep{An2023}, Solar Orbiter (STIX; [4 - 150 keV], \cite{Xiao2022}), Spektr-RG (ART-XC;  [4-30 keV], \cite{Lapshov2022}), GRBAlpha ([70-890 keV], \cite{Ripa2023}), SIRI-2 ([300-7000 keV], \cite{Mitchell2022}), GECAM-C ([6-300 keV] and [0.4 - 6 MeV]\cite{An2023}, and BepiColombo (MGNS; [280-460 keV], \cite{Kozyrev2022})].

Among them, only Konus-Wind and AGILE observe the $\gamma$-ray sky extending to the the energy range where we found the  the Gaussian feature in \textit{Fermi}/GBM spectral data. Therefore, they could confirm the emission line. The detector S2 on board the Konus-Wind satellite was triggered by the burst but it is affected by saturation.\cite{Frederiks2023} reported the time bin 249-257 s as the latest time interval, after the trigger time, analyzed and fitted with the Band model. No further spectral information are available for the emission beyond 257 s.
The AGILE/MCAL instrument observes in the energy range 0.4-100 MeV. While the results of the specral analysis have not been published yet, \cite{Ursi2022} reported that the data covering the time intervals of the brightest pulse of the burst were affected by pile-up and count rate saturation, hampering a reliable evaluation of the energy spectrum during the time intervals relevant for the observation of the line. Moreover, it is worth it noticing that AGILE/MCAL has a lower sensitivity and energy resolution than \textit{Fermi}/GBM at the same energies where we saw the line ($\sim 10-12 $ MeV). This may preclude the identification and characterization of the line \cite{Labanti2009}.\\

The Fermi Collaboration has performed checks on the reliability of data collected by GBM and has released caveat about their use, flagging specific time intervals as BTI (bad time interval) and discouraging their analysis. 
In performing the spectral analysis presented in this work, we  selected time intervals either before or after the announced BTI and analyzed them with the standard procedures and public software provided by the Fermi collaboration.
Moreover, we performed some tests to assess the presence of any potential evidence of an instrumental artifact causing the appearance of the Gaussian line at MeV energies. 

We first visually inspected the background spectra extracted during the time intervals showing the emergence of the line. In the left panels of Extended Data Figure~\ref{fig:countsspectrum}, we show the spectra accumulated by the BGO1 detector during the time interval 280-300 s (top) and 300-320 s (bottom). The black crosses correspond to the background counts rate per energy, while the red crosses show the corresponding source plus background spectrum. Background counts spectra do not show the presence of the same excess that is instead visible in the red data (source+background) at $\sim$ 10 MeV. 
We also visually inspected the response matrices used in the same time intervals, and they also did not show any particular sign of unusual behaviour at MeV energies. 
Based on these checks we exclude that the line is produced by any artifact in the either the background spectrum or the response matrix.

We also visually inspected the counts rate spectra, which are instead model-independent and can be thought as raw data. In the right panels of Extended Data Figure~\ref{fig:countsspectrum}, we report the counts spectra only of the source (red crosses), namely with the background subtracted. The excess at $\sim$ 10-12 MeV is clearly evident also in the counts rate spectra, in both time intervals considered.

The selection of time intervals for the background subtraction can also have an impact on the source spectra analyzed\footnote{In the spirit of reproducibility, we report here the background time intervals selected for each detectors, on which the analysis presented in this work is based: for NaI4: -126 -- -19 s, 1724 -- 2056 s, 2101 -- 2334 s; for NaI6: -309 -- -35 s, 1015 -- 1433 s; for NaI8: -285 -- -23 s, 50 -- 107, 1490 -- 1671; for BGO 1: -428 -- -10 s, 1670 -- 2104 s; for BGO 0: -423 -- -19, 1592 -- 1707 s.}
To test the solidity of the presence of the line, we extracted the spectra by performing a different selection of the time windows for the background spectrum computation. After checking the reliability of the new background spectra and the response matrices  against the presence of possible features in the MeV energy range, the analysis of the source spectra clearly showed the excess at MeV energies, confirming the presence of the Gaussian line in the BGO data.

In general, if multiple detectors observed the burst and a peculiar spectral feature is found in one of them, then it is worth  to test the presence of a such feature by comparing the results of each detector. In fact, if the feature were a statistical fluctuation, having the same fluctuation in a different detector is improbable.
Given the extraordinary brightness of the burst, the other BGO detector on board Fermi, namely BGO 0, which is mounted on the opposite side of the spacecraft with respect to BGO 1 (source viewing angle = 80$^{\circ}$), had also detected the burst (with an angle of 100$^{\circ}$) and registered a large count rate. The lightcurve of BGO 0 is similar to that of BGO1. We analysed the spectra of BGO 0 extracted over the same time intervals considered for BGO 1. 
Extended Data Figure~\ref{fig:spectrum5_bothBGOs} shows the spectral data, together with the best-fitting model, corresponding to the time interval 280-300 s. The BGO 0 data (yellow symbols) are consistent with those of BGO 1 and clearly confirm the presence of the line. In particular, the line, present in both BGO detectors and statistically required by a $\Delta \rm AIC = 146 $, has a fitted position of $E_{\rm gauss} = 13.05_{-0.24}^{+0.26}$ MeV with a $\sigma = 1.78_{-0.25}^{+0.27}$ MeV, fully consistent with what previously found.

Finally, notice that we observe the line becoming softer in energy and dimmer in luminosity by a factor $\sim$ 2 and $\sim$ 5 respectively, over $\sim 80$ s. While this evolution can find plausible explanations in a naturally evolving astrophysical system as a GRB, it is not straightforward to explain it as an instrumental artifact. As a comparison, the Iodine K-edge is a well-known instrumental feature present in GBM data at 33.17 keV \cite{Bissaldi2009}, and the Fermi Collaboration encouraged to exclude the corresponding affected channels from the spectral analysis, but it does not evolve in energy position, nor in normalization, over time.

In conclusion, given that time intervals showing the presence of the line are safely placed outside those affected by pileup and saturation effects, and given that our tests described above returned no evidence for any potential instrumental issue producing the line, we are confident in the astrophysical origin of this feature.

\bmhead{Comparison with other bright GRBs}
The extreme brightness of GRB 221009A provided us with unprecedented high-quality data at MeV energies  collected by the BGO detectors on board the Fermi satellite. These data allowed us to  discover a statistically significant emission line and asses its temporal evolution. However, such a high quality of BGO data is rare and prerogative of extremely bright bursts. The typically larger error bars in the BGO spectral data of Fermi-detected GRBs may prevent the discovery of such a line, it is were a common feature in GRBs. To show this, we performed simulations of spectrum 6 (300-320 s), where the line is most significant, exploring down to which flux it would still have been possible to significantly detect the narrow feature (setting a conservative threshold of $\Delta \rm AIC > 4$, roughly corresponding to one-sigma significance). The spectral simulations were performed with the built-in command \texttt{fakeit} in {\sc xspec} assuming the best-fitting model and parameters of spectrum 6, namely SBPL+Gaussian, but reducing the normalization by a factor of 2, 10, 20, 50, 80, and 100. For each value of the normalization, we simulated 100 realizations of the spectrum, and then fitted each simulated spectrum with both the SBPL and SBPL+Gaussian to compute the AIC. Extended Data Figure~\ref{fig:simul} shows the symmetric one-sigma ranges (16$^\mathrm{th}$ to 84$^\mathrm{th}$ percentiles) of the resulting $\Delta \rm AIC$ as a function of the flux reduction factor. The result shows that the significance of the narrow feature decreases rapidly with the flux, and that a reduction in brightness by a factor between 20 and 40 would make the identification of the feature essentially impossible.

This shows that the best candidate GRBs to search for a similar narrow feature are extremely bright bursts.
Therefore, we investigated the presence of narrow features in the three brightest (most fluent) GRBs in the Fermi Catalog after GRB 221009A (reported in the Catalog to have a fluence of $F = 0.04 \, \rm erg/cm^2$, in the energy range 10-1000 keV\footnote{This value is the one reported in the Catalog, and it is not affected by the revised fluence of GRB~221009A published in \cite{Burns2023}, which is much larger ($F = 0.2 \, \rm erg/cm^2$).}), that is GRB~230307A ($F = 3.15 \times 10^{-3} \,\rm erg/cm^2 $), GRB~130427A ($F = 2.46 \times 10^{-3} \, \rm erg/cm^2 $) and GRB~160625B ($F = 0.64 \times 10^{-3} \, \rm erg/cm^2 $). A posteriori, the spectra analyzed (described below) are found to have a flux in the energy range 10 keV--40 MeV from 1.4 to 29 brighter than the one found in spectrum 6 used for the simulations.\\

Extended Data Fig.~\ref{fig:lc_comparison} shows the comparison of the count rate light curves of the selected GRBs. In the top panel,  the light curve observed by the most illuminated NaI (between 8 and 900 keV) is shown, while in the bottom panel that of the most illuminated BGO (between 300 keV and 40 MeV) is shown. In absence of a univocal interpretation of the feature it is difficult to predict where to expect it, but given the observed position of the line in our burst ($\sim 10$ MeV), the search is most promising in bursts with counts rates in the BGO data as high as in GRB 221009A. For each burst we extracted the spectrum at the peak of the BGO light curve, and a few spectra (minimum 2, depending on the light curve shape) during the steep decay following the peak.

The extreme brightness of GRB~130427A also caused pile-up effects in the detectors after 2.4 s from the trigger time \cite{Preece2014}. Despite this time interval does not include the main peak of the light curve in the BGO, nor the decaying part of it, we restricted the extraction of spectra to the first 2.4 s. We found that in the spectrum of GRB 130407A in this time interval there is no evidence for a line-like excess at high energies. Its spectrum is well fitted by the 2SBPL model, with spectral indices remarkably consistent with predictions from synchrotron in marginally fast cooling regime.
In both GRB~230307A and GRB~160625B, we did not find signatures of a line at high energies both in the peak spectrum and in the spectra analysed during the decaying phase. In both GRBs, the spectra are well modelled by the 2SBPL continuum function. Only in one time interval of GRB~160625B, namely 200.74 -- 204.83 s, there is a hint of an excess around 16 MeV, but the addition of a Gaussian to the continuum does not significantly improve the fit ($\Delta \rm AIC = 0.47$).

Given the lack of compelling evidence of the presence of a similar feature in these three bright GRBs, we extended the search to GRBs with a viewing angle from the axis of the BGO detectors $\theta_{\rm BGO} < 40^\degree$, and among them we selected the 3 brightest GRBs of this sub-sample, namely GRB~170409A, GRB~171227A and GRB~130606B, to ensure also a high signal to noise ratio of their spectra.
We considered time-bins around the main pulse peak and the following decaying phase in their light curves. We extracted a number of spectra varying from 3 to 6, within temporal bins of 2 to 13 s.
We fitted these spectra with a 2SBPL function in the same way as for GRB~221009A. 
Despite the optimal observational conditions, neither a significant excess over the continuum model at high energies was observed in the spectra of this sub-sample, nor a significant systematic trend in the residuals.
In two time-bins before and after the peak of GRB 170409A (32-38 s), brightest GRB of the sub-sample, there are hints of an excess around $\sim 20$ MeV. Nonetheless, the AIC test does not indicate a strong preference for the addition of a Gaussian component on top of the best-fitting spectral model ($\Delta \rm AIC = 3.5$). 

We stress that our search was focused on a few bright GRBs, likely representing the best candidates to find a line (if present) in the spectrum, but a more detailed and comprehensive search throughout the Fermi Catalog is required to better confirm or exclude this possibility. This is beyond the scope of this work.

\bmhead{Alternative interpretations} As mentioned in the main text, a possibile avenue to interpret the narrow spectral feature is that of an intrinsically low-energy spectral line (for instance a 6.4 keV fluorescent K-$\alpha$ iron line) emitted by a sort of `narrow line region' (which could be possibly part of the supernova ejecta) which is up-scattered by the relativistic jet. For the line not to be broadened significantly in the process, the electrons that scatter the photons must be cold: this kind of `bulk' Comptonisation has been proposed to operate in the jets of blazars, boosting the continuum and the broad-line photons \cite{Sikora1994}. The most efficient configuration requires the seed photons to be produced or isotropized by e.g.\ Compton scattering, above, but close to, the relativistic jet Thomson photosphere. The boosted photon energy of the iron K-$\alpha$ line (on which we focus, but the discussion would remain similar for Nickel or Cobalt lines) is $E_\mathrm{line} \approx \Gamma^{2} E_\mathrm{Fe}$, where $\Gamma$ is the jet bulk Lorentz factor and $E_\mathrm{Fe}\sim 6.4\,\mathrm{keV}$. Given the observed line typical photon energy $E_\mathrm{line}\sim 10\,\mathrm{MeV}$, this would require a jet bulk Lorentz factor $\Gamma\sim 40$ for the (un-shocked) cold plasma. The observed luminosity of the boosted iron line would then be $L_\mathrm{line} \sim \tau_\mathrm{T}\Gamma^4 L_\mathrm{Fe}$, where $L_\mathrm{Fe}$ is the luminosity (in fluorescent iron) of the narrow line region and $\tau_\mathrm{T}\sim 0.5\sigma_\mathrm{T}L_\mathrm{jet}/8\pi R \Gamma^3 m_\mathrm{p}c^3\approx 0.46\,L_\mathrm{jet,52}R_{14}^{-1}\Gamma_{1.6}^{-3}$ \cite{Daigne2002} is the Thomson optical depth of the portion of the jet where most of the bulk Comptonization occurs. This would require a narrow line region luminosity $L_\mathrm{Fe}\sim 8.5\times 10^{43}\,L^{-1}_\mathrm{jet,52}R_{14}\Gamma_{1.6}^{-1}\,\mathrm{erg/s}$, which is equivalent to a recombination rate $\dot N_\mathrm{Fe}=L_\mathrm{Fe}/E_\mathrm{Fe}\sim 8.7\times 10^{51}\,L^{-1}_\mathrm{jet,52}R_{14}\Gamma_{1.6}^{-1}\,\mathrm{s^{-1}}$. This can be compared to the expected optimal iron recombination rate in the simplest reflection model (see eq.\ 4 from \cite{Vietri2001}), from which we have that the required iron mass (assuming the narrow line region to be a spherical shell of thickness $\Delta R_\mathrm{Fe}$ and temperature $T$) is $M_\mathrm{Fe}\sim 2.8\times 10^{-3}\,L^{-1}_\mathrm{jet,52}R_{14}\Gamma_{1.6}^{-1}T_7^{3/4}\Delta R_\mathrm{Fe,14}\,\mathrm{M_\odot}$. 

The difficulties with this scenario come from the large required radius for the narrow line region, $R\gtrsim \Delta R_\mathrm{Fe}$, which must be equal or larger than the jet photospheric radius (where $\tau_\mathrm{T}=1$), $R_\mathrm{ph}=4.6\times 10^{13}\,L_\mathrm{jet,52}\Gamma_{1.6}^{-3}\,\mathrm{cm}$, and from the relatively low Lorentz factor of the jet $\Gamma\sim 40$ at the photosphere. The former constitutes a problem because it would require the supernova (SN) ejecta to reach  $R_\mathrm{ph}$ in a very short time (assuming the supernova exploded at the time marked by the GRB precursor), $t_\mathrm{SN}\sim t_\mathrm{line}/(1+z) \sim 300 (1+z)^{-1}\,\mathrm{s}$, which would require relativistic expansion of the SN ejecta with a Lorentz factor $\Gamma_\mathrm{SN}\gtrsim \sqrt{R_\mathrm{ph}/c t_\mathrm{SN}}\sim 5 (1+z)^{1/2} L_\mathrm{jet,52}^{1/2}\Gamma_{1.6}^{-3/2}t_\mathrm{line,2.5}^{-1/2}$, which seems highly unlikely. This problem could be alleviated if the line photons were emitted at lower radii and then isotropized by scattering by the stellar wind material. The low jet bulk Lorentz factor $\Gamma\sim 40$ is also an issue: while the Lorentz factor of the jet is expected to be variable, and at times reach relatively low values such as this one, the average value is likely substantially larger, and the variability timescale is likely much faster than the observed duration of the spectral feature (on a side note, the Lorentz factor variability would substantially broaden the up-scattered line). For these reasons, we believe this scenario is disfavoured.  

A third scenario we explored involves high-latitude emission (HLE) from the shell that produced the most luminous pulse in the GRB light curve. Photons produced at latitudes (angle between the jet expansion direction and the line of sight) $\theta>1/\Gamma$ reach the observer over a time longer than the dynamical timescale, producing a tail of progressively less Doppler boosted emission that can be observed if the emission `turns off abruptly' (i.e.\ the luminosity drops sufficiently rapidly) \cite{Kumar2000,Oganesyan2020,Ascenzi2020,Panaitescu2020}. In this scenario, pairs are produced within the dissipation region that produced the peak of the GRB\,221009A luminosity, for the same arguments as in the first scenario above, but the shell has a larger bulk Lorentz factor, $\Gamma\sim 1000$. When the dissipation ends, the luminosity drops (as observed around $t\sim 260\,\mathrm{s}$ after the trigger), and the HLE tail could in principle become visible. In the tail, photons emitted over a dynamical time and within a solid angle $\mathrm{d}\Omega=2\pi\sin\theta\,\mathrm{d}\theta$ reach the observer over a time interval $\mathrm{d}t_\mathrm{obs}\sim (1+z)R\sin\theta\mathrm{d}\theta/c=(1+z)R\,\mathrm{d}\Omega/2\pi c$ \cite{Oganesyan2020,Ascenzi2020}, where $R$ is the emission (turn-off) radius. The energy in these photons, as measured by the observer, is related to the emitted energy \cite{Salafia2015} through $\mathrm{d}E_\mathrm{obs}=(\delta^3/\Gamma) (\partial E_\mathrm{em}/\partial\Omega)\mathrm{d}\Omega$, where $\delta=\Gamma^{-1}(1-\beta\cos\theta)\propto t_\mathrm{obs}^{-1}$ is the Doppler factor. This leads to the HLE luminosity evolution

\begin{equation}
L_\mathrm{HLE}=\frac{\mathrm{d}E_\mathrm{obs}}{\mathrm{d}t_\mathrm{obs}}=\frac{1}{2}(1+z)^2\frac{R^2}{c^2\beta^3\Gamma^4(t_\mathrm{obs}-t_\mathrm{obs,peak})^3}E_\mathrm{em},
\label{eq:HLE_luminosity}
\end{equation}

\noindent
where $E_\mathrm{em}$ is the isotropic-equivalent energy emitted during a dynamical time (assumed equal at all latitudes), and $t_\mathrm{obs,peak}$ represents the emission turn-off time (which should correspond to the peak time). The advantages of this scenario are that (i) it can accommodate a large Lorentz factor, since the line observed photon energy is given by $E_\mathrm{line}=\delta m_\mathrm{e}c^2$ with $\delta\lesssim \Gamma$, and (ii) it naturally predicts a decrease in both the line luminosity and energy, with $L_\mathrm{line}/E_\mathrm{line}\propto (t_\mathrm{obs}-t_\mathrm{obs,peak})^{-2}$. Taking $t_\mathrm{obs,peak}\sim 230\,\mathrm{s}$ this gives an evolution of $L_\mathrm{line}/E_\mathrm{line}$ that roughly matches the observed one, as demonstrated in Extended Data Figure \ref{fig:Lline_Eline_HLE}.

The pairs in this scenario would have to be created by photon-photon annihilation within the shell, similarly as in the first scenario. A difficulty is that the compactness parameter $\ell'\propto \Gamma^{-3}$ could be much lower (if the shell were to have a much larger Lorentz factor), which could potentially prevent efficient pair creation, despite the very large observed luminosity.

Another potential difficulty is that the entire prompt emission spectrum would be affected by the same HLE mechanism, which raises the question whether the line would emerge sufficiently above the combined HLE broadband emission from the bright pulse to be detectable, without the HLE broadband component being dominant over the `on-axis' emission (otherwise we would see the peak of the broadband component tracking the evolution of the line, which is not the case). A detailed investigation of the conditions for this scenario to be feasible are beyond the scope of this work and will be investigated elsewhere.

\clearpage
\bmhead{Extended Data}

\renewcommand{\figurename}{Extended Data Figure}
\setcounter{figure}{0}
\renewcommand{\tablename}{Extended Data Table}
\setcounter{table}{0}
\begin{table*}[b]
\tiny
\addtolength{\tabcolsep}{-3pt}
\begin{center}
\begin{minipage}{\textwidth}
\caption{Spectral parameters of the best-fitting models for the continuum spectral shapes observed in GRB 221009A (either the SBPL, 2SBPL or PL models), for each time interval analyzed in this work.}\label{tab2:prompt}
\begin{tabular*}{\textwidth}{cccccccc}
\toprule%
Time interval & Model & $\rm L_{\rm iso}$ \footnotemark[1] & $\alpha_1$ ($\Gamma_{\rm PL}$) & $E_{\rm break}$ & $\alpha$ ($\alpha_2$) & $E_{\rm peak}$ & $\beta$  \\
\, [s] & & [$10^{51}$ erg/s] & & [keV] & & [keV] \\
\midrule
0 - 9 [1]  & SBPL & $0.14_{-0.01}^{+0.02}$  & - & - & $-1.68_{-0.01}^{+0.01}$ & $1256.51_{-304.89}^{+418.61}$ & $-3.14_{-0.25}^{+0.36}$\\
\midrule
184 - 196 [2]  & 2SBPL & $3.59_{-0.03}^{+0.03}$   & $-0.93_{-0.02}^{+0.02}$ & $231.32_{-31.11}^{+32.29}$ & $-1.66_{-0.05}^{+0.05}$  & $1188.94_{-39.61}^{+41.14}$ & $-3.03_{-0.06}^{+0.07}$\\
\midrule
196 - 206 [3]  & 2SBPL & $1.1_{-0.02}^{+0.02}$   & $-1.05_{-0.02}^{+0.02}$ & $104.7_{-8.84}^{+9.33}$ & $-1.88_{-0.04}^{+0.02}$  & $278.62_{-36.31}^{+36.30}$ & $-2.76_{-0.08}^{+0.08}$\\
\midrule
206 - 216 [4]  & 2SBPL & $1.49_{-0.03}^{+0.03}$   & $-1.08_{-0.02}^{+0.02}$ & $114.47_{-10.28}^{+9.14}$ & $-1.91_{-0.03}^{+0.02}$  & $946.99_{-120.56}^{+141.91}$ & $-3.27_{-0.16}^{+0.20}$\\
\midrule
280 - 300 [5]  & SBPL & $3.83_{-0.03}^{+0.03}$  & - & - & $-1.51_{-0.01}^{+0.01}$ & $682.91_{-11.67}^{+12.91}$ & $-2.42_{-0.01}^{+0.01}$ \\
\midrule
300 - 320 [6]  & SBPL & $0.91_{-0.04}^{+0.02}$  & - & - & $-1.68_{-0.01}^{+0.02}$  & $543.26_{-214.91}^{+167.32}$ & $-2.06_{-0.04}^{+0.01}$ \\ 
\midrule
320 - 340 [7]  & SBPL & $0.20_{-0.02}^{+0.02}$ & - & - & $-1.68_{-0.03}^{+0.04}$  & $66.36_{-2.83}^{+2.99}$ & $-2.96_{-0.15}^{+0.14}$ \\
\,  & PL & $0.47_{-0.05}^{+0.05}$ & $-1.86_{-0.02}^{+0.02}$ & - & -  & - & - \\
\midrule
340 - 360 [8]  & SBPL & $0.17_{-0.02}^{+0.02}$  & - & - & $-1.74_{-0.03}^{+0.03}$  & $64.84_{-2.56}^{+2.82}$ & $-3.57_{-0.20}^{+0.17}$ \\
\,  & PL & $0.40_{-0.03}^{+0.03}$ & $-1.88_{-0.03}^{+0.02}$ & - & -  & - & - \\
\midrule
360 - 380 [9]  & SBPL & $0.35_{-0.01}^{+0.01}$  & - & - & $-1.42_{-0.05}^{+0.05}$ & $43.88_{-1.59}^{+1.55}$  &  $-2.25_{-0.02}^{+0.02}$\\
\midrule
380 - 400 [10]  & SBPL & $0.90_{-0.02}^{+0.02}$  & - & - & $-1.64_{-0.01}^{+0.01}$ & $435.57_{-16.71}^{+18.67}$  & $-2.48_{-0.04}^{+0.04}$ \\
\midrule
400 - 420 [11]  & SBPL & $0.97_{-0.02}^{+0.02}$  & - & - & $-1.59_{-0.01}^{+0.01}$ & $314.71_{-9.27}^{+10.29}$ & $-2.43_{-0.03}^{+0.03}$ \\
\midrule
420 - 440 [12]  & SBPL & $0.82_{-0.02}^{+0.02}$  & - & - &  $-1.61_{-0.01}^{+0.01}$ & $246.05_{-7.70}^{+8.21}$ &  $-2.38_{-0.03}^{+0.03}$ \\
\midrule
440 - 460 [13]  & SBPL & $1.59_{-0.02}^{+0.02}$  & - & -& $-1.55_{-0.01}^{+0.01}$ & $332.83_{-7.11}^{+7.11}$ & $-2.41_{-0.02}^{+0.02}$  \\
\botrule
\end{tabular*}
\footnotetext[1]{The luminosity for the prompt spectrum is calculated over the energy range 10 keV - 30 MeV.}
\end{minipage}
\end{center}
\end{table*}

\begin{figure}
    \centering
    \includegraphics[width=\textwidth]{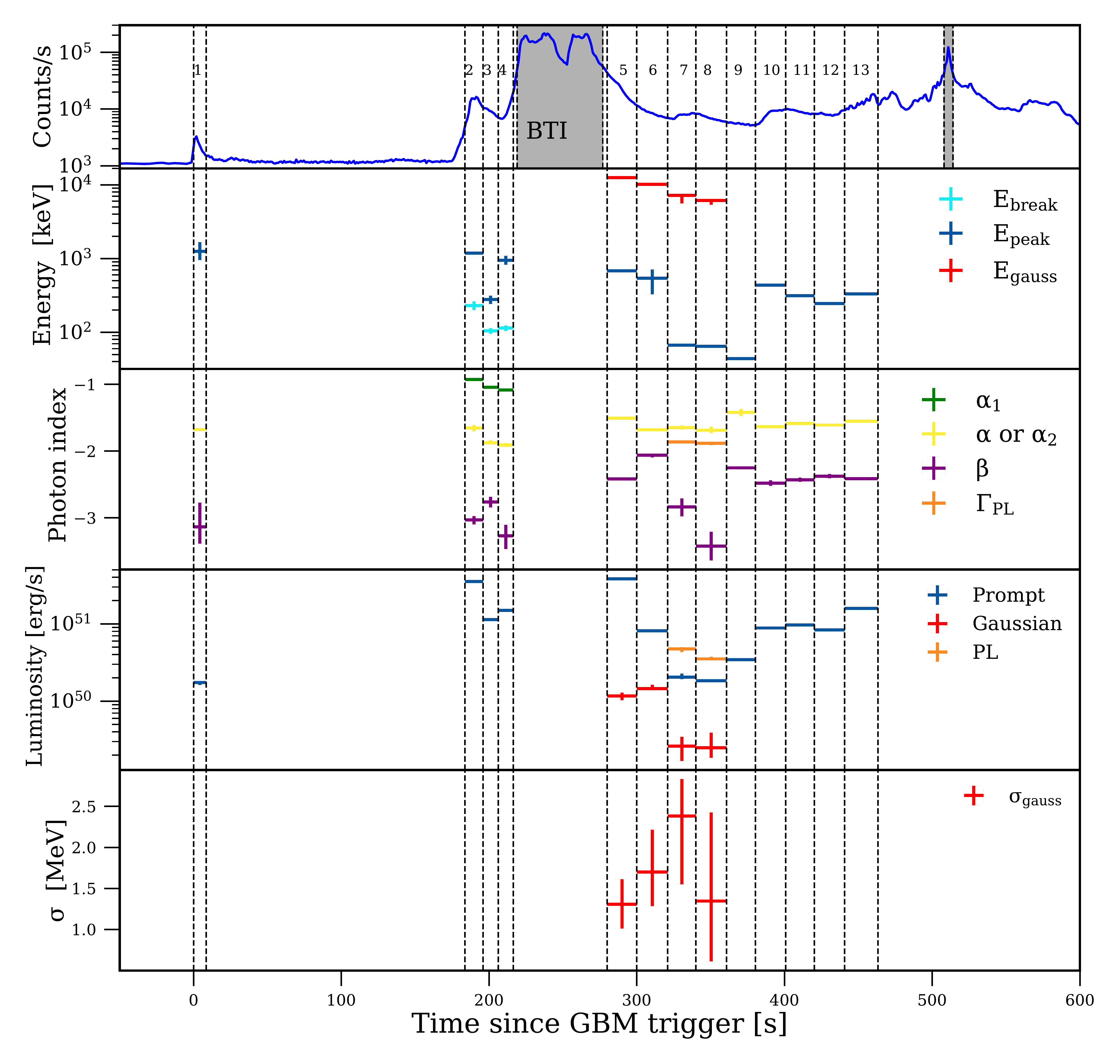} 
    \caption{Light curve and temporal evolution of the spectral parameters of the best 
    models for each of the 13 time intervals analyzed (except for the 7$^\mathrm{th}$ and the 8$^\mathrm{th}$ time intervals, where we assume the Gaussian to be present despite not being statistically required). The top panel shows the light curve of GRB~221009A in the 8-900 keV, together with the 13 time intervals analyzed (dashed vertical black lines).
    In the panels below, all the parameters of the Gaussian model are shown as red points, those related to the power law are shown in orange, while different colours have been used to represent the parameters of the continuum prompt models (SBPL or 2SBPL). From top to bottom, in the second panel there are the characteristic energies $E_{\rm break}$ (turquoise), $E_{\rm peak}$ (blue points) and the central energy of the Gaussian $E_{\rm gauss}$, in the third panel the photon indices $\alpha_1$ (green points), $\alpha$ or $\alpha_2$ (of the SBPL or 2SBPL function, respectively, yellow points), $\beta$ (purple points) and $\Gamma_{PL}$. The fourth panel shows the luminosity of the typical prompt spectral function (either 2SBPL or SBPL, blue points), of the Gaussian and of the power-law functions. The fifth panel shows the evolution of the width $\sigma$ of the Gaussian component.}
    \label{fig:evolution_params}
\end{figure}

\begin{figure}[ht]
    \centering
    \includegraphics[width=0.9\textwidth]{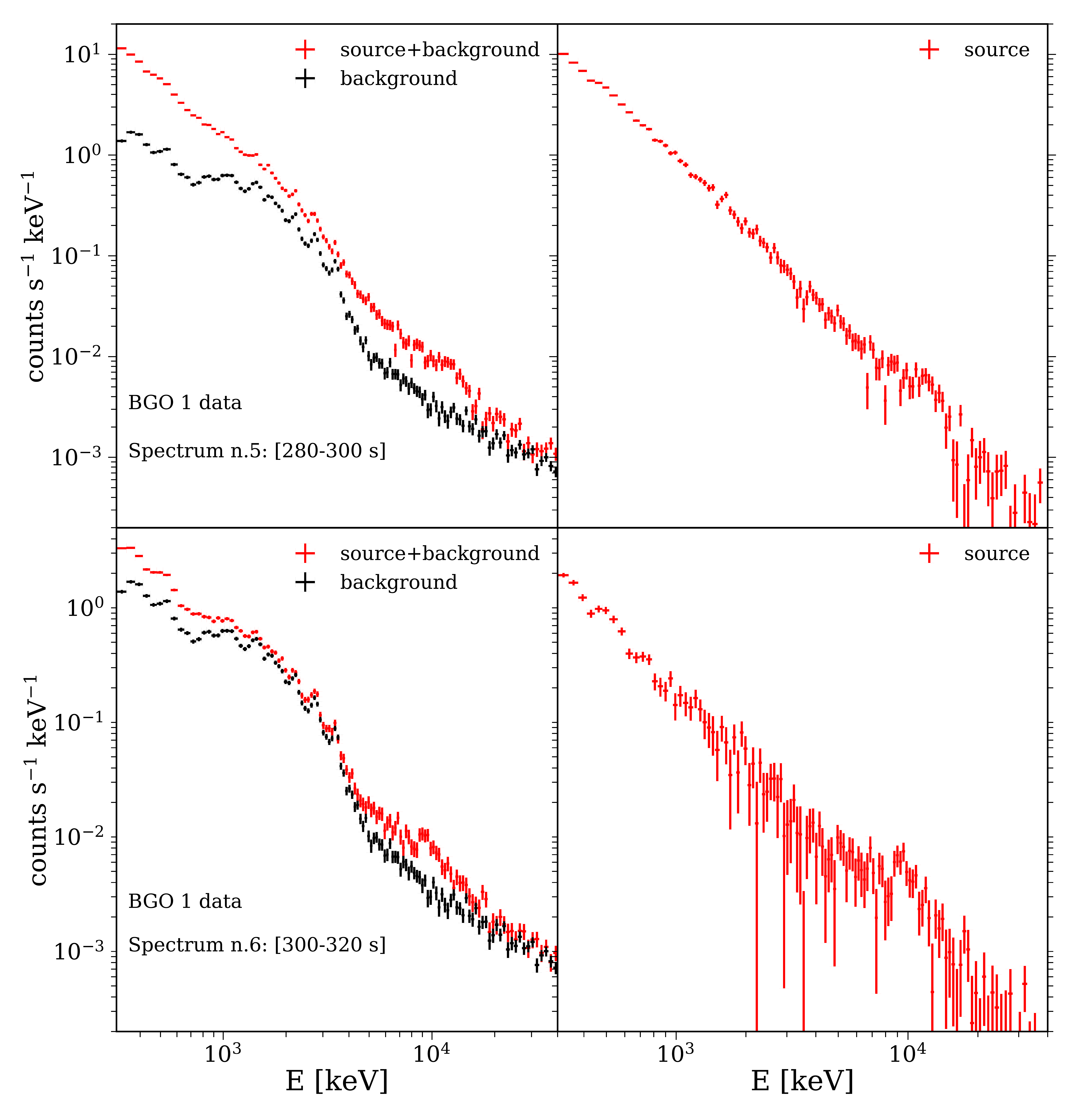}
    \caption{Right panels: count rate spectra of BGO1 detector from GRB 221000A in the time interval 280-300 s (top) and 300-320 s (bottom). The background (in black) and the total (source plus background, in red) counts rate spectra are displayed. Right panels: background subtracted count rate spectrum (i.e. obtained by subtracting the red and black spectra shown in the left panels). Data are shown with the natural instrumental data energy binning.
    }
    \label{fig:countsspectrum}
\end{figure}

\begin{figure}[h]
    \centering
    \includegraphics[width=0.9\textwidth]{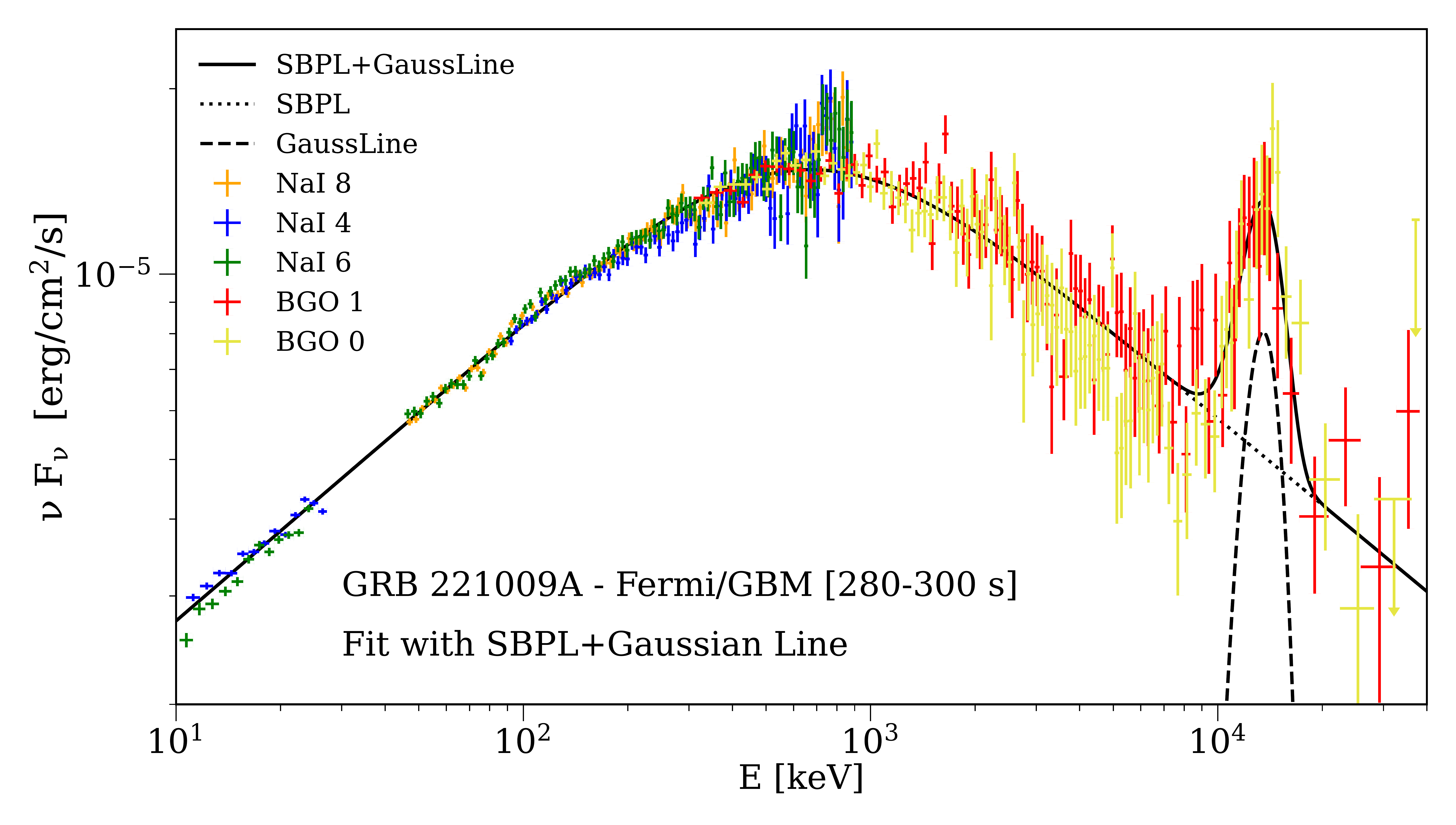}
    \caption{Spectrum of GRB 221000A in the time interval 280-300 s, as detected by \textit{Fermi}/GBM, along with the best-fitting model SBPL+Gaussian line (solid line). This plot shows that the emission line is present in both BGO detectors (BGO 0 and BG0 1) data.
    }
    \label{fig:spectrum5_bothBGOs}
\end{figure}

\begin{figure}[h]
    \centering
    \includegraphics[width=\textwidth]{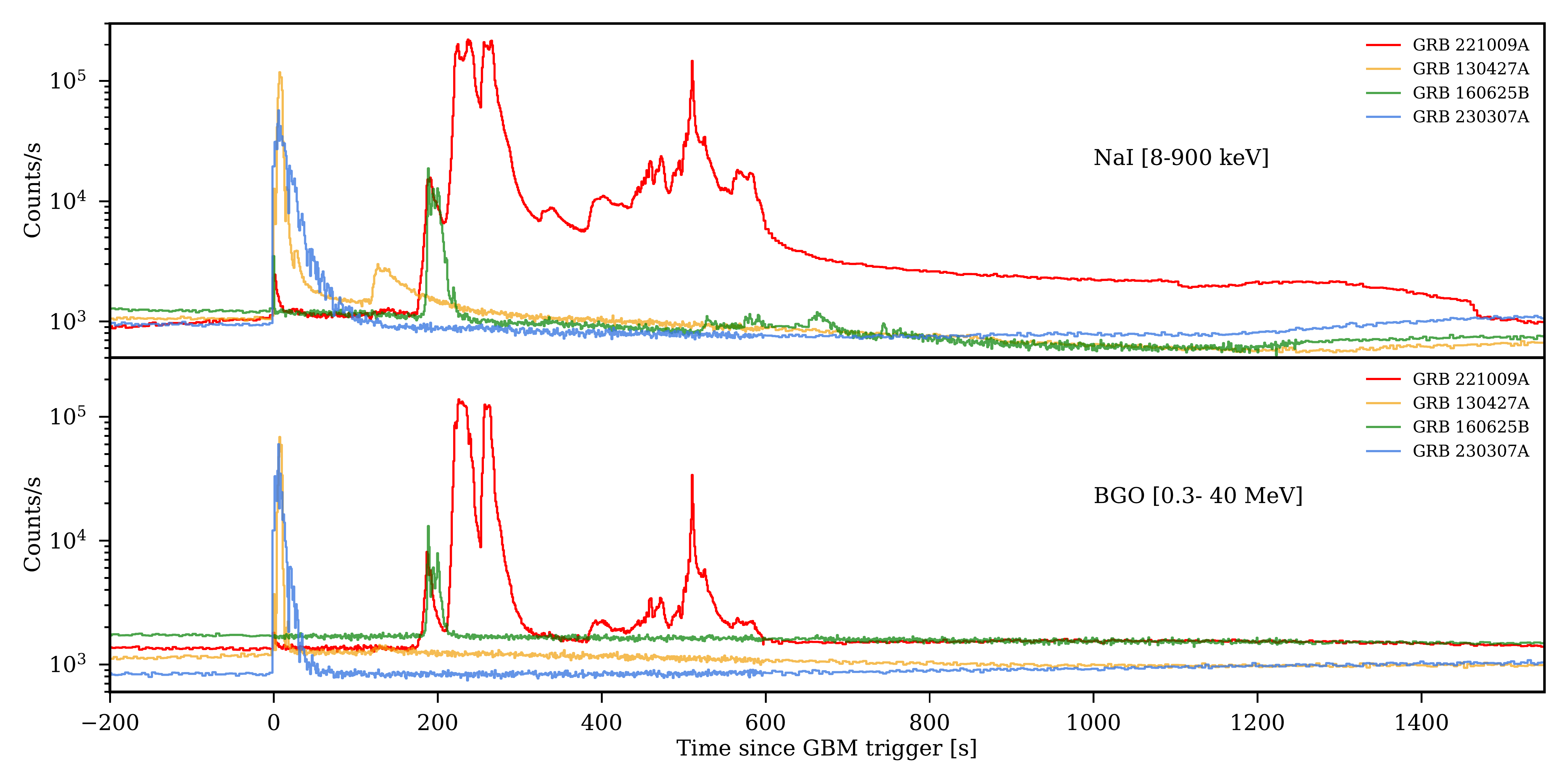} 
    \caption{Comparison of the lightcurves of the 4 brightest (most fluent) GRBs ever detected by Fermi in 15 years of activity. The top panel shows the lightcurve detected in the energy range 8-900 keV by the most illuminated NaI detector, while the bottom panel shows the lightcurves of the same GRBs as detected in the high-energy band 300 keV - 40 MeV by the most illuminated BGO detector. Lightcurves are not background-subtracted.}
    \label{fig:lc_comparison}
\end{figure}

\begin{figure}[h]
    \centering
    \includegraphics[width=0.8\textwidth]{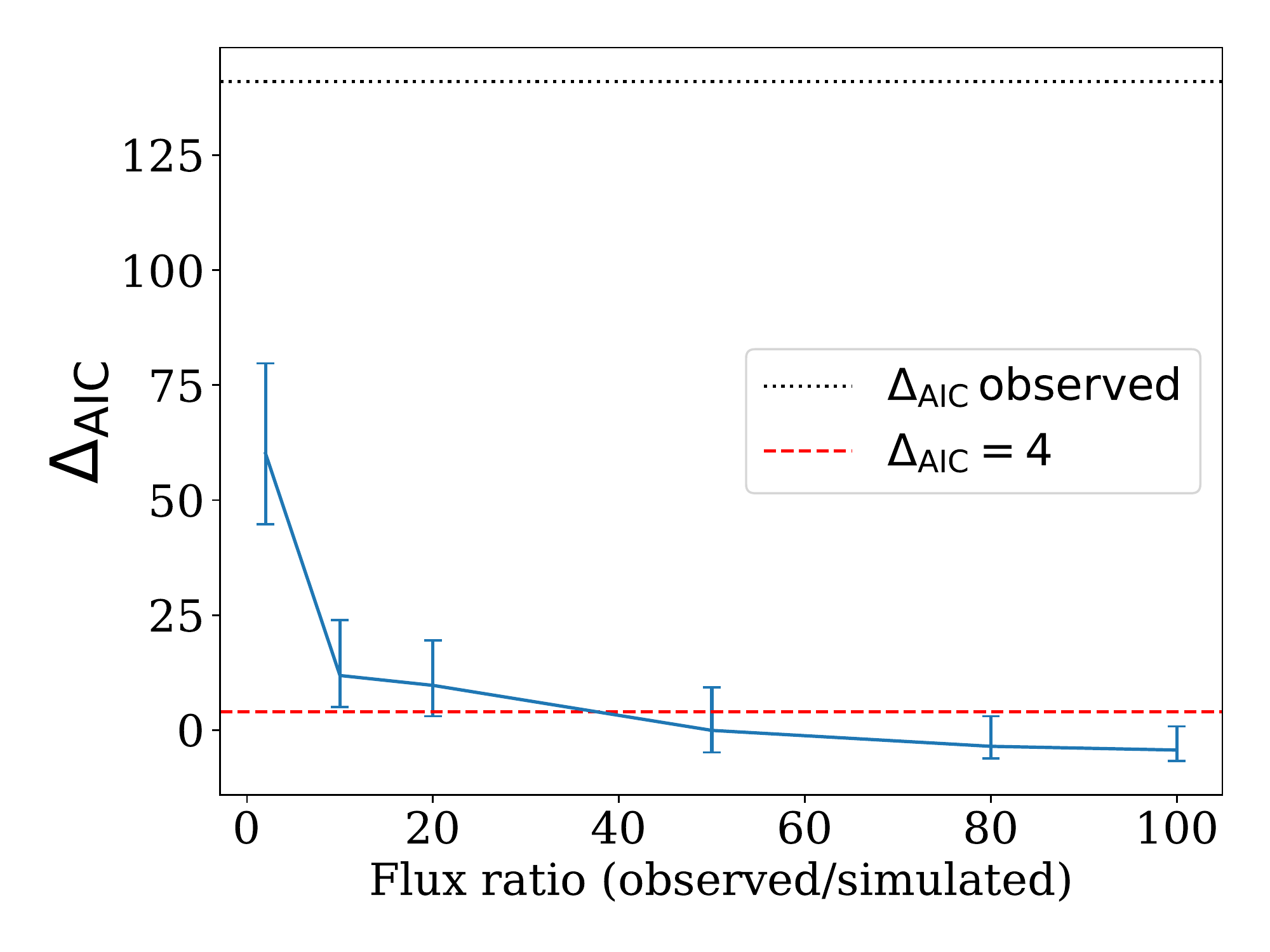} 
    \caption{The $\Delta \rm AIC$ as a function of the flux ratio, resulting from the simulations of a spectrum with the same parameters of the one observed in the time interval 300-320 s (n.6), but with a normalization reduced by a factor from 2 to 100 (see the text for details). The dotted black line represents the value of  $\Delta \rm AIC$ found from the analysis of the spectrum n.6, while the dashed red line represents the threshold $\Delta \rm AIC = 4$ that roughly corresponds to a one-sigma significance.}
    \label{fig:simul}
\end{figure}

\begin{figure}
    \centering
    \includegraphics[width=\textwidth]{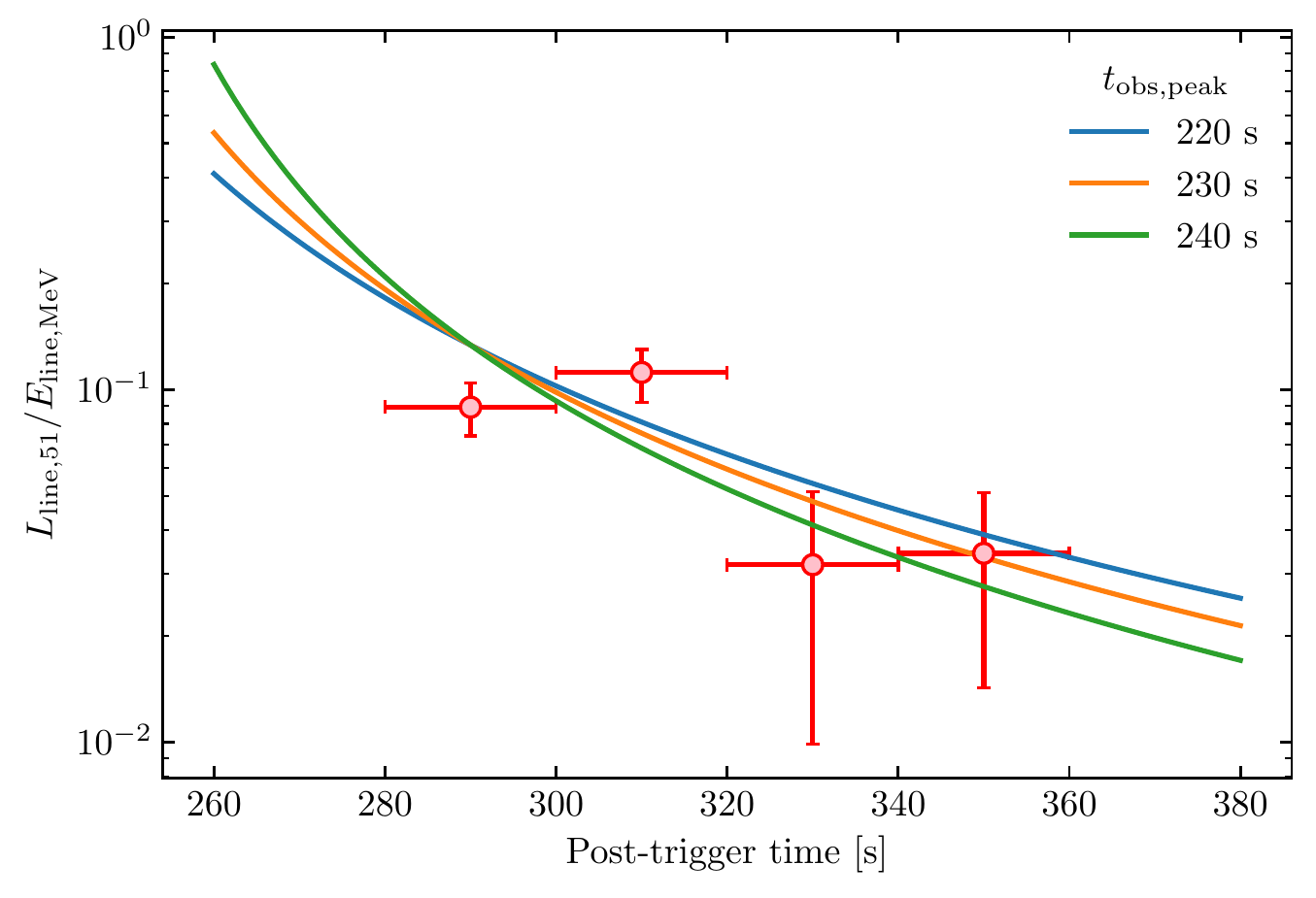}
    \caption{Time evolution of the observed ratio of the line luminosity to its central photon energy (red error bars), compared with the evolution predicted in the HLE scenario.}
    \label{fig:Lline_Eline_HLE}
\end{figure}

\clearpage
\bibliography{sn-bibliography}

\end{document}

%% file: journal-abbrev.tex
\newcommand\aj{{AJ}}%
\newcommand\actaa{{Acta Astron.}}%
\newcommand\araa{{ARA\&A}}%
\newcommand\apj{{ApJ}}%
\newcommand\apjl{{ApJ}}%
\newcommand\apjs{{ApJS}}%
\newcommand\ao{{Appl.~Opt.}}%
\newcommand\apss{{Ap\&SS}}%
\newcommand\aap{{A\&A}}%
\newcommand\aapr{{A\&A~Rev.}}%
\newcommand\aaps{{A\&AS}}%
\newcommand\azh{{AZh}}%
\newcommand\baas{{BAAS}}%
\newcommand\caa{{Chinese Astron. Astrophys.}}%
\newcommand\cjaa{{Chinese J. Astron. Astrophys.}}%
\newcommand\icarus{{Icarus}}%
\newcommand\jcap{{J. Cosmology Astropart. Phys.}}%
\newcommand\jrasc{{JRASC}}%
\newcommand\memras{{MmRAS}}%
\newcommand\mnras{{MNRAS}}%
\newcommand\na{{New A}}%
\newcommand\nar{{New A Rev.}}%
\newcommand\pra{{Phys.~Rev.~A}}%
\newcommand\prb{{Phys.~Rev.~B}}%
\newcommand\prc{{Phys.~Rev.~C}}%
\newcommand\prd{{Phys.~Rev.~D}}%
\newcommand\pre{{Phys.~Rev.~E}}%
\newcommand\prl{{Phys.~Rev.~Lett.}}%
\newcommand\pasa{{PASA}}%
\newcommand\pasp{{PASP}}%
\newcommand\pasj{{PASJ}}%
\newcommand\qjras{{QJRAS}}%
\newcommand\rmxaa{{Rev. Mexicana Astron. Astrofis.}}%
\newcommand\skytel{{S\&T}}%
\newcommand\solphys{{Sol.~Phys.}}%
\newcommand\sovast{{Soviet~Ast.}}%
\newcommand\ssr{{Space~Sci.~Rev.}}%
\newcommand\zap{{ZAp}}%
\newcommand\nat{{Nature}}%
\newcommand\iaucirc{{IAU~Circ.}}%
\newcommand\aplett{{Astrophys.~Lett.}}%
\newcommand\apspr{{Astrophys.~Space~Phys.~Res.}}%
\newcommand\bain{{Bull.~Astron.~Inst.~Netherlands}}%
\newcommand\fcp{{Fund.~Cosmic~Phys.}}%
\newcommand\gca{{Geochim.~Cosmochim.~Acta}}%
\newcommand\grl{{Geophys.~Res.~Lett.}}%
\newcommand\jcp{{J.~Chem.~Phys.}}%
\newcommand\jgr{{J.~Geophys.~Res.}}%
\newcommand\jqsrt{{J.~Quant.~Spec.~Radiat.~Transf.}}%
\newcommand\memsai{{Mem.~Soc.~Astron.~Italiana}}%
\newcommand\nphysa{{Nucl.~Phys.~A}}%
\newcommand\physrep{{Phys.~Rep.}}%
\newcommand\physscr{{Phys.~Scr}}%
\newcommand\planss{{Planet.~Space~Sci.}}%
\newcommand\procspie{{Proc.~SPIE}}%

%% file: sn-article.bbl

\begin{thebibliography}{68}
\ifx \bisbn   \undefined \def \bisbn  #1{ISBN #1}\fi
\ifx \binits  \undefined \def \binits#1{#1}\fi
\ifx \bauthor  \undefined \def \bauthor#1{#1}\fi
\ifx \batitle  \undefined \def \batitle#1{#1}\fi
\ifx \bjtitle  \undefined \def \bjtitle#1{#1}\fi
\ifx \bvolume  \undefined \def \bvolume#1{\textbf{#1}}\fi
\ifx \byear  \undefined \def \byear#1{#1}\fi
\ifx \bissue  \undefined \def \bissue#1{#1}\fi
\ifx \bfpage  \undefined \def \bfpage#1{#1}\fi
\ifx \blpage  \undefined \def \blpage #1{#1}\fi
\ifx \burl  \undefined \def \burl#1{\textsf{#1}}\fi
\ifx \doiurl  \undefined \def \doiurl#1{\url{https://doi.org/#1}}\fi
\ifx \betal  \undefined \def \betal{\textit{et al.}}\fi
\ifx \binstitute  \undefined \def \binstitute#1{#1}\fi
\ifx \binstitutionaled  \undefined \def \binstitutionaled#1{#1}\fi
\ifx \bctitle  \undefined \def \bctitle#1{#1}\fi
\ifx \beditor  \undefined \def \beditor#1{#1}\fi
\ifx \bpublisher  \undefined \def \bpublisher#1{#1}\fi
\ifx \bbtitle  \undefined \def \bbtitle#1{#1}\fi
\ifx \bedition  \undefined \def \bedition#1{#1}\fi
\ifx \bseriesno  \undefined \def \bseriesno#1{#1}\fi
\ifx \blocation  \undefined \def \blocation#1{#1}\fi
\ifx \bsertitle  \undefined \def \bsertitle#1{#1}\fi
\ifx \bsnm \undefined \def \bsnm#1{#1}\fi
\ifx \bsuffix \undefined \def \bsuffix#1{#1}\fi
\ifx \bparticle \undefined \def \bparticle#1{#1}\fi
\ifx \barticle \undefined \def \barticle#1{#1}\fi
\bibcommenthead
\ifx \bconfdate \undefined \def \bconfdate #1{#1}\fi
\ifx \botherref \undefined \def \botherref #1{#1}\fi
\ifx \url \undefined \def \url#1{\textsf{#1}}\fi
\ifx \bchapter \undefined \def \bchapter#1{#1}\fi
\ifx \bbook \undefined \def \bbook#1{#1}\fi
\ifx \bcomment \undefined \def \bcomment#1{#1}\fi
\ifx \oauthor \undefined \def \oauthor#1{#1}\fi
\ifx \citeauthoryear \undefined \def \citeauthoryear#1{#1}\fi
\ifx \endbibitem  \undefined \def \endbibitem {}\fi
\ifx \bconflocation  \undefined \def \bconflocation#1{#1}\fi
\ifx \arxivurl  \undefined \def \arxivurl#1{\textsf{#1}}\fi
\csname PreBibitemsHook\endcsname

\bibitem{Rees1994}
\begin{barticle}
\bauthor{\bsnm{{Rees}}, \binits{M.J.}},
\bauthor{\bsnm{{Meszaros}}, \binits{P.}}:
\batitle{{Unsteady Outflow Models for Cosmological Gamma-Ray Bursts}}.
\bjtitle{\apjl}
\bvolume{430},
\bfpage{93}
(\byear{1994})
{\href{https://arxiv.org/abs/astro-ph/9404038}{{arXiv:astro-ph/9404038}}}
{[astro-ph]}.
\doiurl{10.1086/187446}
\end{barticle}
\endbibitem

\bibitem{Sari1996}
\begin{barticle}
\bauthor{\bsnm{{Sari}}, \binits{R.}},
\bauthor{\bsnm{{Narayan}}, \binits{R.}},
\bauthor{\bsnm{{Piran}}, \binits{T.}}:
\batitle{{Cooling Timescales and Temporal Structure of Gamma-Ray Bursts}}.
\bjtitle{\apj}
\bvolume{473},
\bfpage{204}
(\byear{1996})
{\href{https://arxiv.org/abs/astro-ph/9605005}{{arXiv:astro-ph/9605005}}}
{[astro-ph]}.
\doiurl{10.1086/178136}
\end{barticle}
\endbibitem

\bibitem{Oganesyan2018}
\begin{barticle}
\bauthor{\bsnm{{Oganesyan}}, \binits{G.}},
\bauthor{\bsnm{{Nava}}, \binits{L.}},
\bauthor{\bsnm{{Ghirlanda}}, \binits{G.}},
\bauthor{\bsnm{{Celotti}}, \binits{A.}}:
\batitle{{Characterization of gamma-ray burst prompt emission spectra down to
  soft X-rays}}.
\bjtitle{\aap}
\bvolume{616},
\bfpage{138}
(\byear{2018})
{\href{https://arxiv.org/abs/1710.09383}{{arXiv:1710.09383}}}
{[astro-ph.HE]}.
\doiurl{10.1051/0004-6361/201732172}
\end{barticle}
\endbibitem

\bibitem{Ravasio2019}
\begin{barticle}
\bauthor{\bsnm{{Ravasio}}, \binits{M.E.}},
\bauthor{\bsnm{{Ghirlanda}}, \binits{G.}},
\bauthor{\bsnm{{Nava}}, \binits{L.}},
\bauthor{\bsnm{{Ghisellini}}, \binits{G.}}:
\batitle{{Evidence of two spectral breaks in the prompt emission of gamma-ray
  bursts}}.
\bjtitle{\aap}
\bvolume{625},
\bfpage{60}
(\byear{2019})
{\href{https://arxiv.org/abs/1903.02555}{{arXiv:1903.02555}}}
{[astro-ph.HE]}.
\doiurl{10.1051/0004-6361/201834987}
\end{barticle}
\endbibitem

\bibitem{Ackermann2012}
\begin{barticle}
\bauthor{\bsnm{{Fermi Large Area Telescope Team}}},
\bauthor{\bsnm{{Ackermann}}, \binits{M.}},
\bauthor{\bsnm{{Ajello}}, \binits{M.}},
\bauthor{\bsnm{{Baldini}}, \binits{L.}},
\bauthor{\bsnm{{Barbiellini}}, \binits{G.}},
\bauthor{\bsnm{{Baring}}, \binits{M.G.}},
\bauthor{\bsnm{{Bechtol}}, \binits{K.}},
\bauthor{\bsnm{{Bellazzini}}, \binits{R.}},
\bauthor{\bsnm{{Blandford}}, \binits{R.D.}},
\bauthor{\bsnm{{Bloom}}, \binits{E.D.}},
\bauthor{\bsnm{{Bonamente}}, \binits{E.}},
\bauthor{\bsnm{{Borgland}}, \binits{A.W.}},
\bauthor{\bsnm{{Bottacini}}, \binits{E.}},
\bauthor{\bsnm{{Bouvier}}, \binits{A.}},
\bauthor{\bsnm{{Brigida}}, \binits{M.}},
\bauthor{\bsnm{{Buehler}}, \binits{R.}},
\bauthor{\bsnm{{Buson}}, \binits{S.}},
\bauthor{\bsnm{{Caliandro}}, \binits{G.A.}},
\bauthor{\bsnm{{Cameron}}, \binits{R.A.}},
\bauthor{\bsnm{{Cecchi}}, \binits{C.}},
\bauthor{\bsnm{{Charles}}, \binits{E.}},
\bauthor{\bsnm{{Chekhtman}}, \binits{A.}},
\bauthor{\bsnm{{Chiang}}, \binits{J.}},
\bauthor{\bsnm{{Ciprini}}, \binits{S.}},
\bauthor{\bsnm{{Claus}}, \binits{R.}},
\bauthor{\bsnm{{Cohen-Tanugi}}, \binits{J.}},
\bauthor{\bsnm{{Cutini}}, \binits{S.}},
\bauthor{\bsnm{{D'Ammando}}, \binits{F.}},
\bauthor{\bsnm{{de Palma}}, \binits{F.}},
\bauthor{\bsnm{{Dermer}}, \binits{C.D.}},
\bauthor{\bsnm{{Silva}}, \binits{E.d.C.e.}},
\bauthor{\bsnm{{Drell}}, \binits{P.S.}},
\bauthor{\bsnm{{Drlica-Wagner}}, \binits{A.}},
\bauthor{\bsnm{{Favuzzi}}, \binits{C.}},
\bauthor{\bsnm{{Fukazawa}}, \binits{Y.}},
\bauthor{\bsnm{{Fusco}}, \binits{P.}},
\bauthor{\bsnm{{Gargano}}, \binits{F.}},
\bauthor{\bsnm{{Gasparrini}}, \binits{D.}},
\bauthor{\bsnm{{Gehrels}}, \binits{N.}},
\bauthor{\bsnm{{Germani}}, \binits{S.}},
\bauthor{\bsnm{{Giglietto}}, \binits{N.}},
\bauthor{\bsnm{{Giordano}}, \binits{F.}},
\bauthor{\bsnm{{Giroletti}}, \binits{M.}},
\bauthor{\bsnm{{Glanzman}}, \binits{T.}},
\bauthor{\bsnm{{Granot}}, \binits{J.}},
\bauthor{\bsnm{{Grenier}}, \binits{I.A.}},
\bauthor{\bsnm{{Grove}}, \binits{J.E.}},
\bauthor{\bsnm{{Hadasch}}, \binits{D.}},
\bauthor{\bsnm{{Hanabata}}, \binits{Y.}},
\bauthor{\bsnm{{Harding}}, \binits{A.K.}},
\bauthor{\bsnm{{Hays}}, \binits{E.}},
\bauthor{\bsnm{{Horan}}, \binits{D.}},
\bauthor{\bsnm{{J{\'o}hannesson}}, \binits{G.}},
\bauthor{\bsnm{{Kataoka}}, \binits{J.}},
\bauthor{\bsnm{{Kn{\"o}dlseder}}, \binits{J.}},
\bauthor{\bsnm{{Kocevski}}, \binits{D.}},
\bauthor{\bsnm{{Kuss}}, \binits{M.}},
\bauthor{\bsnm{{Lande}}, \binits{J.}},
\bauthor{\bsnm{{Longo}}, \binits{F.}},
\bauthor{\bsnm{{Loparco}}, \binits{F.}},
\bauthor{\bsnm{{Lovellette}}, \binits{M.N.}},
\bauthor{\bsnm{{Lubrano}}, \binits{P.}},
\bauthor{\bsnm{{Mazziotta}}, \binits{M.N.}},
\bauthor{\bsnm{{McEnery}}, \binits{J.}},
\bauthor{\bsnm{{McGlynn}}, \binits{S.}},
\bauthor{\bsnm{{Michelson}}, \binits{P.F.}},
\bauthor{\bsnm{{Mitthumsiri}}, \binits{W.}},
\bauthor{\bsnm{{Monzani}}, \binits{M.E.}},
\bauthor{\bsnm{{Moretti}}, \binits{E.}},
\bauthor{\bsnm{{Morselli}}, \binits{A.}},
\bauthor{\bsnm{{Moskalenko}}, \binits{I.V.}},
\bauthor{\bsnm{{Murgia}}, \binits{S.}},
\bauthor{\bsnm{{Naumann-Godo}}, \binits{M.}},
\bauthor{\bsnm{{Norris}}, \binits{J.P.}},
\bauthor{\bsnm{{Nuss}}, \binits{E.}},
\bauthor{\bsnm{{Nymark}}, \binits{T.}},
\bauthor{\bsnm{{Ohsugi}}, \binits{T.}},
\bauthor{\bsnm{{Okumura}}, \binits{A.}},
\bauthor{\bsnm{{Omodei}}, \binits{N.}},
\bauthor{\bsnm{{Orlando}}, \binits{E.}},
\bauthor{\bsnm{{Panetta}}, \binits{J.H.}},
\bauthor{\bsnm{{Parent}}, \binits{D.}},
\bauthor{\bsnm{{Pelassa}}, \binits{V.}},
\bauthor{\bsnm{{Pesce-Rollins}}, \binits{M.}},
\bauthor{\bsnm{{Piron}}, \binits{F.}},
\bauthor{\bsnm{{Pivato}}, \binits{G.}},
\bauthor{\bsnm{{Racusin}}, \binits{J.L.}},
\bauthor{\bsnm{{Rain{\`o}}}, \binits{S.}},
\bauthor{\bsnm{{Rando}}, \binits{R.}},
\bauthor{\bsnm{{Razzaque}}, \binits{S.}},
\bauthor{\bsnm{{Reimer}}, \binits{A.}},
\bauthor{\bsnm{{Reimer}}, \binits{O.}},
\bauthor{\bsnm{{Ritz}}, \binits{S.}},
\bauthor{\bsnm{{Ryde}}, \binits{F.}},
\bauthor{\bsnm{{Sgr{\`o}}}, \binits{C.}},
\bauthor{\bsnm{{Siskind}}, \binits{E.J.}},
\bauthor{\bsnm{{Sonbas}}, \binits{E.}},
\bauthor{\bsnm{{Spandre}}, \binits{G.}},
\bauthor{\bsnm{{Spinelli}}, \binits{P.}},
\bauthor{\bsnm{{Stamatikos}}, \binits{M.}},
\bauthor{\bsnm{{Stawarz}}, \binits{{\L}.}},
\bauthor{\bsnm{{Suson}}, \binits{D.J.}},
\bauthor{\bsnm{{Takahashi}}, \binits{H.}},
\bauthor{\bsnm{{Tanaka}}, \binits{T.}},
\bauthor{\bsnm{{Thayer}}, \binits{J.G.}},
\bauthor{\bsnm{{Thayer}}, \binits{J.B.}},
\bauthor{\bsnm{{Tibaldo}}, \binits{L.}},
\bauthor{\bsnm{{Tinivella}}, \binits{M.}},
\bauthor{\bsnm{{Tosti}}, \binits{G.}},
\bauthor{\bsnm{{Uehara}}, \binits{T.}},
\bauthor{\bsnm{{Vandenbroucke}}, \binits{J.}},
\bauthor{\bsnm{{Vasileiou}}, \binits{V.}},
\bauthor{\bsnm{{Vianello}}, \binits{G.}},
\bauthor{\bsnm{{Vitale}}, \binits{V.}},
\bauthor{\bsnm{{Waite}}, \binits{A.P.}},
\bauthor{\bsnm{{Fermi Gamma-ray Burst Monitor Team}}},
\bauthor{\bsnm{{Connaughton}}, \binits{V.}},
\bauthor{\bsnm{{Briggs}}, \binits{M.S.}},
\bauthor{\bsnm{{Guirec}}, \binits{S.}},
\bauthor{\bsnm{{Goldstein}}, \binits{A.}},
\bauthor{\bsnm{{Burgess}}, \binits{J.M.}},
\bauthor{\bsnm{{Bhat}}, \binits{P.N.}},
\bauthor{\bsnm{{Bissaldi}}, \binits{E.}},
\bauthor{\bsnm{{Camero-Arranz}}, \binits{A.}},
\bauthor{\bsnm{{Fishman}}, \binits{J.}},
\bauthor{\bsnm{{Fitzpatrick}}, \binits{G.}},
\bauthor{\bsnm{{Foley}}, \binits{S.}},
\bauthor{\bsnm{{Gruber}}, \binits{D.}},
\bauthor{\bsnm{{Jenke}}, \binits{P.}},
\bauthor{\bsnm{{Kippen}}, \binits{R.M.}},
\bauthor{\bsnm{{Kouveliotou}}, \binits{C.}},
\bauthor{\bsnm{{McBreen}}, \binits{S.}},
\bauthor{\bsnm{{Meegan}}, \binits{C.}},
\bauthor{\bsnm{{Paciesas}}, \binits{W.S.}},
\bauthor{\bsnm{{Preece}}, \binits{R.}},
\bauthor{\bsnm{{Rau}}, \binits{A.}},
\bauthor{\bsnm{{Tierney}}, \binits{D.}},
\bauthor{\bsnm{{van der Horst}}, \binits{A.J.}},
\bauthor{\bsnm{{von Kienlin}}, \binits{A.}},
\bauthor{\bsnm{{Wilson-Hodge}}, \binits{C.}},
\bauthor{\bsnm{{Xiong}}, \binits{S.}}:
\batitle{{Constraining the High-energy Emission from Gamma-Ray Bursts with
  Fermi}}.
\bjtitle{\apj}
\bvolume{754}(\bissue{2}),
\bfpage{121}
(\byear{2012})
{\href{https://arxiv.org/abs/1201.3948}{{arXiv:1201.3948}}}
{[astro-ph.HE]}.
\doiurl{10.1088/0004-637X/754/2/121}
\end{barticle}
\endbibitem

\bibitem{Vianello2018}
\begin{barticle}
\bauthor{\bsnm{{Vianello}}, \binits{G.}},
\bauthor{\bsnm{{Gill}}, \binits{R.}},
\bauthor{\bsnm{{Granot}}, \binits{J.}},
\bauthor{\bsnm{{Omodei}}, \binits{N.}},
\bauthor{\bsnm{{Cohen-Tanugi}}, \binits{J.}},
\bauthor{\bsnm{{Longo}}, \binits{F.}}:
\batitle{{The Bright and the Slow{\textemdash}GRBs 100724B and 160509A with
  High-energy Cutoffs at {\ensuremath{\lesssim}}100 MeV}}.
\bjtitle{\apj}
\bvolume{864}(\bissue{2}),
\bfpage{163}
(\byear{2018})
{\href{https://arxiv.org/abs/1706.01481}{{arXiv:1706.01481}}}
{[astro-ph.HE]}.
\doiurl{10.3847/1538-4357/aad6ea}
\end{barticle}
\endbibitem

\bibitem{Frederiks2023}
\begin{botherref}
\oauthor{\bsnm{{Frederiks}}, \binits{D.}},
\oauthor{\bsnm{{Svinkin}}, \binits{D.}},
\oauthor{\bsnm{{Lysenko}}, \binits{A.L.}},
\oauthor{\bsnm{{Molkov}}, \binits{S.}},
\oauthor{\bsnm{{Tsvetkova}}, \binits{A.}},
\oauthor{\bsnm{{Ulanov}}, \binits{M.}},
\oauthor{\bsnm{{Ridnaia}}, \binits{A.}},
\oauthor{\bsnm{{Lutovinov}}, \binits{A.A.}},
\oauthor{\bsnm{{Lapshov}}, \binits{I.}},
\oauthor{\bsnm{{Tkachenko}}, \binits{A.}},
\oauthor{\bsnm{{Levin}}, \binits{V.}}:
{Properties of the extremely energetic GRB\,221009A from Konus-\textit{WIND}
  and \textit{SRG}/ART-XC observations}.
arXiv e-prints,
2302--13383
(2023)
{\href{https://arxiv.org/abs/2302.13383}{{arXiv:2302.13383}}}
{[astro-ph.HE]}.
\doiurl{10.48550/arXiv.2302.13383}
\end{botherref}
\endbibitem

\bibitem{An2023}
\begin{botherref}
\oauthor{\bsnm{{An}}, \binits{Z.-H.}},
\oauthor{\bsnm{{Antier}}, \binits{S.}},
\oauthor{\bsnm{{Bi}}, \binits{X.-Z.}},
\oauthor{\bsnm{{Bu}}, \binits{Q.-C.}},
\oauthor{\bsnm{{Cai}}, \binits{C.}},
\oauthor{\bsnm{{Cao}}, \binits{X.-L.}},
\oauthor{\bsnm{{Camisasca}}, \binits{A.-E.}},
\oauthor{\bsnm{{Chang}}, \binits{Z.}},
\oauthor{\bsnm{{Chen}}, \binits{G.}},
\oauthor{\bsnm{{Chen}}, \binits{L.}},
\oauthor{\bsnm{{Chen}}, \binits{T.-X.}},
\oauthor{\bsnm{{Chen}}, \binits{W.}},
\oauthor{\bsnm{{Chen}}, \binits{Y.-B.}},
\oauthor{\bsnm{{Chen}}, \binits{Y.}},
\oauthor{\bsnm{{Chen}}, \binits{Y.-P.}},
\oauthor{\bsnm{{Coughlin}}, \binits{M.W.}},
\oauthor{\bsnm{{Cui}}, \binits{W.-W.}},
\oauthor{\bsnm{{Dai}}, \binits{Z.-G.}},
\oauthor{\bsnm{{Hussenot-Desenonges}}, \binits{T.}},
\oauthor{\bsnm{{Du}}, \binits{Y.-Q.}},
\oauthor{\bsnm{{Du}}, \binits{Y.-Y.}},
\oauthor{\bsnm{{Du}}, \binits{Y.-F.}},
\oauthor{\bsnm{{Fan}}, \binits{C.-C.}},
\oauthor{\bsnm{{Frontera}}, \binits{F.}},
\oauthor{\bsnm{{Gao}}, \binits{H.}},
\oauthor{\bsnm{{Gao}}, \binits{M.}},
\oauthor{\bsnm{{Ge}}, \binits{M.-Y.}},
\oauthor{\bsnm{{Gong}}, \binits{K.}},
\oauthor{\bsnm{{Gu}}, \binits{Y.-D.}},
\oauthor{\bsnm{{Guan}}, \binits{J.}},
\oauthor{\bsnm{{Guo}}, \binits{D.-Y.}},
\oauthor{\bsnm{{Guo}}, \binits{Z.-W.}},
\oauthor{\bsnm{{Guidorzi}}, \binits{C.}},
\oauthor{\bsnm{{Han}}, \binits{D.-W.}},
\oauthor{\bsnm{{He}}, \binits{J.-J.}},
\oauthor{\bsnm{{He}}, \binits{J.-W.}},
\oauthor{\bsnm{{Hou}}, \binits{D.-J.}},
\oauthor{\bsnm{{Huang}}, \binits{Y.}},
\oauthor{\bsnm{{Huo}}, \binits{J.}},
\oauthor{\bsnm{{Ji}}, \binits{Z.}},
\oauthor{\bsnm{{Jia}}, \binits{S.-M.}},
\oauthor{\bsnm{{Jiang}}, \binits{W.-C.}},
\oauthor{\bsnm{{Kann}}, \binits{D.A.}},
\oauthor{\bsnm{{Klotz}}, \binits{A.}},
\oauthor{\bsnm{{Kong}}, \binits{L.-D.}},
\oauthor{\bsnm{{Lan}}, \binits{L.}},
\oauthor{\bsnm{{Li}}, \binits{A.}},
\oauthor{\bsnm{{Li}}, \binits{B.}},
\oauthor{\bsnm{{Li}}, \binits{C.-Y.}},
\oauthor{\bsnm{{Li}}, \binits{C.-K.}},
\oauthor{\bsnm{{Li}}, \binits{G.}},
\oauthor{\bsnm{{Li}}, \binits{M.-S.}},
\oauthor{\bsnm{{Li}}, \binits{T.-P.}},
\oauthor{\bsnm{{Li}}, \binits{W.}},
\oauthor{\bsnm{{Li}}, \binits{X.-B.}},
\oauthor{\bsnm{{Li}}, \binits{X.-Q.}},
\oauthor{\bsnm{{Li}}, \binits{X.-F.}},
\oauthor{\bsnm{{Li}}, \binits{Y.-G.}},
\oauthor{\bsnm{{Li}}, \binits{Z.-W.}},
\oauthor{\bsnm{{Liang}}, \binits{J.}},
\oauthor{\bsnm{{Liang}}, \binits{X.-H.}},
\oauthor{\bsnm{{Liao}}, \binits{J.-Y.}},
\oauthor{\bsnm{{Lin}}, \binits{L.}},
\oauthor{\bsnm{{Liu}}, \binits{C.-Z.}},
\oauthor{\bsnm{{Liu}}, \binits{H.-X.}},
\oauthor{\bsnm{{Liu}}, \binits{H.-W.}},
\oauthor{\bsnm{{Liu}}, \binits{J.-C.}},
\oauthor{\bsnm{{Liu}}, \binits{X.-J.}},
\oauthor{\bsnm{{Liu}}, \binits{Y.-Q.}},
\oauthor{\bsnm{{Liu}}, \binits{Y.-R.}},
\oauthor{\bsnm{{Lu}}, \binits{F.-J.}},
\oauthor{\bsnm{{Lu}}, \binits{H.}},
\oauthor{\bsnm{{Lu}}, \binits{X.-F.}},
\oauthor{\bsnm{{Luo}}, \binits{Q.}},
\oauthor{\bsnm{{Luo}}, \binits{T.}},
\oauthor{\bsnm{{Ma}}, \binits{B.-Y.}},
\oauthor{\bsnm{{Ma}}, \binits{F.-L.}},
\oauthor{\bsnm{{Ma}}, \binits{R.-C.}},
\oauthor{\bsnm{{Ma}}, \binits{X.}},
\oauthor{\bsnm{{Maccary}}, \binits{R.}},
\oauthor{\bsnm{{Mao}}, \binits{J.-R.}},
\oauthor{\bsnm{{Meng}}, \binits{B.}},
\oauthor{\bsnm{{Nie}}, \binits{J.-Y.}},
\oauthor{\bsnm{{Orlandini}}, \binits{M.}},
\oauthor{\bsnm{{Ou}}, \binits{G.}},
\oauthor{\bsnm{{Peng}}, \binits{J.-Q.}},
\oauthor{\bsnm{{Peng}}, \binits{W.-X.}},
\oauthor{\bsnm{{Qiao}}, \binits{R.}},
\oauthor{\bsnm{{Qu}}, \binits{J.-L.}},
\oauthor{\bsnm{{Ren}}, \binits{X.-Q.}},
\oauthor{\bsnm{{Shi}}, \binits{J.-Y.}},
\oauthor{\bsnm{{Shi}}, \binits{Q.}},
\oauthor{\bsnm{{Song}}, \binits{L.-M.}},
\oauthor{\bsnm{{Song}}, \binits{X.-Y.}},
\oauthor{\bsnm{{Su}}, \binits{J.}},
\oauthor{\bsnm{{Sun}}, \binits{G.-X.}},
\oauthor{\bsnm{{Sun}}, \binits{L.}},
\oauthor{\bsnm{{Sun}}, \binits{X.-L.}},
\oauthor{\bsnm{{Tan}}, \binits{W.-J.}},
\oauthor{\bsnm{{Tan}}, \binits{Y.}},
\oauthor{\bsnm{{Tao}}, \binits{L.}},
\oauthor{\bsnm{{Tuo}}, \binits{Y.-L.}},
\oauthor{\bsnm{{Turpin}}, \binits{D.}},
\oauthor{\bsnm{{Wang}}, \binits{J.-Z.}},
\oauthor{\bsnm{{Wang}}, \binits{C.}},
\oauthor{\bsnm{{Wang}}, \binits{C.-W.}},
\oauthor{\bsnm{{Wang}}, \binits{H.-J.}},
\oauthor{\bsnm{{Wang}}, \binits{H.}},
\oauthor{\bsnm{{Wang}}, \binits{J.}},
\oauthor{\bsnm{{Wang}}, \binits{L.-J.}},
\oauthor{\bsnm{{Wang}}, \binits{P.-J.}},
\oauthor{\bsnm{{Wang}}, \binits{P.}},
\oauthor{\bsnm{{Wang}}, \binits{W.-S.}},
\oauthor{\bsnm{{Wang}}, \binits{X.-Y.}},
\oauthor{\bsnm{{Wang}}, \binits{X.-L.}},
\oauthor{\bsnm{{Wang}}, \binits{Y.-S.}},
\oauthor{\bsnm{{Wang}}, \binits{Y.}},
\oauthor{\bsnm{{Wen}}, \binits{X.-Y.}},
\oauthor{\bsnm{{Wu}}, \binits{B.-B.}},
\oauthor{\bsnm{{Wu}}, \binits{B.-Y.}},
\oauthor{\bsnm{{Wu}}, \binits{H.}},
\oauthor{\bsnm{{Xiao}}, \binits{S.-H.}},
\oauthor{\bsnm{{Xiao}}, \binits{S.}},
\oauthor{\bsnm{{Xiao}}, \binits{Y.-X.}},
\oauthor{\bsnm{{Xie}}, \binits{S.-L.}},
\oauthor{\bsnm{{Xiong}}, \binits{S.-L.}},
\oauthor{\bsnm{{Xiong}}, \binits{S.-L.}},
\oauthor{\bsnm{{Xu}}, \binits{D.}},
\oauthor{\bsnm{{Xu}}, \binits{H.}},
\oauthor{\bsnm{{Xu}}, \binits{Y.-J.}},
\oauthor{\bsnm{{Xu}}, \binits{Y.-B.}},
\oauthor{\bsnm{{Xu}}, \binits{Y.-C.}},
\oauthor{\bsnm{{Xu}}, \binits{Y.-P.}},
\oauthor{\bsnm{{Xue}}, \binits{W.-C.}},
\oauthor{\bsnm{{Yang}}, \binits{S.}},
\oauthor{\bsnm{{Yang}}, \binits{Y.-J.}},
\oauthor{\bsnm{{Yang}}, \binits{Z.-X.}},
\oauthor{\bsnm{{Ye}}, \binits{W.-T.}},
\oauthor{\bsnm{{Yi}}, \binits{Q.-B.}},
\oauthor{\bsnm{{Yi}}, \binits{S.-X.}},
\oauthor{\bsnm{{Yin}}, \binits{Q.-Q.}},
\oauthor{\bsnm{{You}}, \binits{Y.}},
\oauthor{\bsnm{{Yu}}, \binits{Y.-W.}},
\oauthor{\bsnm{{Yu}}, \binits{W.}},
\oauthor{\bsnm{{Yu}}, \binits{W.-H.}},
\oauthor{\bsnm{{Zeng}}, \binits{M.}},
\oauthor{\bsnm{{Zhang}}, \binits{B.}},
\oauthor{\bsnm{{Zhang}}, \binits{B.-B.}},
\oauthor{\bsnm{{Zhang}}, \binits{D.-L.}},
\oauthor{\bsnm{{Zhang}}, \binits{F.}},
\oauthor{\bsnm{{Zhang}}, \binits{H.-M.}},
\oauthor{\bsnm{{Zhang}}, \binits{J.}},
\oauthor{\bsnm{{Zhang}}, \binits{L.}},
\oauthor{\bsnm{{Zhang}}, \binits{P.}},
\oauthor{\bsnm{{Zhang}}, \binits{P.}},
\oauthor{\bsnm{{Zhang}}, \binits{S.}},
\oauthor{\bsnm{{Zhang}}, \binits{S.-N.}},
\oauthor{\bsnm{{Zhang}}, \binits{W.-C.}},
\oauthor{\bsnm{{Zhang}}, \binits{X.-F.}},
\oauthor{\bsnm{{Zhang}}, \binits{X.-L.}},
\oauthor{\bsnm{{Zhang}}, \binits{Y.-Q.}},
\oauthor{\bsnm{{Zhang}}, \binits{Y.-T.}},
\oauthor{\bsnm{{Zhang}}, \binits{Y.-F.}},
\oauthor{\bsnm{{Zhang}}, \binits{Y.-H.}},
\oauthor{\bsnm{{Zhang}}, \binits{Z.}},
\oauthor{\bsnm{{Zhao}}, \binits{G.-Y.}},
\oauthor{\bsnm{{Zhao}}, \binits{H.-S.}},
\oauthor{\bsnm{{Zhao}}, \binits{H.-Y.}},
\oauthor{\bsnm{{Zhao}}, \binits{Q.-X.}},
\oauthor{\bsnm{{Zhao}}, \binits{S.-J.}},
\oauthor{\bsnm{{Zhao}}, \binits{X.-Y.}},
\oauthor{\bsnm{{Zhao}}, \binits{X.-F.}},
\oauthor{\bsnm{{Zhao}}, \binits{Y.}},
\oauthor{\bsnm{{Zheng}}, \binits{C.}},
\oauthor{\bsnm{{Zheng}}, \binits{S.-J.}},
\oauthor{\bsnm{{Zhou}}, \binits{D.-K.}},
\oauthor{\bsnm{{Zhou}}, \binits{X.}},
\oauthor{\bsnm{{Zhu}}, \binits{X.-C.}}:
{Insight-HXMT and GECAM-C observations of the brightest-of-all-time GRB
  221009A}.
arXiv e-prints,
2303--01203
(2023)
{\href{https://arxiv.org/abs/2303.01203}{{arXiv:2303.01203}}}
{[astro-ph.HE]}.
\doiurl{10.48550/arXiv.2303.01203}
\end{botherref}
\endbibitem

\bibitem{Burns2023}
\begin{botherref}
\oauthor{\bsnm{{Burns}}, \binits{E.}},
\oauthor{\bsnm{{Svinkin}}, \binits{D.}},
\oauthor{\bsnm{{Fenimore}}, \binits{E.}},
\oauthor{\bsnm{{Ag{\"u}{\'\i} Fern{\'a}ndez}}, \binits{J.F.}},
\oauthor{\bsnm{{Frederiks}}, \binits{D.}},
\oauthor{\bsnm{{Kann}}, \binits{D.A.}},
\oauthor{\bsnm{{Hamburg}}, \binits{R.}},
\oauthor{\bsnm{{Lesage}}, \binits{S.}},
\oauthor{\bsnm{{Temiraev}}, \binits{Y.}},
\oauthor{\bsnm{{Tsvetkova}}, \binits{A.}},
\oauthor{\bsnm{{Bissaldi}}, \binits{E.}},
\oauthor{\bsnm{{Briggs}}, \binits{M.S.}},
\oauthor{\bsnm{{Fletcher}}, \binits{C.}},
\oauthor{\bsnm{{Goldstein}}, \binits{A.}},
\oauthor{\bsnm{{Hui}}, \binits{C.M.}},
\oauthor{\bsnm{{Hristov}}, \binits{B.A.}},
\oauthor{\bsnm{{Kocevski}}, \binits{D.}},
\oauthor{\bsnm{{Lysenko}}, \binits{A.L.}},
\oauthor{\bsnm{{Mailyan}}, \binits{B.}},
\oauthor{\bsnm{{Racusin}}, \binits{J.}},
\oauthor{\bsnm{{Ridnaia}}, \binits{A.}},
\oauthor{\bsnm{{Roberts}}, \binits{O.J.}},
\oauthor{\bsnm{{Ulanov}}, \binits{M.}},
\oauthor{\bsnm{{Veres}}, \binits{P.}},
\oauthor{\bsnm{{Wilson-Hodge}}, \binits{C.A.}},
\oauthor{\bsnm{{Wood}}, \binits{J.}}:
{GRB 221009A, The BOAT}.
arXiv e-prints,
2302--14037
(2023)
{\href{https://arxiv.org/abs/2302.14037}{{arXiv:2302.14037}}}
{[astro-ph.HE]}.
\doiurl{10.48550/arXiv.2302.14037}
\end{botherref}
\endbibitem

\bibitem{Malesani2023}
\begin{botherref}
\oauthor{\bsnm{{Malesani}}, \binits{D.B.}},
\oauthor{\bsnm{{Levan}}, \binits{A.J.}},
\oauthor{\bsnm{{Izzo}}, \binits{L.}},
\oauthor{\bsnm{{de Ugarte Postigo}}, \binits{A.}},
\oauthor{\bsnm{{Ghirlanda}}, \binits{G.}},
\oauthor{\bsnm{{Heintz}}, \binits{K.E.}},
\oauthor{\bsnm{{Kann}}, \binits{D.A.}},
\oauthor{\bsnm{{Lamb}}, \binits{G.P.}},
\oauthor{\bsnm{{Palmerio}}, \binits{J.}},
\oauthor{\bsnm{{Salafia}}, \binits{O.S.}},
\oauthor{\bsnm{{Salvaterra}}, \binits{R.}},
\oauthor{\bsnm{{Tanvir}}, \binits{N.R.}},
\oauthor{\bsnm{{Ag{\"u}{\'\i} Fern{\'a}ndez}}, \binits{J.F.}},
\oauthor{\bsnm{{Campana}}, \binits{S.}},
\oauthor{\bsnm{{Chrimes}}, \binits{A.A.}},
\oauthor{\bsnm{{D'Avanzo}}, \binits{P.}},
\oauthor{\bsnm{{D'Elia}}, \binits{V.}},
\oauthor{\bsnm{{Della Valle}}, \binits{M.}},
\oauthor{\bsnm{{De Pasquale}}, \binits{M.}},
\oauthor{\bsnm{{Fynbo}}, \binits{J.P.U.}},
\oauthor{\bsnm{{Gaspari}}, \binits{N.}},
\oauthor{\bsnm{{Gompertz}}, \binits{B.P.}},
\oauthor{\bsnm{{Hartmann}}, \binits{D.H.}},
\oauthor{\bsnm{{Hjorth}}, \binits{J.}},
\oauthor{\bsnm{{Jakobsson}}, \binits{P.}},
\oauthor{\bsnm{{Palazzi}}, \binits{E.}},
\oauthor{\bsnm{{Pian}}, \binits{E.}},
\oauthor{\bsnm{{Pugliese}}, \binits{G.}},
\oauthor{\bsnm{{Ravasio}}, \binits{M.E.}},
\oauthor{\bsnm{{Rossi}}, \binits{A.}},
\oauthor{\bsnm{{Saccardi}}, \binits{A.}},
\oauthor{\bsnm{{Schady}}, \binits{P.}},
\oauthor{\bsnm{{Schneider}}, \binits{B.}},
\oauthor{\bsnm{{Sollerman}}, \binits{J.}},
\oauthor{\bsnm{{Starling}}, \binits{R.L.C.}},
\oauthor{\bsnm{{Th{\"o}ne}}, \binits{C.C.}},
\oauthor{\bsnm{{van der Horst}}, \binits{A.J.}},
\oauthor{\bsnm{{Vergani}}, \binits{S.D.}},
\oauthor{\bsnm{{Watson}}, \binits{D.}},
\oauthor{\bsnm{{Wiersema}}, \binits{K.}},
\oauthor{\bsnm{{Xu}}, \binits{D.}},
\oauthor{\bsnm{{Zafar}}, \binits{T.}}:
{The brightest GRB ever detected: GRB 221009A as a highly luminous event at z =
  0.151}.
arXiv e-prints,
2302--07891
(2023)
{\href{https://arxiv.org/abs/2302.07891}{{arXiv:2302.07891}}}
{[astro-ph.HE]}.
\doiurl{10.48550/arXiv.2302.07891}
\end{botherref}
\endbibitem

\bibitem{Akaike1974}
\begin{barticle}
\bauthor{\bsnm{{Akaike}}, \binits{H.}}:
\batitle{{A New Look at the Statistical Model Identification}}.
\bjtitle{IEEE Transactions on Automatic Control}
\bvolume{19},
\bfpage{716}--\blpage{723}
(\byear{1974})
\end{barticle}
\endbibitem

\bibitem{Burnham2004}
\begin{barticle}
\bauthor{\bsnm{Burnham}, \binits{K.P.}},
\bauthor{\bsnm{Anderson}, \binits{D.R.}}:
\batitle{Multimodel inference: Understanding aic and bic in model selection}.
\bjtitle{Sociological Methods \& Research}
\bvolume{33}(\bissue{2}),
\bfpage{261}--\blpage{304}
(\byear{2004})
{\href{https://arxiv.org/abs/https://doi.org/10.1177/0049124104268644}{{https://doi.org/10.1177/0049124104268644}}}.
\doiurl{10.1177/0049124104268644}
\end{barticle}
\endbibitem

\bibitem{Meszaros1993}
\begin{barticle}
\bauthor{\bsnm{{Meszaros}}, \binits{P.}},
\bauthor{\bsnm{{Rees}}, \binits{M.J.}}:
\batitle{{Relativistic Fireballs and Their Impact on External Matter: Models
  for Cosmological Gamma-Ray Bursts}}.
\bjtitle{\apj}
\bvolume{405},
\bfpage{278}
(\byear{1993}).
\doiurl{10.1086/172360}
\end{barticle}
\endbibitem

\bibitem{Sari1998}
\begin{barticle}
\bauthor{\bsnm{{Sari}}, \binits{R.}},
\bauthor{\bsnm{{Piran}}, \binits{T.}},
\bauthor{\bsnm{{Narayan}}, \binits{R.}}:
\batitle{{Spectra and Light Curves of Gamma-Ray Burst Afterglows}}.
\bjtitle{\apjl}
\bvolume{497}(\bissue{1}),
\bfpage{17}--\blpage{20}
(\byear{1998})
{\href{https://arxiv.org/abs/astro-ph/9712005}{{arXiv:astro-ph/9712005}}}
{[astro-ph]}.
\doiurl{10.1086/311269}
\end{barticle}
\endbibitem

\bibitem{Ravasio2019b}
\begin{barticle}
\bauthor{\bsnm{{Ravasio}}, \binits{M.E.}},
\bauthor{\bsnm{{Oganesyan}}, \binits{G.}},
\bauthor{\bsnm{{Salafia}}, \binits{O.S.}},
\bauthor{\bsnm{{Ghirlanda}}, \binits{G.}},
\bauthor{\bsnm{{Ghisellini}}, \binits{G.}},
\bauthor{\bsnm{{Branchesi}}, \binits{M.}},
\bauthor{\bsnm{{Campana}}, \binits{S.}},
\bauthor{\bsnm{{Covino}}, \binits{S.}},
\bauthor{\bsnm{{Salvaterra}}, \binits{R.}}:
\batitle{{GRB 190114C: from prompt to afterglow?}}
\bjtitle{\aap}
\bvolume{626},
\bfpage{12}
(\byear{2019})
{\href{https://arxiv.org/abs/1902.01861}{{arXiv:1902.01861}}}
{[astro-ph.HE]}.
\doiurl{10.1051/0004-6361/201935214}
\end{barticle}
\endbibitem

\bibitem{Lesage2023}
\begin{botherref}
\oauthor{\bsnm{{Lesage}}, \binits{S.}},
\oauthor{\bsnm{{Veres}}, \binits{P.}},
\oauthor{\bsnm{{Briggs}}, \binits{M.S.}},
\oauthor{\bsnm{{Goldstein}}, \binits{A.}},
\oauthor{\bsnm{{Kocevski}}, \binits{D.}},
\oauthor{\bsnm{{Burns}}, \binits{E.}},
\oauthor{\bsnm{{Wilson-Hodge}}, \binits{C.A.}},
\oauthor{\bsnm{{Bhat}}, \binits{P.N.}},
\oauthor{\bsnm{{Huppenkothen}}, \binits{D.}},
\oauthor{\bsnm{{Fryer}}, \binits{C.L.}},
\oauthor{\bsnm{{Hamburg}}, \binits{R.}},
\oauthor{\bsnm{{Racusin}}, \binits{J.}},
\oauthor{\bsnm{{Bissaldi}}, \binits{E.}},
\oauthor{\bsnm{{Cleveland}}, \binits{W.H.}},
\oauthor{\bsnm{{Dalessi}}, \binits{S.}},
\oauthor{\bsnm{{Fletcher}}, \binits{C.}},
\oauthor{\bsnm{{Giles}}, \binits{M.M.}},
\oauthor{\bsnm{{Hristov}}, \binits{B.A.}},
\oauthor{\bsnm{{Hui}}, \binits{C.M.}},
\oauthor{\bsnm{{Mailyan}}, \binits{B.}},
\oauthor{\bsnm{{Poolakkil}}, \binits{S.}},
\oauthor{\bsnm{{Roberts}}, \binits{O.J.}},
\oauthor{\bsnm{{von Kienlin}}, \binits{A.}},
\oauthor{\bsnm{{Wood}}, \binits{J.}},
\oauthor{\bsnm{{Ajello}}, \binits{M.}},
\oauthor{\bsnm{{Arimoto}}, \binits{M.}},
\oauthor{\bsnm{{Baldini}}, \binits{L.}},
\oauthor{\bsnm{{Ballet}}, \binits{J.}},
\oauthor{\bsnm{{Baring}}, \binits{M.G.}},
\oauthor{\bsnm{{Bastieri}}, \binits{D.}},
\oauthor{\bsnm{{Becerra Gonzalez}}, \binits{J.}},
\oauthor{\bsnm{{Bellazzini}}, \binits{R.}},
\oauthor{\bsnm{{Bissaldi}}, \binits{E.}},
\oauthor{\bsnm{{Blandford}}, \binits{R.D.}},
\oauthor{\bsnm{{Bonino}}, \binits{R.}},
\oauthor{\bsnm{{Bruel}}, \binits{P.}},
\oauthor{\bsnm{{Buson}}, \binits{S.}},
\oauthor{\bsnm{{Cameron}}, \binits{R.A.}},
\oauthor{\bsnm{{Caputo}}, \binits{R.}},
\oauthor{\bsnm{{Caraveo}}, \binits{P.A.}},
\oauthor{\bsnm{{Cavazzuti}}, \binits{E.}},
\oauthor{\bsnm{{Chiaro}}, \binits{G.}},
\oauthor{\bsnm{{Cibrario}}, \binits{N.}},
\oauthor{\bsnm{{Ciprini}}, \binits{S.}},
\oauthor{\bsnm{{Cristarella Orestano}}, \binits{P.}},
\oauthor{\bsnm{{Crnogorcevic}}, \binits{M.}},
\oauthor{\bsnm{{Cuoco}}, \binits{A.}},
\oauthor{\bsnm{{Cutini}}, \binits{S.}},
\oauthor{\bsnm{{DAmmando}}, \binits{F.}},
\oauthor{\bsnm{{De Gaetano}}, \binits{S.}},
\oauthor{\bsnm{{Di Lalla}}, \binits{N.}},
\oauthor{\bsnm{{Di Venere}}, \binits{L.}},
\oauthor{\bsnm{{Dominguez}}, \binits{A.}},
\oauthor{\bsnm{{Fegan}}, \binits{S.J.}},
\oauthor{\bsnm{{Ferrara}}, \binits{E.C.}},
\oauthor{\bsnm{{Fleischhack}}, \binits{H.}},
\oauthor{\bsnm{{Fukazawa}}, \binits{Y.}},
\oauthor{\bsnm{{Funk}}, \binits{S.}},
\oauthor{\bsnm{{Fusco}}, \binits{P.}},
\oauthor{\bsnm{{Galanti}}, \binits{G.}},
\oauthor{\bsnm{{Gammaldi}}, \binits{V.}},
\oauthor{\bsnm{{Gargano}}, \binits{F.}},
\oauthor{\bsnm{{Gasbarra}}, \binits{C.}},
\oauthor{\bsnm{{Gasparrini}}, \binits{D.}},
\oauthor{\bsnm{{Germani}}, \binits{S.}},
\oauthor{\bsnm{{Giacchino}}, \binits{F.}},
\oauthor{\bsnm{{Giglietto}}, \binits{N.}},
\oauthor{\bsnm{{Gill}}, \binits{R.}},
\oauthor{\bsnm{{Giroletti}}, \binits{M.}},
\oauthor{\bsnm{{Granot}}, \binits{J.}},
\oauthor{\bsnm{{Green}}, \binits{D.}},
\oauthor{\bsnm{{Grenier}}, \binits{I.A.}},
\oauthor{\bsnm{{Guiriec}}, \binits{S.}},
\oauthor{\bsnm{{Gustafsson}}, \binits{M.}},
\oauthor{\bsnm{{Hays}}, \binits{E.}},
\oauthor{\bsnm{{Hewitt}}, \binits{J.W.}},
\oauthor{\bsnm{{Horan}}, \binits{D.}},
\oauthor{\bsnm{{Hou}}, \binits{X.}},
\oauthor{\bsnm{{Kuss}}, \binits{M.}},
\oauthor{\bsnm{{Latronico}}, \binits{L.}},
\oauthor{\bsnm{{Laviron}}, \binits{A.}},
\oauthor{\bsnm{{Lemoine-Goumard}}, \binits{M.}},
\oauthor{\bsnm{{Li}}, \binits{J.}},
\oauthor{\bsnm{{Liodakis}}, \binits{I.}},
\oauthor{\bsnm{{Longo}}, \binits{F.}},
\oauthor{\bsnm{{Loparco}}, \binits{F.}},
\oauthor{\bsnm{{Lorusso}}, \binits{L.}},
\oauthor{\bsnm{{Lovellette}}, \binits{M.N.}},
\oauthor{\bsnm{{Lubrano}}, \binits{P.}},
\oauthor{\bsnm{{Maldera}}, \binits{S.}},
\oauthor{\bsnm{{Manfreda}}, \binits{A.}},
\oauthor{\bsnm{{Marti-Devesa}}, \binits{G.}},
\oauthor{\bsnm{{Mazziotta}}, \binits{M.N.}},
\oauthor{\bsnm{{McEnery}}, \binits{J.E.}},
\oauthor{\bsnm{{Mereu}}, \binits{I.}},
\oauthor{\bsnm{{Meyer}}, \binits{M.}},
\oauthor{\bsnm{{Michelson}}, \binits{P.F.}},
\oauthor{\bsnm{{Mizuno}}, \binits{T.}},
\oauthor{\bsnm{{Monzani}}, \binits{M.E.}},
\oauthor{\bsnm{{Morselli}}, \binits{A.}},
\oauthor{\bsnm{{Moskalenko}}, \binits{I.V.}},
\oauthor{\bsnm{{Negro}}, \binits{M.}},
\oauthor{\bsnm{{Nuss}}, \binits{E.}},
\oauthor{\bsnm{{Omodei}}, \binits{N.}},
\oauthor{\bsnm{{Orlando}}, \binits{E.}},
\oauthor{\bsnm{{Ormes}}, \binits{J.F.}},
\oauthor{\bsnm{{Paneque}}, \binits{D.}},
\oauthor{\bsnm{{Panzarini}}, \binits{G.}},
\oauthor{\bsnm{{Persic}}, \binits{M.}},
\oauthor{\bsnm{{Pesce-Rollins}}, \binits{M.}},
\oauthor{\bsnm{{Pillera}}, \binits{R.}},
\oauthor{\bsnm{{Piron}}, \binits{F.}},
\oauthor{\bsnm{{Poon}}, \binits{H.}},
\oauthor{\bsnm{{Porter}}, \binits{T.A.}},
\oauthor{\bsnm{{Principe}}, \binits{G.}},
\oauthor{\bsnm{{Raino}}, \binits{S.}},
\oauthor{\bsnm{{Rando}}, \binits{R.}},
\oauthor{\bsnm{{Rani}}, \binits{B.}},
\oauthor{\bsnm{{Razzano}}, \binits{M.}},
\oauthor{\bsnm{{Razzaque}}, \binits{S.}},
\oauthor{\bsnm{{Reimer}}, \binits{A.}},
\oauthor{\bsnm{{Reimer}}, \binits{O.}},
\oauthor{\bsnm{{Ryde}}, \binits{F.}},
\oauthor{\bsnm{{Sanchez-Conde}}, \binits{M.}},
\oauthor{\bsnm{{Saz Parkinson}}, \binits{P.M.}},
\oauthor{\bsnm{{Scotton}}, \binits{L.}},
\oauthor{\bsnm{{Serini}}, \binits{D.}},
\oauthor{\bsnm{{Sgro}}, \binits{C.}},
\oauthor{\bsnm{{Sharma}}, \binits{V.}},
\oauthor{\bsnm{{Siskind}}, \binits{E.J.}},
\oauthor{\bsnm{{Spandre}}, \binits{G.}},
\oauthor{\bsnm{{Spinelli}}, \binits{P.}},
\oauthor{\bsnm{{Tajima}}, \binits{H.}},
\oauthor{\bsnm{{Torres}}, \binits{D.F.}},
\oauthor{\bsnm{{Valverde}}, \binits{J.}},
\oauthor{\bsnm{{Venters}}, \binits{T.}},
\oauthor{\bsnm{{Wadiasingh}}, \binits{Z.}},
\oauthor{\bsnm{{Wood}}, \binits{K.}},
\oauthor{\bsnm{{Zaharijas}}, \binits{G.}}:
{Fermi-GBM Discovery of GRB 221009A: An Extraordinarily Bright GRB from Onset
  to Afterglow}.
arXiv e-prints,
2303--14172
(2023)
{\href{https://arxiv.org/abs/2303.14172}{{arXiv:2303.14172}}}
{[astro-ph.HE]}
\end{botherref}
\endbibitem

\bibitem{Mazets1981}
\begin{barticle}
\bauthor{\bsnm{{Mazets}}, \binits{E.P.}},
\bauthor{\bsnm{{Golenetskii}}, \binits{S.V.}},
\bauthor{\bsnm{{Aptekar}}, \binits{R.L.}},
\bauthor{\bsnm{{Gurian}}, \binits{I.A.}},
\bauthor{\bsnm{{Ilinskii}}, \binits{V.N.}}:
\batitle{{Cyclotron and annihilation lines in {\ensuremath{\gamma}}-ray
  bursts}}.
\bjtitle{\nat}
\bvolume{290}(\bissue{5805}),
\bfpage{378}--\blpage{382}
(\byear{1981}).
\doiurl{10.1038/290378a0}
\end{barticle}
\endbibitem

\bibitem{Murakami1988}
\begin{barticle}
\bauthor{\bsnm{{Murakami}}, \binits{T.}},
\bauthor{\bsnm{{Fujii}}, \binits{M.}},
\bauthor{\bsnm{{Hayashida}}, \binits{K.}},
\bauthor{\bsnm{{Itoh}}, \binits{M.}},
\bauthor{\bsnm{{Nishimura}}, \binits{J.}}:
\batitle{{Evidence for cyclotron absorption from spectral features in gamma-ray
  bursts seen with Ginga}}.
\bjtitle{\nat}
\bvolume{335}(\bissue{6187}),
\bfpage{234}--\blpage{235}
(\byear{1988}).
\doiurl{10.1038/335234a0}
\end{barticle}
\endbibitem

\bibitem{Band1996}
\begin{barticle}
\bauthor{\bsnm{{Band}}, \binits{D.L.}},
\bauthor{\bsnm{{Ryder}}, \binits{S.}},
\bauthor{\bsnm{{Ford}}, \binits{L.A.}},
\bauthor{\bsnm{{Matteson}}, \binits{J.L.}},
\bauthor{\bsnm{{Palmer}}, \binits{D.M.}},
\bauthor{\bsnm{{Teegarden}}, \binits{B.J.}},
\bauthor{\bsnm{{Briggs}}, \binits{M.S.}},
\bauthor{\bsnm{{Paciesas}}, \binits{W.S.}},
\bauthor{\bsnm{{Pendleton}}, \binits{G.N.}},
\bauthor{\bsnm{{Preece}}, \binits{R.D.}}:
\batitle{{BATSE Gamma-Ray Burst Line Search. IV. Line Candidates from the
  Visual Search}}.
\bjtitle{\apj}
\bvolume{458},
\bfpage{746}
(\byear{1996})
{\href{https://arxiv.org/abs/astro-ph/9508158}{{arXiv:astro-ph/9508158}}}
{[astro-ph]}.
\doiurl{10.1086/176855}
\end{barticle}
\endbibitem

\bibitem{Palmer1994}
\begin{barticle}
\bauthor{\bsnm{{Palmer}}, \binits{D.M.}},
\bauthor{\bsnm{{Teegarden}}, \binits{B.J.}},
\bauthor{\bsnm{{Schaefer}}, \binits{B.E.}},
\bauthor{\bsnm{{Cline}}, \binits{T.L.}},
\bauthor{\bsnm{{Band}}, \binits{D.L.}},
\bauthor{\bsnm{{Ford}}, \binits{L.A.}},
\bauthor{\bsnm{{Matteson}}, \binits{J.L.}},
\bauthor{\bsnm{{Paciesas}}, \binits{W.S.}},
\bauthor{\bsnm{{Pendleton}}, \binits{G.N.}},
\bauthor{\bsnm{{Briggs}}, \binits{M.S.}},
\bauthor{\bsnm{{Preece}}, \binits{R.D.}},
\bauthor{\bsnm{{Fishman}}, \binits{G.J.}},
\bauthor{\bsnm{{Meegan}}, \binits{C.A.}},
\bauthor{\bsnm{{Wilson}}, \binits{R.B.}},
\bauthor{\bsnm{{Lestrade}}, \binits{J.P.}}:
\batitle{{BATSE Gamma-Ray Burst Line Search. I. Search for Narrow Lines in
  Spectroscopy Detector Data}}.
\bjtitle{\apjl}
\bvolume{433},
\bfpage{77}
(\byear{1994}).
\doiurl{10.1086/187552}
\end{barticle}
\endbibitem

\bibitem{Amati2000}
\begin{barticle}
\bauthor{\bsnm{{Amati}}, \binits{L.}},
\bauthor{\bsnm{{Frontera}}, \binits{F.}},
\bauthor{\bsnm{{Vietri}}, \binits{M.}},
\bauthor{\bsnm{{in't Zand}}, \binits{J.J.M.}},
\bauthor{\bsnm{{Soffitta}}, \binits{P.}},
\bauthor{\bsnm{{Costa}}, \binits{E.}},
\bauthor{\bsnm{{Del Sordo}}, \binits{S.}},
\bauthor{\bsnm{{Pian}}, \binits{E.}},
\bauthor{\bsnm{{Piro}}, \binits{L.}},
\bauthor{\bsnm{{Antonelli}}, \binits{L.A.}},
\bauthor{\bsnm{{Fiume}}, \binits{D.D.}},
\bauthor{\bsnm{{Feroci}}, \binits{M.}},
\bauthor{\bsnm{{Gandolfi}}, \binits{G.}},
\bauthor{\bsnm{{Guidorzi}}, \binits{C.}},
\bauthor{\bsnm{{Heise}}, \binits{J.}},
\bauthor{\bsnm{{Kuulkers}}, \binits{E.}},
\bauthor{\bsnm{{Masetti}}, \binits{N.}},
\bauthor{\bsnm{{Montanari}}, \binits{E.}},
\bauthor{\bsnm{{Nicastro}}, \binits{L.}},
\bauthor{\bsnm{{Orlandini}}, \binits{M.}},
\bauthor{\bsnm{{Palazzi}}, \binits{E.}}:
\batitle{{Discovery of a Transient Absorption Edge in the X-ray Spectrum of GRB
  990705}}.
\bjtitle{Science}
\bvolume{290}(\bissue{5493}),
\bfpage{953}--\blpage{955}
(\byear{2000})
{\href{https://arxiv.org/abs/astro-ph/0012318}{{arXiv:astro-ph/0012318}}}
{[astro-ph]}.
\doiurl{10.1126/science.290.5493.953}
\end{barticle}
\endbibitem

\bibitem{Frontera2004}
\begin{barticle}
\bauthor{\bsnm{{Frontera}}, \binits{F.}},
\bauthor{\bsnm{{Amati}}, \binits{L.}},
\bauthor{\bsnm{{in 't Zand}}, \binits{J.J.M.}},
\bauthor{\bsnm{{Lazzati}}, \binits{D.}},
\bauthor{\bsnm{{K{\"o}nigl}}, \binits{A.}},
\bauthor{\bsnm{{Vietri}}, \binits{M.}},
\bauthor{\bsnm{{Costa}}, \binits{E.}},
\bauthor{\bsnm{{Feroci}}, \binits{M.}},
\bauthor{\bsnm{{Guidorzi}}, \binits{C.}},
\bauthor{\bsnm{{Montanari}}, \binits{E.}},
\bauthor{\bsnm{{Orlandini}}, \binits{M.}},
\bauthor{\bsnm{{Pian}}, \binits{E.}},
\bauthor{\bsnm{{Piro}}, \binits{L.}}:
\batitle{{The Prompt X-Ray Emission of GRB 011211: Possible Evidence of a
  Transient Absorption Feature}}.
\bjtitle{\apj}
\bvolume{616}(\bissue{2}),
\bfpage{1078}--\blpage{1085}
(\byear{2004})
{\href{https://arxiv.org/abs/astro-ph/0408436}{{arXiv:astro-ph/0408436}}}
{[astro-ph]}.
\doiurl{10.1086/425066}
\end{barticle}
\endbibitem

\bibitem{Piro1999}
\begin{barticle}
\bauthor{\bsnm{{Piro}}, \binits{L.}},
\bauthor{\bsnm{{Costa}}, \binits{E.}},
\bauthor{\bsnm{{Feroci}}, \binits{M.}},
\bauthor{\bsnm{{Stratta}}, \binits{G.}},
\bauthor{\bsnm{{Frontera}}, \binits{F.}},
\bauthor{\bsnm{{Amati}}, \binits{L.}},
\bauthor{\bsnm{{dal Fiume}}, \binits{D.}},
\bauthor{\bsnm{{Antonelli}}, \binits{L.A.}},
\bauthor{\bsnm{{Heise}}, \binits{J.}},
\bauthor{\bsnm{{in 't Zand}}, \binits{J.}},
\bauthor{\bsnm{{Owens}}, \binits{A.}},
\bauthor{\bsnm{{Parmar}}, \binits{A.N.}},
\bauthor{\bsnm{{Cusumano}}, \binits{G.}},
\bauthor{\bsnm{{Vietri}}, \binits{M.}},
\bauthor{\bsnm{{Perola}}, \binits{G.C.}}:
\batitle{{Iron line signatures in X-ray afterglows of GRB by BeppoSAX}}.
\bjtitle{\aaps}
\bvolume{138},
\bfpage{431}--\blpage{432}
(\byear{1999})
{\href{https://arxiv.org/abs/astro-ph/9906363}{{arXiv:astro-ph/9906363}}}
{[astro-ph]}.
\doiurl{10.1051/aas:1999296}
\end{barticle}
\endbibitem

\bibitem{Antonelli2000}
\begin{barticle}
\bauthor{\bsnm{{Antonelli}}, \binits{L.A.}},
\bauthor{\bsnm{{Piro}}, \binits{L.}},
\bauthor{\bsnm{{Vietri}}, \binits{M.}},
\bauthor{\bsnm{{Costa}}, \binits{E.}},
\bauthor{\bsnm{{Soffitta}}, \binits{P.}},
\bauthor{\bsnm{{Feroci}}, \binits{M.}},
\bauthor{\bsnm{{Amati}}, \binits{L.}},
\bauthor{\bsnm{{Frontera}}, \binits{F.}},
\bauthor{\bsnm{{Pian}}, \binits{E.}},
\bauthor{\bsnm{{in 't Zand}}, \binits{J.J.M.}},
\bauthor{\bsnm{{Heise}}, \binits{J.}},
\bauthor{\bsnm{{Kuulkers}}, \binits{E.}},
\bauthor{\bsnm{{Nicastro}}, \binits{L.}},
\bauthor{\bsnm{{Butler}}, \binits{R.C.}},
\bauthor{\bsnm{{Stella}}, \binits{L.}},
\bauthor{\bsnm{{Perola}}, \binits{G.C.}}:
\batitle{{Discovery of a Redshifted Iron K Line in the X-Ray Afterglow of GRB
  000214}}.
\bjtitle{\apjl}
\bvolume{545}(\bissue{1}),
\bfpage{39}--\blpage{42}
(\byear{2000})
{\href{https://arxiv.org/abs/astro-ph/0010221}{{arXiv:astro-ph/0010221}}}
{[astro-ph]}.
\doiurl{10.1086/317328}
\end{barticle}
\endbibitem

\bibitem{Yoshida2001}
\begin{barticle}
\bauthor{\bsnm{{Yoshida}}, \binits{A.}},
\bauthor{\bsnm{{Namiki}}, \binits{M.}},
\bauthor{\bsnm{{Yonetoku}}, \binits{D.}},
\bauthor{\bsnm{{Murakami}}, \binits{T.}},
\bauthor{\bsnm{{Otani}}, \binits{C.}},
\bauthor{\bsnm{{Kawai}}, \binits{N.}},
\bauthor{\bsnm{{Ueda}}, \binits{Y.}},
\bauthor{\bsnm{{Shibata}}, \binits{R.}},
\bauthor{\bsnm{{Uno}}, \binits{S.}}:
\batitle{{A Possible Emission Feature in an X-Ray Afterglow of GRB 970828 as a
  Radiative Recombination Edge}}.
\bjtitle{\apjl}
\bvolume{557}(\bissue{1}),
\bfpage{27}--\blpage{30}
(\byear{2001})
{\href{https://arxiv.org/abs/astro-ph/0107331}{{arXiv:astro-ph/0107331}}}
{[astro-ph]}.
\doiurl{10.1086/323143}
\end{barticle}
\endbibitem

\bibitem{Reeves2002}
\begin{barticle}
\bauthor{\bsnm{{Reeves}}, \binits{J.N.}},
\bauthor{\bsnm{{Watson}}, \binits{D.}},
\bauthor{\bsnm{{Osborne}}, \binits{J.P.}},
\bauthor{\bsnm{{Pounds}}, \binits{K.A.}},
\bauthor{\bsnm{{O'Brien}}, \binits{P.T.}},
\bauthor{\bsnm{{Short}}, \binits{A.D.T.}},
\bauthor{\bsnm{{Turner}}, \binits{M.J.L.}},
\bauthor{\bsnm{{Watson}}, \binits{M.G.}},
\bauthor{\bsnm{{Mason}}, \binits{K.O.}},
\bauthor{\bsnm{{Ehle}}, \binits{M.}},
\bauthor{\bsnm{{Schartel}}, \binits{N.}}:
\batitle{{The signature of supernova ejecta in the X-ray afterglow of the
  {\ensuremath{\gamma}}-ray burst 011211}}.
\bjtitle{\nat}
\bvolume{416}(\bissue{6880}),
\bfpage{512}--\blpage{515}
(\byear{2002})
{\href{https://arxiv.org/abs/astro-ph/0204075}{{arXiv:astro-ph/0204075}}}
{[astro-ph]}.
\doiurl{10.1038/416512a}
\end{barticle}
\endbibitem

\bibitem{Watson2003}
\begin{barticle}
\bauthor{\bsnm{{Watson}}, \binits{D.}},
\bauthor{\bsnm{{Reeves}}, \binits{J.N.}},
\bauthor{\bsnm{{Hjorth}}, \binits{J.}},
\bauthor{\bsnm{{Jakobsson}}, \binits{P.}},
\bauthor{\bsnm{{Pedersen}}, \binits{K.}}:
\batitle{{Delayed Soft X-Ray Emission Lines in the Afterglow of GRB 030227}}.
\bjtitle{\apjl}
\bvolume{595}(\bissue{1}),
\bfpage{29}--\blpage{32}
(\byear{2003})
{\href{https://arxiv.org/abs/astro-ph/0306284}{{arXiv:astro-ph/0306284}}}
{[astro-ph]}.
\doiurl{10.1086/378790}
\end{barticle}
\endbibitem

\bibitem{Piro2000}
\begin{barticle}
\bauthor{\bsnm{{Piro}}, \binits{L.}},
\bauthor{\bsnm{{Garmire}}, \binits{G.}},
\bauthor{\bsnm{{Garcia}}, \binits{M.}},
\bauthor{\bsnm{{Stratta}}, \binits{G.}},
\bauthor{\bsnm{{Costa}}, \binits{E.}},
\bauthor{\bsnm{{Feroci}}, \binits{M.}},
\bauthor{\bsnm{{M{\'e}sz{\'a}ros}}, \binits{P.}},
\bauthor{\bsnm{{Vietri}}, \binits{M.}},
\bauthor{\bsnm{{Bradt}}, \binits{H.}},
\bauthor{\bsnm{{Frail}}, \binits{D.}},
\bauthor{\bsnm{{Frontera}}, \binits{F.}},
\bauthor{\bsnm{{Halpern}}, \binits{J.}},
\bauthor{\bsnm{{Heise}}, \binits{J.}},
\bauthor{\bsnm{{Hurley}}, \binits{K.}},
\bauthor{\bsnm{{Kawai}}, \binits{N.}},
\bauthor{\bsnm{{Kippen}}, \binits{R.M.}},
\bauthor{\bsnm{{Marshall}}, \binits{F.}},
\bauthor{\bsnm{{Murakami}}, \binits{T.}},
\bauthor{\bsnm{{Sokolov}}, \binits{V.V.}},
\bauthor{\bsnm{{Takeshima}}, \binits{T.}},
\bauthor{\bsnm{{Yoshida}}, \binits{A.}}:
\batitle{{Observation of X-ray Lines from a Gamma-Ray Burst (GRB991216):
  Evidence of Moving Ejecta from the Progenitor}}.
\bjtitle{Science}
\bvolume{290}(\bissue{5493}),
\bfpage{955}--\blpage{958}
(\byear{2000})
{\href{https://arxiv.org/abs/astro-ph/0011337}{{arXiv:astro-ph/0011337}}}
{[astro-ph]}.
\doiurl{10.1126/science.290.5493.955}
\end{barticle}
\endbibitem

\bibitem{Butler2003}
\begin{barticle}
\bauthor{\bsnm{{Butler}}, \binits{N.R.}},
\bauthor{\bsnm{{Marshall}}, \binits{H.L.}},
\bauthor{\bsnm{{Ricker}}, \binits{G.R.}},
\bauthor{\bsnm{{Vanderspek}}, \binits{R.K.}},
\bauthor{\bsnm{{Ford}}, \binits{P.G.}},
\bauthor{\bsnm{{Crew}}, \binits{G.B.}},
\bauthor{\bsnm{{Lamb}}, \binits{D.Q.}},
\bauthor{\bsnm{{Jernigan}}, \binits{J.G.}}:
\batitle{{The X-Ray Afterglows of GRB 020813 and GRB 021004 with Chandra HETGS:
  Possible Evidence for a Supernova Prior to GRB 020813}}.
\bjtitle{\apj}
\bvolume{597}(\bissue{2}),
\bfpage{1010}--\blpage{1016}
(\byear{2003})
{\href{https://arxiv.org/abs/astro-ph/0303539}{{arXiv:astro-ph/0303539}}}
{[astro-ph]}.
\doiurl{10.1086/378511}
\end{barticle}
\endbibitem

\bibitem{Sako2005}
\begin{barticle}
\bauthor{\bsnm{{Sako}}, \binits{M.}},
\bauthor{\bsnm{{Harrison}}, \binits{F.A.}},
\bauthor{\bsnm{{Rutledge}}, \binits{R.E.}}:
\batitle{{A Search for Discrete X-Ray Spectral Features in a Sample of Bright
  {\ensuremath{\gamma}}-Ray Burst Afterglows}}.
\bjtitle{\apj}
\bvolume{623}(\bissue{2}),
\bfpage{973}--\blpage{999}
(\byear{2005})
{\href{https://arxiv.org/abs/astro-ph/0406210}{{arXiv:astro-ph/0406210}}}
{[astro-ph]}.
\doiurl{10.1086/425644}
\end{barticle}
\endbibitem

\bibitem{Campana2016}
\begin{barticle}
\bauthor{\bsnm{{Campana}}, \binits{S.}},
\bauthor{\bsnm{{Braito}}, \binits{V.}},
\bauthor{\bsnm{{D'Avanzo}}, \binits{P.}},
\bauthor{\bsnm{{Ghirlanda}}, \binits{G.}},
\bauthor{\bsnm{{Melandri}}, \binits{A.}},
\bauthor{\bsnm{{Pescalli}}, \binits{A.}},
\bauthor{\bsnm{{Salafia}}, \binits{O.S.}},
\bauthor{\bsnm{{Salvaterra}}, \binits{R.}},
\bauthor{\bsnm{{Tagliaferri}}, \binits{G.}},
\bauthor{\bsnm{{Vergani}}, \binits{S.D.}}:
\batitle{{Searching for narrow absorption and emission lines in XMM-Newton
  spectra of gamma-ray bursts}}.
\bjtitle{\aap}
\bvolume{592},
\bfpage{85}
(\byear{2016})
{\href{https://arxiv.org/abs/1606.03876}{{arXiv:1606.03876}}}
{[astro-ph.HE]}.
\doiurl{10.1051/0004-6361/201628402}
\end{barticle}
\endbibitem

\bibitem{Drenkhahn2002}
\begin{barticle}
\bauthor{\bsnm{{Drenkhahn}}, \binits{G.}},
\bauthor{\bsnm{{Spruit}}, \binits{H.C.}}:
\batitle{{Efficient acceleration and radiation in Poynting flux powered GRB
  outflows}}.
\bjtitle{\aap}
\bvolume{391},
\bfpage{1141}--\blpage{1153}
(\byear{2002})
{\href{https://arxiv.org/abs/astro-ph/0202387}{{arXiv:astro-ph/0202387}}}
{[astro-ph]}.
\doiurl{10.1051/0004-6361:20020839}
\end{barticle}
\endbibitem

\bibitem{Lazzati2009}
\begin{barticle}
\bauthor{\bsnm{{Lazzati}}, \binits{D.}},
\bauthor{\bsnm{{Morsony}}, \binits{B.J.}},
\bauthor{\bsnm{{Begelman}}, \binits{M.C.}}:
\batitle{{Very High Efficiency Photospheric Emission in Long-Duration
  {\ensuremath{\gamma}}-Ray Bursts}}.
\bjtitle{\apjl}
\bvolume{700}(\bissue{1}),
\bfpage{47}--\blpage{50}
(\byear{2009})
{\href{https://arxiv.org/abs/0904.2779}{{arXiv:0904.2779}}}
{[astro-ph.HE]}.
\doiurl{10.1088/0004-637X/700/1/L47}
\end{barticle}
\endbibitem

\bibitem{Zhang2011}
\begin{barticle}
\bauthor{\bsnm{{Zhang}}, \binits{B.}},
\bauthor{\bsnm{{Yan}}, \binits{H.}}:
\batitle{{The Internal-collision-induced Magnetic Reconnection and Turbulence
  (ICMART) Model of Gamma-ray Bursts}}.
\bjtitle{\apj}
\bvolume{726}(\bissue{2}),
\bfpage{90}
(\byear{2011})
{\href{https://arxiv.org/abs/1011.1197}{{arXiv:1011.1197}}}
{[astro-ph.HE]}.
\doiurl{10.1088/0004-637X/726/2/90}
\end{barticle}
\endbibitem

\bibitem{Beloborodov2003}
\begin{barticle}
\bauthor{\bsnm{{Beloborodov}}, \binits{A.M.}}:
\batitle{{Nuclear Composition of Gamma-Ray Burst Fireballs}}.
\bjtitle{\apj}
\bvolume{588}(\bissue{2}),
\bfpage{931}--\blpage{944}
(\byear{2003})
{\href{https://arxiv.org/abs/astro-ph/0210522}{{arXiv:astro-ph/0210522}}}
{[astro-ph]}.
\doiurl{10.1086/374217}
\end{barticle}
\endbibitem

\bibitem{Sikora1994}
\begin{barticle}
\bauthor{\bsnm{{Sikora}}, \binits{M.}},
\bauthor{\bsnm{{Begelman}}, \binits{M.C.}},
\bauthor{\bsnm{{Rees}}, \binits{M.J.}}:
\batitle{{Comptonization of Diffuse Ambient Radiation by a Relativistic Jet:
  The Source of Gamma Rays from Blazars?}}
\bjtitle{\apj}
\bvolume{421},
\bfpage{153}
(\byear{1994}).
\doiurl{10.1086/173633}
\end{barticle}
\endbibitem

\bibitem{Vietri2001}
\begin{barticle}
\bauthor{\bsnm{{Vietri}}, \binits{M.}},
\bauthor{\bsnm{{Ghisellini}}, \binits{G.}},
\bauthor{\bsnm{{Lazzati}}, \binits{D.}},
\bauthor{\bsnm{{Fiore}}, \binits{F.}},
\bauthor{\bsnm{{Stella}}, \binits{L.}}:
\batitle{{Illuminated, and Enlightened, by GRB 991216}}.
\bjtitle{\apjl}
\bvolume{550}(\bissue{1}),
\bfpage{43}--\blpage{46}
(\byear{2001})
{\href{https://arxiv.org/abs/astro-ph/0011580}{{arXiv:astro-ph/0011580}}}
{[astro-ph]}.
\doiurl{10.1086/319475}
\end{barticle}
\endbibitem

\bibitem{Peer2004}
\begin{barticle}
\bauthor{\bsnm{{Pe'er}}, \binits{A.}},
\bauthor{\bsnm{{Waxman}}, \binits{E.}}:
\batitle{{Prompt Gamma-Ray Burst Spectra: Detailed Calculations and the Effect
  of Pair Production}}.
\bjtitle{\apj}
\bvolume{613}(\bissue{1}),
\bfpage{448}--\blpage{459}
(\byear{2004})
{\href{https://arxiv.org/abs/astro-ph/0311252}{{arXiv:astro-ph/0311252}}}
{[astro-ph]}.
\doiurl{10.1086/422989}
\end{barticle}
\endbibitem

\bibitem{Kumar2000}
\begin{barticle}
\bauthor{\bsnm{{Kumar}}, \binits{P.}},
\bauthor{\bsnm{{Panaitescu}}, \binits{A.}}:
\batitle{{Afterglow Emission from Naked Gamma-Ray Bursts}}.
\bjtitle{\apjl}
\bvolume{541}(\bissue{2}),
\bfpage{51}--\blpage{54}
(\byear{2000})
{\href{https://arxiv.org/abs/astro-ph/0006317}{{arXiv:astro-ph/0006317}}}
{[astro-ph]}.
\doiurl{10.1086/312905}
\end{barticle}
\endbibitem

\bibitem{Oganesyan2020}
\begin{barticle}
\bauthor{\bsnm{{Oganesyan}}, \binits{G.}},
\bauthor{\bsnm{{Ascenzi}}, \binits{S.}},
\bauthor{\bsnm{{Branchesi}}, \binits{M.}},
\bauthor{\bsnm{{Salafia}}, \binits{O.S.}},
\bauthor{\bsnm{{Dall'Osso}}, \binits{S.}},
\bauthor{\bsnm{{Ghirlanda}}, \binits{G.}}:
\batitle{{Structured Jets and X-Ray Plateaus in Gamma-Ray Burst Phenomena}}.
\bjtitle{\apj}
\bvolume{893}(\bissue{2}),
\bfpage{88}
(\byear{2020})
{\href{https://arxiv.org/abs/1904.08786}{{arXiv:1904.08786}}}
{[astro-ph.HE]}.
\doiurl{10.3847/1538-4357/ab8221}
\end{barticle}
\endbibitem

\bibitem{Ascenzi2020}
\begin{barticle}
\bauthor{\bsnm{{Ascenzi}}, \binits{S.}},
\bauthor{\bsnm{{Oganesyan}}, \binits{G.}},
\bauthor{\bsnm{{Salafia}}, \binits{O.S.}},
\bauthor{\bsnm{{Branchesi}}, \binits{M.}},
\bauthor{\bsnm{{Ghirlanda}}, \binits{G.}},
\bauthor{\bsnm{{Dall'Osso}}, \binits{S.}}:
\batitle{{High-latitude emission from the structured jet of
  {\ensuremath{\gamma}}-ray bursts observed off-axis}}.
\bjtitle{\aap}
\bvolume{641},
\bfpage{61}
(\byear{2020})
{\href{https://arxiv.org/abs/2004.12215}{{arXiv:2004.12215}}}
{[astro-ph.HE]}.
\doiurl{10.1051/0004-6361/202038265}
\end{barticle}
\endbibitem

\bibitem{Meegan2009}
\begin{barticle}
\bauthor{\bsnm{{Meegan}}, \binits{C.}},
\bauthor{\bsnm{{Lichti}}, \binits{G.}},
\bauthor{\bsnm{{Bhat}}, \binits{P.N.}},
\bauthor{\bsnm{{Bissaldi}}, \binits{E.}},
\bauthor{\bsnm{{Briggs}}, \binits{M.S.}},
\bauthor{\bsnm{{Connaughton}}, \binits{V.}},
\bauthor{\bsnm{{Diehl}}, \binits{R.}},
\bauthor{\bsnm{{Fishman}}, \binits{G.}},
\bauthor{\bsnm{{Greiner}}, \binits{J.}},
\bauthor{\bsnm{{Hoover}}, \binits{A.S.}},
\bauthor{\bsnm{{van der Horst}}, \binits{A.J.}},
\bauthor{\bsnm{{von Kienlin}}, \binits{A.}},
\bauthor{\bsnm{{Kippen}}, \binits{R.M.}},
\bauthor{\bsnm{{Kouveliotou}}, \binits{C.}},
\bauthor{\bsnm{{McBreen}}, \binits{S.}},
\bauthor{\bsnm{{Paciesas}}, \binits{W.S.}},
\bauthor{\bsnm{{Preece}}, \binits{R.}},
\bauthor{\bsnm{{Steinle}}, \binits{H.}},
\bauthor{\bsnm{{Wallace}}, \binits{M.S.}},
\bauthor{\bsnm{{Wilson}}, \binits{R.B.}},
\bauthor{\bsnm{{Wilson-Hodge}}, \binits{C.}}:
\batitle{{The Fermi Gamma-ray Burst Monitor}}.
\bjtitle{\apj}
\bvolume{702}(\bissue{1}),
\bfpage{791}--\blpage{804}
(\byear{2009})
{\href{https://arxiv.org/abs/0908.0450}{{arXiv:0908.0450}}}
{[astro-ph.IM]}.
\doiurl{10.1088/0004-637X/702/1/791}
\end{barticle}
\endbibitem

\bibitem{Gruber2014}
\begin{barticle}
\bauthor{\bsnm{{Gruber}}, \binits{D.}},
\bauthor{\bsnm{{Goldstein}}, \binits{A.}},
\bauthor{\bsnm{{Weller von Ahlefeld}}, \binits{V.}},
\bauthor{\bsnm{{Narayana Bhat}}, \binits{P.}},
\bauthor{\bsnm{{Bissaldi}}, \binits{E.}},
\bauthor{\bsnm{{Briggs}}, \binits{M.S.}},
\bauthor{\bsnm{{Byrne}}, \binits{D.}},
\bauthor{\bsnm{{Cleveland}}, \binits{W.H.}},
\bauthor{\bsnm{{Connaughton}}, \binits{V.}},
\bauthor{\bsnm{{Diehl}}, \binits{R.}},
\bauthor{\bsnm{{Fishman}}, \binits{G.J.}},
\bauthor{\bsnm{{Fitzpatrick}}, \binits{G.}},
\bauthor{\bsnm{{Foley}}, \binits{S.}},
\bauthor{\bsnm{{Gibby}}, \binits{M.}},
\bauthor{\bsnm{{Giles}}, \binits{M.M.}},
\bauthor{\bsnm{{Greiner}}, \binits{J.}},
\bauthor{\bsnm{{Guiriec}}, \binits{S.}},
\bauthor{\bsnm{{van der Horst}}, \binits{A.J.}},
\bauthor{\bsnm{{von Kienlin}}, \binits{A.}},
\bauthor{\bsnm{{Kouveliotou}}, \binits{C.}},
\bauthor{\bsnm{{Layden}}, \binits{E.}},
\bauthor{\bsnm{{Lin}}, \binits{L.}},
\bauthor{\bsnm{{Meegan}}, \binits{C.A.}},
\bauthor{\bsnm{{McGlynn}}, \binits{S.}},
\bauthor{\bsnm{{Paciesas}}, \binits{W.S.}},
\bauthor{\bsnm{{Pelassa}}, \binits{V.}},
\bauthor{\bsnm{{Preece}}, \binits{R.D.}},
\bauthor{\bsnm{{Rau}}, \binits{A.}},
\bauthor{\bsnm{{Wilson-Hodge}}, \binits{C.A.}},
\bauthor{\bsnm{{Xiong}}, \binits{S.}},
\bauthor{\bsnm{{Younes}}, \binits{G.}},
\bauthor{\bsnm{{Yu}}, \binits{H.-F.}}:
\batitle{{The Fermi GBM Gamma-Ray Burst Spectral Catalog: Four Years of Data}}.
\bjtitle{\apjs}
\bvolume{211}(\bissue{1}),
\bfpage{12}
(\byear{2014})
{\href{https://arxiv.org/abs/1401.5069}{{arXiv:1401.5069}}}
{[astro-ph.HE]}.
\doiurl{10.1088/0067-0049/211/1/12}
\end{barticle}
\endbibitem

\bibitem{Gompertz2023}
\begin{barticle}
\bauthor{\bsnm{{Gompertz}}, \binits{B.P.}},
\bauthor{\bsnm{{Ravasio}}, \binits{M.E.}},
\bauthor{\bsnm{{Nicholl}}, \binits{M.}},
\bauthor{\bsnm{{Levan}}, \binits{A.J.}},
\bauthor{\bsnm{{Metzger}}, \binits{B.D.}},
\bauthor{\bsnm{{Oates}}, \binits{S.R.}},
\bauthor{\bsnm{{Lamb}}, \binits{G.P.}},
\bauthor{\bsnm{{Fong}}, \binits{W.-f.}},
\bauthor{\bsnm{{Malesani}}, \binits{D.B.}},
\bauthor{\bsnm{{Rastinejad}}, \binits{J.C.}},
\bauthor{\bsnm{{Tanvir}}, \binits{N.R.}},
\bauthor{\bsnm{{Evans}}, \binits{P.A.}},
\bauthor{\bsnm{{Jonker}}, \binits{P.G.}},
\bauthor{\bsnm{{Page}}, \binits{K.L.}},
\bauthor{\bsnm{{Pe'er}}, \binits{A.}}:
\batitle{{The case for a minute-long merger-driven gamma-ray burst from
  fast-cooling synchrotron emission}}.
\bjtitle{Nature Astronomy}
\bvolume{7},
\bfpage{67}--\blpage{79}
(\byear{2023})
{\href{https://arxiv.org/abs/2205.05008}{{arXiv:2205.05008}}}
{[astro-ph.HE]}.
\doiurl{10.1038/s41550-022-01819-4}
\end{barticle}
\endbibitem

\bibitem{Yang2023}
\begin{botherref}
\oauthor{\bsnm{{Yang}}, \binits{J.}},
\oauthor{\bsnm{{Zhao}}, \binits{X.-H.}},
\oauthor{\bsnm{{Yan}}, \binits{Z.}},
\oauthor{\bsnm{{Wang}}, \binits{X.I.}},
\oauthor{\bsnm{{Zhang}}, \binits{Y.-Q.}},
\oauthor{\bsnm{{An}}, \binits{Z.-H.}},
\oauthor{\bsnm{{Cai}}, \binits{C.}},
\oauthor{\bsnm{{Li}}, \binits{X.-Q.}},
\oauthor{\bsnm{{Li}}, \binits{Z.}},
\oauthor{\bsnm{{Liu}}, \binits{J.-C.}},
\oauthor{\bsnm{{Liu}}, \binits{Z.-K.}},
\oauthor{\bsnm{{Ma}}, \binits{X.}},
\oauthor{\bsnm{{Meng}}, \binits{Y.-Z.}},
\oauthor{\bsnm{{Peng}}, \binits{W.-X.}},
\oauthor{\bsnm{{Qiao}}, \binits{R.}},
\oauthor{\bsnm{{Shao}}, \binits{L.}},
\oauthor{\bsnm{{Song}}, \binits{L.-M.}},
\oauthor{\bsnm{{Tan}}, \binits{W.-J.}},
\oauthor{\bsnm{{Wang}}, \binits{P.}},
\oauthor{\bsnm{{Wang}}, \binits{C.-W.}},
\oauthor{\bsnm{{Wen}}, \binits{X.-Y.}},
\oauthor{\bsnm{{Xiao}}, \binits{S.}},
\oauthor{\bsnm{{Xue}}, \binits{W.-C.}},
\oauthor{\bsnm{{Yang}}, \binits{Y.-h.}},
\oauthor{\bsnm{{Yin}}, \binits{Y.}},
\oauthor{\bsnm{{Zhang}}, \binits{B.}},
\oauthor{\bsnm{{Zhang}}, \binits{F.}},
\oauthor{\bsnm{{Zhang}}, \binits{S.}},
\oauthor{\bsnm{{Zhang}}, \binits{S.-N.}},
\oauthor{\bsnm{{Zheng}}, \binits{C.}},
\oauthor{\bsnm{{Zheng}}, \binits{S.-J.}},
\oauthor{\bsnm{{Xiong}}, \binits{S.-L.}},
\oauthor{\bsnm{{Zhang}}, \binits{B.-B.}}:
{Synchrotron Radiation Dominates the Extremely Bright GRB 221009A}.
arXiv e-prints,
2303--00898
(2023)
{\href{https://arxiv.org/abs/2303.00898}{{arXiv:2303.00898}}}
{[astro-ph.HE]}.
\doiurl{10.48550/arXiv.2303.00898}
\end{botherref}
\endbibitem

\bibitem{Breit1934}
\begin{barticle}
\bauthor{\bsnm{{Breit}}, \binits{G.}},
\bauthor{\bsnm{{Wheeler}}, \binits{J.A.}}:
\batitle{{Collision of Two Light Quanta}}.
\bjtitle{Physical Review}
\bvolume{46}(\bissue{12}),
\bfpage{1087}--\blpage{1091}
(\byear{1934}).
\doiurl{10.1103/PhysRev.46.1087}
\end{barticle}
\endbibitem

\bibitem{Jauch1976}
\begin{bbook}
\bauthor{\bsnm{{Jauch}}, \binits{J.M.}},
\bauthor{\bsnm{{Rohrlich}}, \binits{F.}}:
\bbtitle{The Theory of Photons and Electrons. The Relativistic Quantum Field
  Theory of Charged Particles with Spin One-half},
(\byear{1976})
\end{bbook}
\endbibitem

\bibitem{Rees2005}
\begin{barticle}
\bauthor{\bsnm{{Rees}}, \binits{M.J.}},
\bauthor{\bsnm{{M{\'e}sz{\'a}ros}}, \binits{P.}}:
\batitle{{Dissipative Photosphere Models of Gamma-Ray Bursts and X-Ray
  Flashes}}.
\bjtitle{\apj}
\bvolume{628}(\bissue{2}),
\bfpage{847}--\blpage{852}
(\byear{2005})
{\href{https://arxiv.org/abs/astro-ph/0412702}{{arXiv:astro-ph/0412702}}}
{[astro-ph]}.
\doiurl{10.1086/430818}
\end{barticle}
\endbibitem

\bibitem{Svensson1982}
\begin{barticle}
\bauthor{\bsnm{{Svensson}}, \binits{R.}}:
\batitle{{Electron-Positron Pair Equilibria in Relativistic Plasmas}}.
\bjtitle{\apj}
\bvolume{258},
\bfpage{335}
(\byear{1982}).
\doiurl{10.1086/160082}
\end{barticle}
\endbibitem

\bibitem{Sari1995}
\begin{barticle}
\bauthor{\bsnm{{Sari}}, \binits{R.}},
\bauthor{\bsnm{{Piran}}, \binits{T.}}:
\batitle{{Hydrodynamic Timescales and Temporal Structure of Gamma-Ray Bursts}}.
\bjtitle{\apjl}
\bvolume{455},
\bfpage{143}
(\byear{1995})
{\href{https://arxiv.org/abs/astro-ph/9508081}{{arXiv:astro-ph/9508081}}}
{[astro-ph]}.
\doiurl{10.1086/309835}
\end{barticle}
\endbibitem

\bibitem{Ravasio2018}
\begin{barticle}
\bauthor{\bsnm{{Ravasio}}, \binits{M.E.}},
\bauthor{\bsnm{{Oganesyan}}, \binits{G.}},
\bauthor{\bsnm{{Ghirlanda}}, \binits{G.}},
\bauthor{\bsnm{{Nava}}, \binits{L.}},
\bauthor{\bsnm{{Ghisellini}}, \binits{G.}},
\bauthor{\bsnm{{Pescalli}}, \binits{A.}},
\bauthor{\bsnm{{Celotti}}, \binits{A.}}:
\batitle{{Consistency with synchrotron emission in the bright GRB 160625B
  observed by Fermi}}.
\bjtitle{\aap}
\bvolume{613},
\bfpage{16}
(\byear{2018})
{\href{https://arxiv.org/abs/1711.03106}{{arXiv:1711.03106}}}
{[astro-ph.HE]}.
\doiurl{10.1051/0004-6361/201732245}
\end{barticle}
\endbibitem

\bibitem{Veres2022}
\begin{barticle}
\bauthor{\bsnm{{Veres}}, \binits{P.}},
\bauthor{\bsnm{{Burns}}, \binits{E.}},
\bauthor{\bsnm{{Bissaldi}}, \binits{E.}},
\bauthor{\bsnm{{Lesage}}, \binits{S.}},
\bauthor{\bsnm{{Roberts}}, \binits{O.}},
\bauthor{\bsnm{{Fermi GBM Team}}}:
\batitle{{GRB 221009A: Fermi GBM detection of an extraordinarily bright GRB}}.
\bjtitle{GRB Coordinates Network}
\bvolume{32636},
\bfpage{1}
(\byear{2022})
\end{barticle}
\endbibitem

\bibitem{Bissaldi2022}
\begin{barticle}
\bauthor{\bsnm{{Bissaldi}}, \binits{E.}},
\bauthor{\bsnm{{Omodei}}, \binits{N.}},
\bauthor{\bsnm{{Kerr}}, \binits{M.}},
\bauthor{\bsnm{{Fermi-LAT Team}}}:
\batitle{{GRB 221009A or Swift J1913.1+1946: Fermi-LAT detection}}.
\bjtitle{GRB Coordinates Network}
\bvolume{32637},
\bfpage{1}
(\byear{2022})
\end{barticle}
\endbibitem

\bibitem{Pillera2022}
\begin{barticle}
\bauthor{\bsnm{{Pillera}}, \binits{R.}},
\bauthor{\bsnm{{Bissaldi}}, \binits{E.}},
\bauthor{\bsnm{{Omodei}}, \binits{N.}},
\bauthor{\bsnm{{La Mura}}, \binits{G.}},
\bauthor{\bsnm{{Longo}}, \binits{F.}},
\bauthor{\bsnm{{Fermi-LAT team}}}:
\batitle{{GRB 221009A: Fermi-LAT refined analysis}}.
\bjtitle{GRB Coordinates Network}
\bvolume{32658},
\bfpage{1}
(\byear{2022})
\end{barticle}
\endbibitem

\bibitem{Ursi2022}
\begin{barticle}
\bauthor{\bsnm{{Ursi}}, \binits{A.}},
\bauthor{\bsnm{{Panebianco}}, \binits{G.}},
\bauthor{\bsnm{{Pittori}}, \binits{C.}},
\bauthor{\bsnm{{Verrecchia}}, \binits{F.}},
\bauthor{\bsnm{{Longo}}, \binits{F.}},
\bauthor{\bsnm{{Parmiggiani}}, \binits{N.}},
\bauthor{\bsnm{{Tavani}}, \binits{M.}},
\bauthor{\bsnm{{Argan}}, \binits{A.}},
\bauthor{\bsnm{{Cardillo}}, \binits{M.}},
\bauthor{\bsnm{{Casentini}}, \binits{C.}},
\bauthor{\bsnm{{Evangelista}}, \binits{Y.}},
\bauthor{\bsnm{{Foffano}}, \binits{L.}},
\bauthor{\bsnm{{Menegoni}}, \binits{E.}},
\bauthor{\bsnm{{Piano}}, \binits{G.}},
\bauthor{\bsnm{{Lucarelli}}, \binits{F.}},
\bauthor{\bsnm{{Addis}}, \binits{A.}},
\bauthor{\bsnm{{Baroncelli}}, \binits{L.}},
\bauthor{\bsnm{{Bulgarelli}}, \binits{A.}},
\bauthor{\bsnm{{di Piano}}, \binits{A.}},
\bauthor{\bsnm{{Fioretti}}, \binits{V.}},
\bauthor{\bsnm{{Fuschino}}, \binits{F.}},
\bauthor{\bsnm{{Romani}}, \binits{M.}},
\bauthor{\bsnm{{Marisaldi}}, \binits{M.}},
\bauthor{\bsnm{{Pilia}}, \binits{M.}},
\bauthor{\bsnm{{Trois}}, \binits{A.}},
\bauthor{\bsnm{{Donnarumma}}, \binits{I.}},
\bauthor{\bsnm{{Giuliani}}, \binits{A.}},
\bauthor{\bsnm{{Tempesta}}, \binits{P.}},
\bauthor{\bsnm{{Agile Team}}}:
\batitle{{GRB 221009A (Swift J1913.1+1946): AGILE/MCAL detection}}.
\bjtitle{GRB Coordinates Network}
\bvolume{32650},
\bfpage{1}
(\byear{2022})
\end{barticle}
\endbibitem

\bibitem{Piano2022}
\begin{barticle}
\bauthor{\bsnm{{Piano}}, \binits{G.}},
\bauthor{\bsnm{{Verrecchia}}, \binits{F.}},
\bauthor{\bsnm{{Bulgarelli}}, \binits{A.}},
\bauthor{\bsnm{{Ursi}}, \binits{A.}},
\bauthor{\bsnm{{Panebianco}}, \binits{G.}},
\bauthor{\bsnm{{Pittori}}, \binits{C.}},
\bauthor{\bsnm{{Longo}}, \binits{F.}},
\bauthor{\bsnm{{Parmiggiani}}, \binits{N.}},
\bauthor{\bsnm{{Tavani}}, \binits{M.}},
\bauthor{\bsnm{{Argan}}, \binits{A.}},
\bauthor{\bsnm{{Cardillo}}, \binits{M.}},
\bauthor{\bsnm{{Casentini}}, \binits{C.}},
\bauthor{\bsnm{{Evangelista}}, \binits{Y.}},
\bauthor{\bsnm{{Foffano}}, \binits{L.}},
\bauthor{\bsnm{{Menegoni}}, \binits{E.}},
\bauthor{\bsnm{{Lucarelli}}, \binits{F.}},
\bauthor{\bsnm{{Addis}}, \binits{A.}},
\bauthor{\bsnm{{Baroncelli}}, \binits{L.}},
\bauthor{\bsnm{{di Piano}}, \binits{A.}},
\bauthor{\bsnm{{Fioretti}}, \binits{V.}},
\bauthor{\bsnm{{Fuschino}}, \binits{F.}},
\bauthor{\bsnm{{Romani}}, \binits{M.}},
\bauthor{\bsnm{{Marisaldi}}, \binits{M.}},
\bauthor{\bsnm{{Pilia}}, \binits{M.}},
\bauthor{\bsnm{{Trois}}, \binits{A.}},
\bauthor{\bsnm{{Donnarumma}}, \binits{I.}},
\bauthor{\bsnm{{Giuliani}}, \binits{A.}},
\bauthor{\bsnm{{Tempesta}}, \binits{P.}},
\bauthor{\bsnm{{Agile Team}}}:
\batitle{{GRB 221009A (Swift J1913.1+1946): AGILE/GRID detection}}.
\bjtitle{GRB Coordinates Network}
\bvolume{32657},
\bfpage{1}
(\byear{2022})
\end{barticle}
\endbibitem

\bibitem{Gotz2022}
\begin{barticle}
\bauthor{\bsnm{{Gotz}}, \binits{D.}},
\bauthor{\bsnm{{Mereghetti}}, \binits{S.}},
\bauthor{\bsnm{{Savchenko}}, \binits{V.}},
\bauthor{\bsnm{{Ferrigno}}, \binits{C.}},
\bauthor{\bsnm{{Bozzo}}, \binits{E.}},
\bauthor{\bsnm{{IBAS Team}}}:
\batitle{{GRB221009A/Swift J1913.1+1946: INTEGRAL SPI/ACS observations}}.
\bjtitle{GRB Coordinates Network}
\bvolume{32660},
\bfpage{1}
(\byear{2022})
\end{barticle}
\endbibitem

\bibitem{Xiao2022}
\begin{barticle}
\bauthor{\bsnm{{Xiao}}, \binits{H.}},
\bauthor{\bsnm{{Krucker}}, \binits{S.}},
\bauthor{\bsnm{{Daniel}}, \binits{R.}}:
\batitle{{GRB221009A/Swift J1913.1+1946: Solar Orbiter STIX measurements.}}
\bjtitle{GRB Coordinates Network}
\bvolume{32661},
\bfpage{1}
(\byear{2022})
\end{barticle}
\endbibitem

\bibitem{Lapshov2022}
\begin{barticle}
\bauthor{\bsnm{{Lapshov}}, \binits{I.}},
\bauthor{\bsnm{{Molkov}}, \binits{S.}},
\bauthor{\bsnm{{Mereminsky}}, \binits{I.}},
\bauthor{\bsnm{{Semena}}, \binits{A.}},
\bauthor{\bsnm{{Arefiev}}, \binits{V.}},
\bauthor{\bsnm{{Tkachenko}}, \binits{A.}},
\bauthor{\bsnm{{Lutovinov}}, \binits{A.}},
\bauthor{\bsnm{{SRG/ART-XC Team}}}:
\batitle{{GRB221009A/Swift J1913.1+1946: SRG/ART-XC observation}}.
\bjtitle{GRB Coordinates Network}
\bvolume{32663},
\bfpage{1}
(\byear{2022})
\end{barticle}
\endbibitem

\bibitem{Ripa2023}
\begin{botherref}
\oauthor{\bsnm{{Ripa}}, \binits{J.}},
\oauthor{\bsnm{{Takahashi}}, \binits{H.}},
\oauthor{\bsnm{{Fukazawa}}, \binits{Y.}},
\oauthor{\bsnm{{Werner}}, \binits{N.}},
\oauthor{\bsnm{{Munz}}, \binits{F.}},
\oauthor{\bsnm{{Pal}}, \binits{A.}},
\oauthor{\bsnm{{Ohno}}, \binits{M.}},
\oauthor{\bsnm{{Dafcikova}}, \binits{M.}},
\oauthor{\bsnm{{Meszaros}}, \binits{L.}},
\oauthor{\bsnm{{Csak}}, \binits{B.}},
\oauthor{\bsnm{{Husarikova}}, \binits{N.}},
\oauthor{\bsnm{{Kolar}}, \binits{M.}},
\oauthor{\bsnm{{Galgoczi}}, \binits{G.}},
\oauthor{\bsnm{{Breuer}}, \binits{J.-P.}},
\oauthor{\bsnm{{Hroch}}, \binits{F.}},
\oauthor{\bsnm{{Hudec}}, \binits{J.}},
\oauthor{\bsnm{{Kapus}}, \binits{J.}},
\oauthor{\bsnm{{Frajt}}, \binits{M.}},
\oauthor{\bsnm{{Rezenov}}, \binits{M.}},
\oauthor{\bsnm{{Laszlo}}, \binits{R.}},
\oauthor{\bsnm{{Koleda}}, \binits{M.}},
\oauthor{\bsnm{{Smelko}}, \binits{M.}},
\oauthor{\bsnm{{Hanak}}, \binits{P.}},
\oauthor{\bsnm{{Lipovsky}}, \binits{P.}},
\oauthor{\bsnm{{Urbanec}}, \binits{T.}},
\oauthor{\bsnm{{Kasal}}, \binits{M.}},
\oauthor{\bsnm{{Povalac}}, \binits{A.}},
\oauthor{\bsnm{{Uchida}}, \binits{Y.}},
\oauthor{\bsnm{{Poon}}, \binits{H.}},
\oauthor{\bsnm{{Matake}}, \binits{H.}},
\oauthor{\bsnm{{Nakazawa}}, \binits{K.}},
\oauthor{\bsnm{{Uchida}}, \binits{N.}},
\oauthor{\bsnm{{Bozoki}}, \binits{T.}},
\oauthor{\bsnm{{Dalya}}, \binits{G.}},
\oauthor{\bsnm{{Enoto}}, \binits{T.}},
\oauthor{\bsnm{{Frei}}, \binits{Z.}},
\oauthor{\bsnm{{Friss}}, \binits{G.}},
\oauthor{\bsnm{{Ichinohe}}, \binits{Y.}},
\oauthor{\bsnm{{Kapas}}, \binits{K.}},
\oauthor{\bsnm{{Kiss}}, \binits{L.L.}},
\oauthor{\bsnm{{Mizuno}}, \binits{T.}},
\oauthor{\bsnm{{Odaka}}, \binits{H.}},
\oauthor{\bsnm{{Takatsy}}, \binits{J.}},
\oauthor{\bsnm{{Topinka}}, \binits{M.}},
\oauthor{\bsnm{{Torigoe}}, \binits{K.}}:
{The peak-flux of GRB 221009A measured with GRBAlpha}.
arXiv e-prints,
2302--10047
(2023)
{\href{https://arxiv.org/abs/2302.10047}{{arXiv:2302.10047}}}
{[astro-ph.HE]}.
\doiurl{10.48550/arXiv.2302.10047}
\end{botherref}
\endbibitem

\bibitem{Mitchell2022}
\begin{barticle}
\bauthor{\bsnm{{Mitchell}}, \binits{L.J.}},
\bauthor{\bsnm{{Phlips}}, \binits{B.F.}},
\bauthor{\bsnm{{Johnson}}, \binits{W.N.}}:
\batitle{{GRB 221009A: Gamma-ray Detection by SIRI-2}}.
\bjtitle{GRB Coordinates Network}
\bvolume{32746},
\bfpage{1}
(\byear{2022})
\end{barticle}
\endbibitem

\bibitem{Kozyrev2022}
\begin{barticle}
\bauthor{\bsnm{{Kozyrev}}, \binits{A.S.}},
\bauthor{\bsnm{{Golovin}}, \binits{D.V.}},
\bauthor{\bsnm{{Litvak}}, \binits{M.L.}},
\bauthor{\bsnm{{Mitrofanov}}, \binits{I.G.}},
\bauthor{\bsnm{{Sanin}}, \binits{A.B.}},
\bauthor{\bsnm{{Mgns/Bepicolombo Team}}},
\bauthor{\bsnm{{Hend/Mars Odyssey Team}}},
\bauthor{\bsnm{{Benkhoff}}, \binits{J.}},
\bauthor{\bsnm{{Bepicolombo Team}}},
\bauthor{\bsnm{{Hurley}}, \binits{K.}},
\bauthor{\bsnm{{Ipn}}},
\bauthor{\bsnm{{Svinkin}}, \binits{D.}},
\bauthor{\bsnm{{Golenetskii}}, \binits{S.}},
\bauthor{\bsnm{{Frederiks}}, \binits{D.}},
\bauthor{\bsnm{{Ridnaia}}, \binits{A.}},
\bauthor{\bsnm{{Lysenko}}, \binits{A.}},
\bauthor{\bsnm{{Cline}}, \binits{T.}},
\bauthor{\bsnm{{Konus-Wind Team}}},
\bauthor{\bsnm{{von Kienlin}}, \binits{A.}},
\bauthor{\bsnm{{Zhang}}, \binits{X.}},
\bauthor{\bsnm{{Rau}}, \binits{A.}},
\bauthor{\bsnm{{Savchenko}}, \binits{V.}},
\bauthor{\bsnm{{Bozzo}}, \binits{E.}},
\bauthor{\bsnm{{Ferrigno}}, \binits{C.}},
\bauthor{\bsnm{{INTEGRAL SPI-ACS Grb Team}}},
\bauthor{\bsnm{{Barthelmy}}, \binits{S.}},
\bauthor{\bsnm{{Cummings}}, \binits{J.}},
\bauthor{\bsnm{{Krimm}}, \binits{H.}},
\bauthor{\bsnm{{Palmer}}, \binits{D.}},
\bauthor{\bsnm{{Tohuvavohu}}, \binits{A.}},
\bauthor{\bsnm{{Swift-Bat Team}}},
\bauthor{\bsnm{{Boynton}}, \binits{W.}},
\bauthor{\bsnm{{Fellows}}, \binits{C.}},
\bauthor{\bsnm{{Harshman}}, \binits{K.}},
\bauthor{\bsnm{{Enos}}, \binits{H.}},
\bauthor{\bsnm{{Starr}}, \binits{R.}},
\bauthor{\bsnm{{Gardner}}, \binits{A.S.}},
\bauthor{\bsnm{{Grs-Odyssey Grb Team}}}:
\batitle{{Improved IPN localization for GRB 221009A (BepiColombo-MGNS light
  curve)}}.
\bjtitle{GRB Coordinates Network}
\bvolume{32805},
\bfpage{1}
(\byear{2022})
\end{barticle}
\endbibitem

\bibitem{Labanti2009}
\begin{barticle}
\bauthor{\bsnm{{Labanti}}, \binits{C.}},
\bauthor{\bsnm{{Marisaldi}}, \binits{M.}},
\bauthor{\bsnm{{Fuschino}}, \binits{F.}},
\bauthor{\bsnm{{Galli}}, \binits{M.}},
\bauthor{\bsnm{{Argan}}, \binits{A.}},
\bauthor{\bsnm{{Bulgarelli}}, \binits{A.}},
\bauthor{\bsnm{{Di Cocco}}, \binits{G.}},
\bauthor{\bsnm{{Gianotti}}, \binits{F.}},
\bauthor{\bsnm{{Tavani}}, \binits{M.}},
\bauthor{\bsnm{{Trifoglio}}, \binits{M.}}:
\batitle{{Design and construction of the Mini-Calorimeter of the AGILE
  satellite}}.
\bjtitle{Nuclear Instruments and Methods in Physics Research A}
\bvolume{598}(\bissue{2}),
\bfpage{470}--\blpage{479}
(\byear{2009})
{\href{https://arxiv.org/abs/0810.1842}{{arXiv:0810.1842}}}
{[astro-ph]}.
\doiurl{10.1016/j.nima.2008.09.021}
\end{barticle}
\endbibitem

\bibitem{Bissaldi2009}
\begin{barticle}
\bauthor{\bsnm{{Bissaldi}}, \binits{E.}},
\bauthor{\bsnm{{von Kienlin}}, \binits{A.}},
\bauthor{\bsnm{{Lichti}}, \binits{G.}},
\bauthor{\bsnm{{Steinle}}, \binits{H.}},
\bauthor{\bsnm{{Bhat}}, \binits{P.N.}},
\bauthor{\bsnm{{Briggs}}, \binits{M.S.}},
\bauthor{\bsnm{{Fishman}}, \binits{G.J.}},
\bauthor{\bsnm{{Hoover}}, \binits{A.S.}},
\bauthor{\bsnm{{Kippen}}, \binits{R.M.}},
\bauthor{\bsnm{{Krumrey}}, \binits{M.}},
\bauthor{\bsnm{{Gerlach}}, \binits{M.}},
\bauthor{\bsnm{{Connaughton}}, \binits{V.}},
\bauthor{\bsnm{{Diehl}}, \binits{R.}},
\bauthor{\bsnm{{Greiner}}, \binits{J.}},
\bauthor{\bsnm{{van der Horst}}, \binits{A.J.}},
\bauthor{\bsnm{{Kouveliotou}}, \binits{C.}},
\bauthor{\bsnm{{McBreen}}, \binits{S.}},
\bauthor{\bsnm{{Meegan}}, \binits{C.A.}},
\bauthor{\bsnm{{Paciesas}}, \binits{W.S.}},
\bauthor{\bsnm{{Preece}}, \binits{R.D.}},
\bauthor{\bsnm{{Wilson-Hodge}}, \binits{C.A.}}:
\batitle{{Ground-based calibration and characterization of the Fermi gamma-ray
  burst monitor detectors}}.
\bjtitle{Experimental Astronomy}
\bvolume{24}(\bissue{1-3}),
\bfpage{47}--\blpage{88}
(\byear{2009})
{\href{https://arxiv.org/abs/0812.2908}{{arXiv:0812.2908}}}
{[astro-ph]}.
\doiurl{10.1007/s10686-008-9135-4}
\end{barticle}
\endbibitem

\bibitem{Preece2014}
\begin{barticle}
\bauthor{\bsnm{{Preece}}, \binits{R.}},
\bauthor{\bsnm{{Burgess}}, \binits{J.M.}},
\bauthor{\bsnm{{von Kienlin}}, \binits{A.}},
\bauthor{\bsnm{{Bhat}}, \binits{P.N.}},
\bauthor{\bsnm{{Briggs}}, \binits{M.S.}},
\bauthor{\bsnm{{Byrne}}, \binits{D.}},
\bauthor{\bsnm{{Chaplin}}, \binits{V.}},
\bauthor{\bsnm{{Cleveland}}, \binits{W.}},
\bauthor{\bsnm{{Collazzi}}, \binits{A.C.}},
\bauthor{\bsnm{{Connaughton}}, \binits{V.}},
\bauthor{\bsnm{{Diekmann}}, \binits{A.}},
\bauthor{\bsnm{{Fitzpatrick}}, \binits{G.}},
\bauthor{\bsnm{{Foley}}, \binits{S.}},
\bauthor{\bsnm{{Gibby}}, \binits{M.}},
\bauthor{\bsnm{{Giles}}, \binits{M.}},
\bauthor{\bsnm{{Goldstein}}, \binits{A.}},
\bauthor{\bsnm{{Greiner}}, \binits{J.}},
\bauthor{\bsnm{{Gruber}}, \binits{D.}},
\bauthor{\bsnm{{Jenke}}, \binits{P.}},
\bauthor{\bsnm{{Kippen}}, \binits{R.M.}},
\bauthor{\bsnm{{Kouveliotou}}, \binits{C.}},
\bauthor{\bsnm{{McBreen}}, \binits{S.}},
\bauthor{\bsnm{{Meegan}}, \binits{C.}},
\bauthor{\bsnm{{Paciesas}}, \binits{W.S.}},
\bauthor{\bsnm{{Pelassa}}, \binits{V.}},
\bauthor{\bsnm{{Tierney}}, \binits{D.}},
\bauthor{\bsnm{{van der Horst}}, \binits{A.J.}},
\bauthor{\bsnm{{Wilson-Hodge}}, \binits{C.}},
\bauthor{\bsnm{{Xiong}}, \binits{S.}},
\bauthor{\bsnm{{Younes}}, \binits{G.}},
\bauthor{\bsnm{{Yu}}, \binits{H.-F.}},
\bauthor{\bsnm{{Ackermann}}, \binits{M.}},
\bauthor{\bsnm{{Ajello}}, \binits{M.}},
\bauthor{\bsnm{{Axelsson}}, \binits{M.}},
\bauthor{\bsnm{{Baldini}}, \binits{L.}},
\bauthor{\bsnm{{Barbiellini}}, \binits{G.}},
\bauthor{\bsnm{{Baring}}, \binits{M.G.}},
\bauthor{\bsnm{{Bastieri}}, \binits{D.}},
\bauthor{\bsnm{{Bellazzini}}, \binits{R.}},
\bauthor{\bsnm{{Bissaldi}}, \binits{E.}},
\bauthor{\bsnm{{Bonamente}}, \binits{E.}},
\bauthor{\bsnm{{Bregeon}}, \binits{J.}},
\bauthor{\bsnm{{Brigida}}, \binits{M.}},
\bauthor{\bsnm{{Bruel}}, \binits{P.}},
\bauthor{\bsnm{{Buehler}}, \binits{R.}},
\bauthor{\bsnm{{Buson}}, \binits{S.}},
\bauthor{\bsnm{{Caliandro}}, \binits{G.A.}},
\bauthor{\bsnm{{Cameron}}, \binits{R.A.}},
\bauthor{\bsnm{{Caraveo}}, \binits{P.A.}},
\bauthor{\bsnm{{Cecchi}}, \binits{C.}},
\bauthor{\bsnm{{Charles}}, \binits{E.}},
\bauthor{\bsnm{{Chekhtman}}, \binits{A.}},
\bauthor{\bsnm{{Chiang}}, \binits{J.}},
\bauthor{\bsnm{{Chiaro}}, \binits{G.}},
\bauthor{\bsnm{{Ciprini}}, \binits{S.}},
\bauthor{\bsnm{{Claus}}, \binits{R.}},
\bauthor{\bsnm{{Cohen-Tanugi}}, \binits{J.}},
\bauthor{\bsnm{{Cominsky}}, \binits{L.R.}},
\bauthor{\bsnm{{Conrad}}, \binits{J.}},
\bauthor{\bsnm{{D'Ammando}}, \binits{F.}},
\bauthor{\bsnm{{de Angelis}}, \binits{A.}},
\bauthor{\bsnm{{de Palma}}, \binits{F.}},
\bauthor{\bsnm{{Dermer}}, \binits{C.D.}},
\bauthor{\bsnm{{Desiante}}, \binits{R.}},
\bauthor{\bsnm{{Digel}}, \binits{S.W.}},
\bauthor{\bsnm{{Di Venere}}, \binits{L.}},
\bauthor{\bsnm{{Drell}}, \binits{P.S.}},
\bauthor{\bsnm{{Drlica-Wagner}}, \binits{A.}},
\bauthor{\bsnm{{Favuzzi}}, \binits{C.}},
\bauthor{\bsnm{{Franckowiak}}, \binits{A.}},
\bauthor{\bsnm{{Fukazawa}}, \binits{Y.}},
\bauthor{\bsnm{{Fusco}}, \binits{P.}},
\bauthor{\bsnm{{Gargano}}, \binits{F.}},
\bauthor{\bsnm{{Gehrels}}, \binits{N.}},
\bauthor{\bsnm{{Germani}}, \binits{S.}},
\bauthor{\bsnm{{Giglietto}}, \binits{N.}},
\bauthor{\bsnm{{Giordano}}, \binits{F.}},
\bauthor{\bsnm{{Giroletti}}, \binits{M.}},
\bauthor{\bsnm{{Godfrey}}, \binits{G.}},
\bauthor{\bsnm{{Granot}}, \binits{J.}},
\bauthor{\bsnm{{Grenier}}, \binits{I.A.}},
\bauthor{\bsnm{{Guiriec}}, \binits{S.}},
\bauthor{\bsnm{{Hadasch}}, \binits{D.}},
\bauthor{\bsnm{{Hanabata}}, \binits{Y.}},
\bauthor{\bsnm{{Harding}}, \binits{A.K.}},
\bauthor{\bsnm{{Hayashida}}, \binits{M.}},
\bauthor{\bsnm{{Iyyani}}, \binits{S.}},
\bauthor{\bsnm{{Jogler}}, \binits{T.}},
\bauthor{\bsnm{{J{\'o}hannesson}}, \binits{G.}},
\bauthor{\bsnm{{Kawano}}, \binits{T.}},
\bauthor{\bsnm{{Kn{\"o}dlseder}}, \binits{J.}},
\bauthor{\bsnm{{Kocevski}}, \binits{D.}},
\bauthor{\bsnm{{Kuss}}, \binits{M.}},
\bauthor{\bsnm{{Lande}}, \binits{J.}},
\bauthor{\bsnm{{Larsson}}, \binits{J.}},
\bauthor{\bsnm{{Larsson}}, \binits{S.}},
\bauthor{\bsnm{{Latronico}}, \binits{L.}},
\bauthor{\bsnm{{Longo}}, \binits{F.}},
\bauthor{\bsnm{{Loparco}}, \binits{F.}},
\bauthor{\bsnm{{Lovellette}}, \binits{M.N.}},
\bauthor{\bsnm{{Lubrano}}, \binits{P.}},
\bauthor{\bsnm{{Mayer}}, \binits{M.}},
\bauthor{\bsnm{{Mazziotta}}, \binits{M.N.}},
\bauthor{\bsnm{{Michelson}}, \binits{P.F.}},
\bauthor{\bsnm{{Mizuno}}, \binits{T.}},
\bauthor{\bsnm{{Monzani}}, \binits{M.E.}},
\bauthor{\bsnm{{Moretti}}, \binits{E.}},
\bauthor{\bsnm{{Morselli}}, \binits{A.}},
\bauthor{\bsnm{{Murgia}}, \binits{S.}},
\bauthor{\bsnm{{Nemmen}}, \binits{R.}},
\bauthor{\bsnm{{Nuss}}, \binits{E.}},
\bauthor{\bsnm{{Nymark}}, \binits{T.}},
\bauthor{\bsnm{{Ohno}}, \binits{M.}},
\bauthor{\bsnm{{Ohsugi}}, \binits{T.}},
\bauthor{\bsnm{{Okumura}}, \binits{A.}},
\bauthor{\bsnm{{Omodei}}, \binits{N.}},
\bauthor{\bsnm{{Orienti}}, \binits{M.}},
\bauthor{\bsnm{{Paneque}}, \binits{D.}},
\bauthor{\bsnm{{Perkins}}, \binits{J.S.}},
\bauthor{\bsnm{{Pesce-Rollins}}, \binits{M.}},
\bauthor{\bsnm{{Piron}}, \binits{F.}},
\bauthor{\bsnm{{Pivato}}, \binits{G.}},
\bauthor{\bsnm{{Porter}}, \binits{T.A.}},
\bauthor{\bsnm{{Racusin}}, \binits{J.L.}},
\bauthor{\bsnm{{Rain{\`o}}}, \binits{S.}},
\bauthor{\bsnm{{Rando}}, \binits{R.}},
\bauthor{\bsnm{{Razzano}}, \binits{M.}},
\bauthor{\bsnm{{Razzaque}}, \binits{S.}},
\bauthor{\bsnm{{Reimer}}, \binits{A.}},
\bauthor{\bsnm{{Reimer}}, \binits{O.}},
\bauthor{\bsnm{{Ritz}}, \binits{S.}},
\bauthor{\bsnm{{Roth}}, \binits{M.}},
\bauthor{\bsnm{{Ryde}}, \binits{F.}},
\bauthor{\bsnm{{Sartori}}, \binits{A.}},
\bauthor{\bsnm{{Scargle}}, \binits{J.D.}},
\bauthor{\bsnm{{Schulz}}, \binits{A.}},
\bauthor{\bsnm{{Sgr{\`o}}}, \binits{C.}},
\bauthor{\bsnm{{Siskind}}, \binits{E.J.}},
\bauthor{\bsnm{{Spandre}}, \binits{G.}},
\bauthor{\bsnm{{Spinelli}}, \binits{P.}},
\bauthor{\bsnm{{Suson}}, \binits{D.J.}},
\bauthor{\bsnm{{Tajima}}, \binits{H.}},
\bauthor{\bsnm{{Takahashi}}, \binits{H.}},
\bauthor{\bsnm{{Thayer}}, \binits{J.G.}},
\bauthor{\bsnm{{Thayer}}, \binits{J.B.}},
\bauthor{\bsnm{{Tibaldo}}, \binits{L.}},
\bauthor{\bsnm{{Tinivella}}, \binits{M.}},
\bauthor{\bsnm{{Torres}}, \binits{D.F.}},
\bauthor{\bsnm{{Tosti}}, \binits{G.}},
\bauthor{\bsnm{{Troja}}, \binits{E.}},
\bauthor{\bsnm{{Usher}}, \binits{T.L.}},
\bauthor{\bsnm{{Vandenbroucke}}, \binits{J.}},
\bauthor{\bsnm{{Vasileiou}}, \binits{V.}},
\bauthor{\bsnm{{Vianello}}, \binits{G.}},
\bauthor{\bsnm{{Vitale}}, \binits{V.}},
\bauthor{\bsnm{{Werner}}, \binits{M.}},
\bauthor{\bsnm{{Winer}}, \binits{B.L.}},
\bauthor{\bsnm{{Wood}}, \binits{K.S.}},
\bauthor{\bsnm{{Zhu}}, \binits{S.}}:
\batitle{{The First Pulse of the Extremely Bright GRB 130427A: A Test Lab for
  Synchrotron Shocks}}.
\bjtitle{Science}
\bvolume{343}(\bissue{6166}),
\bfpage{51}--\blpage{54}
(\byear{2014})
{\href{https://arxiv.org/abs/1311.5581}{{arXiv:1311.5581}}}
{[astro-ph.HE]}.
\doiurl{10.1126/science.1242302}
\end{barticle}
\endbibitem

\bibitem{Daigne2002}
\begin{barticle}
\bauthor{\bsnm{{Daigne}}, \binits{F.}},
\bauthor{\bsnm{{Mochkovitch}}, \binits{R.}}:
\batitle{{The expected thermal precursors of gamma-ray bursts in the internal
  shock model}}.
\bjtitle{\mnras}
\bvolume{336}(\bissue{4}),
\bfpage{1271}--\blpage{1280}
(\byear{2002})
{\href{https://arxiv.org/abs/astro-ph/0207456}{{arXiv:astro-ph/0207456}}}
{[astro-ph]}.
\doiurl{10.1046/j.1365-8711.2002.05875.x}
\end{barticle}
\endbibitem

\bibitem{Panaitescu2020}
\begin{barticle}
\bauthor{\bsnm{{Panaitescu}}, \binits{A.}}:
\batitle{{X-Ray Afterglows from the Gamma-Ray Burst ``Large-angle'' Emission}}.
\bjtitle{\apj}
\bvolume{895}(\bissue{1}),
\bfpage{39}
(\byear{2020})
{\href{https://arxiv.org/abs/2005.00104}{{arXiv:2005.00104}}}
{[astro-ph.HE]}.
\doiurl{10.3847/1538-4357/ab8bdf}
\end{barticle}
\endbibitem

\bibitem{Salafia2015}
\begin{barticle}
\bauthor{\bsnm{{Salafia}}, \binits{O.S.}},
\bauthor{\bsnm{{Ghisellini}}, \binits{G.}},
\bauthor{\bsnm{{Pescalli}}, \binits{A.}},
\bauthor{\bsnm{{Ghirlanda}}, \binits{G.}},
\bauthor{\bsnm{{Nappo}}, \binits{F.}}:
\batitle{{Structure of gamma-ray burst jets: intrinsic versus apparent
  properties}}.
\bjtitle{\mnras}
\bvolume{450}(\bissue{4}),
\bfpage{3549}--\blpage{3558}
(\byear{2015})
{\href{https://arxiv.org/abs/1502.06608}{{arXiv:1502.06608}}}
{[astro-ph.HE]}.
\doiurl{10.1093/mnras/stv766}
\end{barticle}
\endbibitem

\end{thebibliography}
